%% file: main.tex
\documentclass[%
 reprint,
 amsmath,amssymb,
 aps,nofootinbib,nobibnotes,onecolumn
]{revtex4-2}

\usepackage{graphicx}
\usepackage{mathtools}
\usepackage{dcolumn}
\usepackage{bm}
\usepackage[dvipsnames, usenames]{xcolor}
\usepackage{acronym}
\usepackage[hidelinks]{hyperref}
\usepackage{physics}
\usepackage{orcidlink}
\usepackage{xspace}
\usepackage{bbm}
\usepackage{tikz}
\usetikzlibrary{bayesnet}

\newcommand{\LIGOlabMIT}{\affiliation{LIGO, Massachusetts Institute of Technology, 77 Massachusetts Avenue, Cambridge, MA 02139, USA}}
\newcommand{\MKI}{\affiliation{Kavli Institute for Astrophysics and Space Research and Department of Physics, Massachusetts Institute of Technology, 77 Massachusetts Avenue, Cambridge, MA 02139, USA}}

\def\theYear{\the\year}

\begin{document}

\setcounter{page}{1}


\title{When (not) to trust Monte Carlo approximations for hierarchical Bayesian inference}

\newcommand{\expec}[1]{\left\langle#1\right\rangle}  
\newcommand{\drm}{{\rm d}}
\newcommand{\irm}{{\rm i}}
\newcommand{\nn}{\nonumber}
\newcommand{\beq}{\begin{equation}}
\newcommand{\eeq}{\end{equation}}
\newcommand{\bdm}{\begin{displaymath}}
\newcommand{\edm}{\end{displaymath}}
\newcommand{\T}[1]{\tilde{#1}}
\newcommand{\wT}[1]{\widetilde{#1}}
\newcommand{\Cdot}{\!\cdot\!}
\newcommand{\SNR}{\textnormal{SNR}}
\newcommand{\pvtwo}{\texttt{IMRPhenomPv2}}
\newcommand{\xphm}{\texttt{IMRPhenomXPHM}}
\newcommand{\pastro}{p_{\rm astro}}
\newcommand{\pastroi}{p_{{\rm astro},i}}
\newcommand{\pastroyj}{p_{{\rm astro}, \gamma(j)}}
\newcommand{\pastroyzj}{p_{{\rm astro}, \gamma_0(j)}}
\newcommand{\mcdet}{\mathcal{M}_{\rm c,det}}
\newcommand{\chieff}{\chi_{\rm eff}}
\newcommand{\dynesty}{\textsc{dynesty}\xspace}
\newcommand{\etabar}{\overline{\eta}}
\newcommand{\muhat}{\hat{\mu}}
\newcommand{\expectlarge}[1]{\left\langle #1 \right\rangle}
\newcommand{\variancelarge}[1]{\mathbb{V}\left[#1\right]}
\newcommand{\expect}[1]{\langle #1\rangle}
\newcommand{\variance}[1]{\mathbb{V}[#1]}
\newcommand{\logsumexp}[3]{{\rm LSE}_{#2}^{#3}\left(#1\right)}
\definecolor{Gray}{gray}{0.9}
\definecolor{orange}{rgb}{0.9,0.5,0}
\newcommand{\jh}[1]{{\textcolor{magenta}{\texttt{JH: #1}}}}
\newcommand{\est}[1]{\hat{\mathfrak{E}}\left[ #1 \right]}
\newcommand{\Nobs}{N_{\rm obs}}
\newcommand{\Nexp}{N_{\rm exp}}
\newcommand{\Neff}{N_{\rm eff}}
\newcommand{\dataset}{\{d_i\}_{i=1}^{\Nobs}}
\newcommand{\Ninj}{N_{\rm inj}}
\newcommand{\Nfound}{N_{\rm found}}
\newcommand{\Nsamp}{N_{\rm samp}}
\newcommand{\NPE}{N_{\rm PE}}
\newcommand{\thresh}{\rho_{\rm thr}}
\newcommand{\fixme}[1]{\textcolor{red}{FIXME: #1}}
\newcommand{\innerp}[1]{\left\langle#1\right\rangle}
\newcommand{\ECMGF}{{\rm ECMGF}}
\newcommand{\EMGF}{{\rm EMGF}}
\newcommand{\KLD}{{\rm KL}}

\acrodef{GW}{gravitational wave}
\newcommand{\GW}{\ac{GW}\xspace}
\newcommand{\GWs}{\acp{GW}\xspace}

\acrodef{ZAMS}{zero age main sequence}
\acrodef{LVK}{LIGO-Virgo-KAGRA collaboration}

\acrodef{NS}{neutron star}
\acrodef{BH}{black hole}
\newcommand{\BH}{\ac{BH}\xspace}
\newcommand{\BHs}{\acp{BH}\xspace}

\acrodef{BBH}{binary black hole}
\newcommand{\BBH}{\ac{BBH}\xspace}
\newcommand{\BBHs}{\acp{BBH}\xspace}

\acrodef{PE}{parameter estimation}
\newcommand{\PE}{\ac{PE}\xspace}

\acrodef{HPD}{highest posterior density}
\newcommand{\HPD}{\ac{HPD}\xspace}

\acrodef{KL}{Kullback--Leibler}
\newcommand{\KL}{\ac{KL}\xspace}

\author{Jack Heinzel\,\orcidlink{0000-0002-5794-821X}}\email{heinzelj@mit.edu}\LIGOlabMIT\MKI

\author{Salvatore Vitale\,\orcidlink{0000-0003-2700-0767}}\LIGOlabMIT\MKI

\date{\today}

\begin{abstract}
The coming years of gravitational wave astrophysics promises thousands of new detections, which can unlock fundamental scientific insights if the information in each observation can be properly synthesized into a coherent picture. State-of-the-art approaches often accomplish this with hierarchical Bayesian inference. However, this typically relies on Monte Carlo approximations that are already very expensive in current data, and may become prohibitively so in the future. In this paper we show how this process can be understood from a first-principles statistical approach. We derive an error estimator $\hat{E}$ for quantifying the amount of information that is lost due to the Monte Carlo approximation and recommend that this error is limited to no more than $\hat{E} \lesssim 0.2$ bits for reliable inference. We also show that the hierarchical likelihood estimator is biased but may be corrected. Finally, we show some practical examples for inference on synthetic gravitational-wave population inference, demonstrating that simple models with strong assumptions can be much more stable to Monte Carlo uncertainty than those with weaker assumptions. We also provide a \texttt{pip} installable package \texttt{population-error} with which analysts can calculate the error statistics $\hat{E}$.
\end{abstract}

\maketitle

\input{introduction}

\input{monte_carlo_integration}
\input{robustness_measures}
\input{gaussian}

\input{gw_example}

\input{conclusion}


\appendix

\input{general_case}
\input{gaussian_appendix}
\input{rate_likelihood_appendix}

\bibliography{main}
\end{document}

%% file: introduction.tex
\section{Introduction and summary of results}

The initial detection of \acp{GW} \citep{LIGOScientific:2016aoc} opened a new and unique probe into fundamental questions in astrophysics, cosmology, nuclear physics and gravitational physics. Since then, the \ac{LVK} has confidently detected nearly 100 fully characterized observations of \acp{GW} \citep{LIGOScientific:2018mvr, LIGOScientific:2020ibl, LIGOScientific:2021usb, KAGRA:2021vkt} and promises to detect many more in the coming years \citep{KAGRA:2013rdx}. Astrophysicists use these data to probe the formation of compact objects like \acp{BH} and \acp{NS} \citep[e.g.][]{Vitale:2015tea, LIGOScientific:2016vpg, Zevin:2017evb, Mandel:2018hfr, Mandel:2021smh, KAGRA:2021duu}, cosmologists use \acp{GW} to constrain the structure and evolution of the universe \citep[e.g.][]{Schutz:1986gp, Holz:2005df, Messenger:2011gi, LIGOScientific:2017adf, Christensen:2018iqi, Ezquiaga:2022zkx, Gair:2022zsa}, nuclear scientists use merging \acp{NS} to understand the equation of state of matter at supranuclear densities \citep[e.g.][]{Baym:2017whm, Capano:2019eae, Dietrich:2020efo, Chatziioannou:2020pqz, Landry:2020vaw}, and gravitational physicists study the dynamic, strongly curved spacetimes around merging \acp{BH} to test and place limits on the fundamental theory of gravity \citep[e.g.][]{Mirshekari:2011yq, Will:2014kxa, LIGOScientific:2016lio, LIGOScientific:2018dkp, LIGOScientific:2021sio}. In all of these disciplines, modern approaches often work by extracting information from the collection of \ac{GW} observations using a powerful statistical method called hierarchical Bayesian inference \citep[e.g.][]{KAGRA:2021duu, DelPozzo:2011vcw, DelPozzo:2013ala, Isi:2019asy}. Hierarchical Bayesian inference posits an unknown underlying population from which each observation originates, and then, using the observed data, infers the underlying population by constructing a probabilistic forward model (and corresponding likelihood) for generating the observations \citep[see, e.g.][]{Essick:2023upv}.

However, the population analysis problem is technically challenging for several reasons. As the number of events included in the catalog grows, the approximations in a typical hierarchical Bayesian analysis begin to break down. For one, the noise is usually approximated as Gaussian and stationary \citep{Moore:2014lga,LIGOScientific:2019hgc}, however, the non-Gaussian and nonstationary reality can occasionally cause biases in the analysis of single events \citep{Pankow:2018qpo,Chatziioannou:2021ezd,Hourihane:2022doe,Davis:2022ird,Payne:2022spz,Ghonge:2023ksb,Macas:2023wiw,Udall:2024ovp}, which can propagate to biases in the population level analyses. Furthermore, non-astrophysical noise transients present in real noise can imitate astrophysical \acp{GW} and sometimes are significant enough to contaminate the catalog of sources included in the \ac{GW} population analysis \citep{Cabero:2019orq, Zevin:2016qwy,LIGO:2021ppb,Soni:2021cjy,Virgo:2022ysc, KAGRA:2020agh,Ashton:2021tvz,Magee:2024xiw}. Third, the results of a population inference may be strongly model dependent, and therefore misspecification of the model can cause additional biases or lead to incorrect interpretations of the population \citep[e.g.][]{Cheng:2023ddt, Alvarez-Lopez:2025ltt}.

In this paper, we focus on an additional challenge in the \ac{GW} population inference problem: we do not have practical access to the exact population-level Bayesian likelihood. 
Instead, it is common to \textit{approximate} the likelihood using a Monte Carlo approach, which estimates the many integrals appearing in the hierarchical likelihood. While this is an acceptable approximation for many simple population inference problems, it may fail for others \citep[e.g.][]{Golomb:2022bon}. 

This problem is well known in the \ac{GW} population analysis community. However the underlying statistical phenomena is still not completely understood from a first-principles approach. In summary, our work focuses on the four following results:

\begin{enumerate}
    \item \textbf{Uncertainty quantification:} We derive an error statistic $\hat{E}$ to estimate the expected amount of information lost due to the Monte Carlo approximations. $\hat{E}$ may be computed in post-processing, and we empirically verify the accuracy of this error estimator.
    \item \textbf{Unbiased likelihood estimator:} We show the \ac{GW} population analysis likelihood estimator is a biased estimator. The selection efficiency integral enters nonlinearly in the likelihood and causes the likelihood estimate to tend to be biased to larger values when the uncertainty is larger. We show this bias can be corrected with a simple correction term during inference.
    \item \textbf{Log-likelihood variance thresholds are safe...:} Using the Cauchy--Schwarz inequality, we show that the log-likelihood variance threshold conjectured in \citet{Talbot:2023pex} \textit{ensures} the error in the posterior estimator is bounded. On the other hand, we show counterexamples that thresholds based on individual Monte Carlo integrals are not necessarily safe (as used in e.g., previous \ac{LVK} population analyses~\cite{KAGRA:2021duu}).
    \item \textbf{...However, they can be stricter than necessary:} In practice, parametric models with only a few parameters are often better behaved than the log-likelihood variance would indicate, but weaker models with many parameters tend to require the strict log-likelihood variance threshold. The analyst can check the validity of an inference using the error statistic $\hat{E}$. This is related to the fact that only uncertainty in the \textit{differences} in the log-likelihood is important \citep{Farr:2019rap, Essick:2022ojx, Talbot:2023pex}.
\end{enumerate}

Our work is inspired by previous studies (\citet{Farr:2019rap, Essick:2022ojx}, and \citet{Talbot:2023pex}) to reconcile two opposing traditions for handling the likelihood uncertainty, one based on ensuring \textit{each} Monte Carlo integral is acceptable (made widespread by \citet{KAGRA:2021duu}), the other based on a combined uncertainty estimate (introduced by \citet{Essick:2022ojx}). 

\citet{Farr:2019rap} showed that whenever the uncertainty in the detection efficiency surpasses a derived threshold, \ac{GW} population inference will break down. \citet{Farr:2019rap} also pointed out that only uncertainty in the \textit{relative} likelihood is important. However, there are other sources of Monte Carlo uncertainty beyond the detection efficiency estimate in \ac{GW} population inference beyond the detection efficiency estimate, so \citet{KAGRA:2021duu} introduced additional convergence checks on each individual Monte Carlo approximation (see  Eq.~\ref{eq: single event estimator} below). \citet{Essick:2022ojx} introduced ideas about likelihood covariances and argued analysts should check the uncertainty in the \textit{relative} likelihood across the support of the posterior. \citet{Essick:2022ojx} also point out that this relative likelihood uncertainty due to the selection function asymptotically scales linearly in the number of observations, and this relative uncertainty due to other Monte Carlo integrals is asymptotically constant in the number of observations. 

\citet{Talbot:2023pex} suggest that the theoretical scaling derived by \citet{Essick:2022ojx} may not yet hold in realistic settings and recommend the adoption of a simple threshold based on the combined uncertainty in the log-likelihood. We show that the threshold proposed in \citet{Talbot:2023pex} is stricter than the requirements on the uncertainty in the relative likelihood proposed in \citet{Essick:2022ojx} and that the threshold in \citet{Talbot:2023pex} is a sufficient condition for bounding the systematic error from Monte Carlo approximations. We also verify the scalings derived by \citet{Essick:2022ojx} hold asymptotically, although we discuss caveats in Sec.~\ref{sec: robustness measures}. We suggest these scalings are useful as theoretical tools for forecasting future approximate hierarchical Bayesian inference, but should not be used as a diagnostic for reliable inference.

The central aim of this paper is to develop a single test to assess the reliability of an approximated posterior. In the \ac{GW} population analysis literature, there are two common tests in use. The first test is based on the convergence of all individual Monte Carlo integrals and can be shown to be a \textit{necessary condition} for a reliable population inference. That is, given a reliable population inference, one is guaranteed to satisfy certain ``effective sample size'' criteria.\footnote{This is not an iron-clad logical statement. The ``guarantee'' is statistical in nature and based on estimators, so it is always possible to have some extremely unlikely counterexample. \label{footnote: stat1}} However, it is possible that an unreliable posterior estimator will still satisfy these tests, as we show in Sec.~\ref{sec: gw example}.

A second approach is based on the combined uncertainty of all the Monte Carlo estimates, discussed by \citet{Talbot:2023pex}. We show in Sec.~\ref{sec: robustness measures} that this a \textit{sufficient condition} for a reliable population inference: given that one satisfies the combined uncertainty threshold, one is guaranteed to have a reliable posterior estimator up to some bound.\footnote{Again, this logical implication is only statistical in nature.\label{footnote: stat2}} However, this ``log-likelihood variance'' test is, in general, much stricter than the effective sample size test and therefore can threaten to limit the power of approximate hierarchical Bayesian inference beyond what is ultimately required. 

In this paper, we develop a single test which we argue is a \textit{sufficient and necessary condition} for a reliable posterior estimator.\footnote{See footnotes \ref{footnote: stat1} and \ref{footnote: stat2}.} We accomplish this with the error statistic $\hat{E}$. If our approximate posterior is given as $\hat{p}(\Lambda)$ and the true posterior is $p(\Lambda)$, then the error statistic estimates
\beq
\hat{E} \approx \expectlarge{\mathrm{KL}(p|\hat{p})} = \expectlarge{\int p(\Lambda) \log_2\left(\frac{p(\Lambda)}{\hat{p}(\Lambda)}\right) \dd\Lambda }, 
\label{eq: error statistic as kl}
\eeq
where the expectation is taken over the ensemble of Monte Carlo realizations of the uncertain posterior estimator, and $\mathrm{KL}$ denotes the \ac{KL} divergence. We discuss the theory behind the error statistic in Sec.~\ref{sec: robustness measures} and show examples of the performance of this test in Sec.~\ref{sec: gaussian example} and \ref{sec: gw example}.
The error statistic is defined as the sum of two terms, which we call the precision statistic and the accuracy statistic. We note that the precision statistic is proportional to the integrated uncertainty in the relative likelihood proposed initially by \citet{Essick:2022ojx} and used in \citet{Talbot:2023pex}. 

The rest of the paper is organized as follows. First, in Sec.~\ref{sec: population analysis} we review population inference and describe the popular approach for approximating the Bayesian posterior with Monte Carlo integration. In Sec.~\ref{sec: monte carlo integrals}, we briefly describe the statistical theory underlying Monte Carlo integration and why the \ac{GW} population likelihood estimator is biased. In Sec. \ref{sec: bias in population inference} we discuss bias in a general population analysis, leading into Sec. \ref{sec: robustness measures}, which introduces statistics for quantifying the uncertainty and bias in a population analysis. Section \ref{sec: gaussian example} applies the theory developed within to a simple Gaussian hierarchical inference problem, and Sec. \ref{sec: gw example} shows examples on a realistic \ac{GW} population analysis problem. We conclude in Sec. \ref{sec: conclusion}.
In this paper, angle brackets $\langle \cdot\rangle$ always denote expectation over the Monte Carlo realizations. 

\section{Gravitational wave population analysis}
\label{sec: population analysis}

Suppose we have a collection of $\Nobs$ data $\{d_i\}$ which each contains a \ac{GW} signal obscured by noise. Our task is to infer the shape of the underlying population of \ac{GW} sources. The most common approach is Bayesian, where we want to infer the Bayesian posterior probability distribution over the underlying population parameters $\Lambda$, based on the observed data. $\Lambda$ are called the \textit{hyperparameters}, and they can refer to physical parameters of the population (e.g. the stellar initial mass function or the efficiency of a common envelope ejection), or they can be purely phenomenological (e.g. the powerlaw index in the \ac{BH} mass function). 

The Bayesian posterior probability distribution is computed with Bayes' theorem
\beq
p(\Lambda | \{d_i\}) = \frac{\mathcal{L}(\{d_i\} | \Lambda) \pi(\Lambda)}{\int \mathcal{L}(\{d_i\} | \Lambda') \pi(\Lambda') \dd\Lambda'},
\label{eq: Bayes theorem}
\eeq
where $\pi(\Lambda)$ is the prior probability distribution over $\Lambda$, and the denominator is called the evidence and ensures the posterior is a normalized probability distribution. The likelihood $\mathcal{L}(\{d_i\} | \Lambda)$ is the probability of observing the dataset, conditioned on the population described by $\Lambda$.

The typical likelihood function used in \ac{GW} population analysis is 
\beq
\mathcal{L}(\{d_i\}|\Lambda) \propto \xi(\Lambda)^{-\Nobs}\prod_{i=1}^{\Nobs}p(d_i|\Lambda).
\label{eq: rate marginalized likelihood}
\eeq
and is derived in Refs.~\citep{Farr:2013yna,Mandel:2018mve,Vitale2020,Essick:2023upv}.
$p(d_i|\Lambda)$ is the probability of obtaining the data $d_i$ from the population $\Lambda$, marginalized over the uncertainty in the parameters of that \ac{GW} event $\theta$
\beq
p(d_i|\Lambda) = \int p(d_i|\theta) p(\theta|\Lambda)d\theta,
\label{eq: single event integral}
\eeq
and $\xi(\Lambda)$ is the expected fraction of events which are detected,
\beq
\xi(\Lambda) = \int p(\theta|\Lambda) p(\mathcal{D}|\theta) \mathbb{D}[\mathcal{D}] \dd\theta \dd\mathcal{D},
\label{eq: selection efficiency}
\eeq
where $\mathbb{D}[\mathcal{D}]$ is an indicator function which is 0 when the data $\mathcal{D}$ is undetected and 1 when the data is detected.

However, the integral quantities $\xi(\Lambda)$ and $p(d_i|\Lambda)$ are usually impossible to compute in closed form, and so instead are estimated with a Monte Carlo approach (we discuss the statistical theory of Monte Carlo integration in Section~\ref{sec: monte carlo integrals}). Specifically, we often use the approximation
\beq
\hat{\mathcal{L}}(\{d\}|\Lambda) \propto \hat{\xi}(\Lambda)^{-\Nobs}\prod_{i=1}^{\Nobs} \hat{p}(d_i|\Lambda),
\label{eq: estimated hierarchical likelihood}
\eeq
where the single event integral quantities are often estimated with 
\beq
\hat{p}(d_i|\Lambda) \propto \frac{1}{\NPE}\sum_{j=1}^{\NPE}\frac{p(\theta_{ij}|\Lambda)}{\pi(\theta_{ij}|{\rm PE})}\bigg|_{\theta_{ij}\sim p(\theta|d_i)}
\label{eq: single event estimator}
\eeq
with the $\theta_{ij}$ drawn from the \PE posterior $p(\theta|d_i)$ using the prior $\pi(\theta|{\rm PE})$; $1\le i\le \Nobs$ indexes the event number and $1 \le j \le \NPE$ indexes the individual event samples. We ignore the constant evidence term $\int p(d_i|\theta)\pi(\theta|{\rm PE})\dd\theta$, as constant factors in the likelihood are normalized out in the Bayesian posterior. 

The selection efficiency is estimated with a large number of simulated sources $\{\theta_j\}_{j=1}^{\Ninj} \sim p(\theta|{\rm draw})$ drawn from a fiducial population. The signals are then added to noise from the \GW detectors. Then, searches \citep[e.g.][]{Allen:2005fk, Usman:2015kfa, Messick:2016aqy} are run over these data, detecting some subset of the simulated events. We estimate the selection efficiency by \citep{Tiwari:2017ndi, Farr:2019rap}
\beq
\hat{\xi}(\Lambda) = \frac{1}{\Ninj}\sum_{j=1}^{\Ninj}\mathbb{D}[d_j]\frac{p(\theta_j|\Lambda)}{p(\theta_j|{\rm draw})} = \frac{1}{\Ninj}\sum_{j=1}^{\Nfound}\frac{p(\theta_j|\Lambda)}{p(\theta_j|{\rm draw})},
\label{eq: selection effects estimator}
\eeq
where the indicator $\mathbb{D}[d_j]$ is 1 when the data segment $d_j$ is detected by the pipelines, and 0 when it is not. Equivalently the sum may be performed over only the found injections. 

We will show in Sec.~\ref{sec: monte carlo integrals} that Eqs.~\ref{eq: single event estimator} and \ref{eq: selection effects estimator} are unbiased estimators for the marginalized $p(d_i|\Lambda)$ in Eq.~\ref{eq: single event integral} and for the selection efficiency integral in Eq.~\ref{eq: selection efficiency}, respectively. A product of uncorrelated unbiased estimators is also an unbiased estimator, and so the bias in the likelihood is not due to the approximation in the single event integrals but due to how the selection efficiency enters the likelihood.
We show in Section~\ref{subsec: monte carlo bias} that running any estimator through a nonlinear function, such as $\hat{\xi}(\Lambda)^{-\Nobs}$, produces a biased estimator. In particular, this means the likelihood estimator is biased, as we discuss further below.

%% file: monte_carlo_integration.tex
\section{Monte Carlo Integration}
\label{sec: monte carlo integrals}

Equations~\ref{eq: single event estimator} and~\ref{eq: selection effects estimator} are examples of Monte Carlo integrals. In this section, we discuss the basic theory of Monte Carlo integration, the uncertainty in estimators, and biased and unbiased estimation.

Monte Carlo integration is a technique for estimating the value of an integral stochastically. Take many random draws from the domain of integration using a known probability density which is easy to sample, then reweight to the integrand function. Specifically, given an integral 
\beq
I = \int f(\theta) \dd\theta,
\eeq
and $M$ independent draws from a distribution $\theta_i \sim p(\theta)$, the Monte Carlo estimate of the integral is (see chapter 29 of \citet{MacKay2003})
\beq
\hat{I} = \frac{1}{M}\sum_{i=1}^M \frac{f(\theta_i)}{p(\theta_i)}.
\eeq
Monte Carlo integration is a general technique for estimating nonelementary or high dimensional integrals, because the uncertainty in the estimate is not explicitly dependent on the dimension of the domain of the integral. 

We can compute the expectation of the estimator
\beq
\langle\hat{I}\rangle = \int \left(\frac{1}{M}\sum_{i=1}^M \frac{f(\theta_i)}{p(\theta_i)}\right) p(\theta_1) \dd\theta_1 p(\theta_2) \dd\theta_2 ... p(\theta_M) \dd\theta_M  = \frac{1}{M}\sum_{i=1}^M \int \left(\frac{f(\theta_i)}{p(\theta_i)}\right)p(\theta_1)\dd\theta_1 p(\theta_2) \dd\theta_2 ... p(\theta_M) \dd\theta_M
\eeq
using the fact that integration is linear. Notice that for the $i^{\rm th}$ term in the sum, the integrand is independent of all but one of the integration variables $\theta_i$, and all others integrate away to 1. Therefore, 
\beq
\langle\hat{I}\rangle = \frac{1}{M}\sum_{i=1}^M \int \left(\frac{f(\theta_i)}{p(\theta_i)}\right)p(\theta_i)\dd\theta_i = \frac{1}{M}\sum_{i=1}^M \int f(\theta_i) \dd\theta_i = \int f(\theta)\dd\theta = I.
\label{eq: expectation of Monte Carlo integral}
\eeq
Because the expectation of the Monte Carlo integral is correct ($\expect{\hat{I}} = I$), we call the Monte Carlo estimator $\hat{I}$ an \textit{unbiased} estimator for $I$. We should note, however, that the simplification in Eq.~\ref{eq: expectation of Monte Carlo integral} only works if $p(\theta) > 0$ wherever $f(\theta) \neq 0$. Otherwise, the estimator is in general \textit{biased}. You cannot re-weight a distribution outside its domain.

Using a similar approach, we can compute the variance of the estimator as 
\beq
\variance{\hat{I}} = \langle(\hat{I} - \expect{I})^2\rangle = \frac{1}{M}\left(\int \frac{f(\theta)^2}{p(\theta)}\dd\theta - I^2\right),
\label{eq: variance of Monte Carlo integral}
\eeq
which is just a constant value suppressed by a factor of $M^{-1}$. This gives the Monte Carlo integral the attractive property that the uncertainty can be decreased as much as we like, simply by drawing more samples. How \textit{much} we must increase the number of samples, however, depends sensitively on how similar $p(\theta)$ is to $f(\theta)$: the integral in Eq.~\ref{eq: variance of Monte Carlo integral} can be extremely large or even infinite for poorly chosen $p(\theta)$. Indeed, there is a similar issue as Eq.~\ref{eq: expectation of Monte Carlo integral}. If there is a finite-measure subset of the domain where $f(\theta) \neq 0$ but $p(\theta) = 0$, then the variance diverges \textit{for all} $M$. Even if the domain where $f(\theta) \neq 0$ also has $p(\theta) \neq 0$, if $p(\theta)$ becomes too small faster than $f(\theta)$, the variance may still diverge.



Just as we are able to estimate $\hat{I} \approx I$ with Monte Carlo integration, we can estimate $\mathbb{V}[\hat{I}]$ with
\beq
\hat{\sigma}^2_{I} = \frac{1}{M} \left( \frac{1}{M-1} \sum\limits_i^M \left(\frac{f(\theta_i)}{p(\theta_i)} - \hat{I} \right)^2 \right) = \frac{1}{M-1}\left(\frac{1}{M}\sum_{i=1}^M\frac{f(\theta_i)^2}{p(\theta_i)^2} - \hat{I}^2\right),
\label{eq: variance estimator}
\eeq
which is also an unbiased estimator: $\expect{\hat{\sigma}^2_I} = \mathbb{V}[\hat{I}]$ (this fact can be shown by exactly the same procedure as Eq.~\ref{eq: expectation of Monte Carlo integral}), provided the variance exists. Note the denominator $M-1$ rather than $M$---known as Bessel's correction---is required for the estimator to be unbiased \citep{kendall1977advanced}.
Since Eqs.~\ref{eq: single event estimator} and~\ref{eq: selection effects estimator} are Monte Carlo integrals, they are unbiased estimators for the corresponding integrals: $\langle\hat{p}(d_i|\Lambda)\rangle = p(d_i|\Lambda)$ and $\langle\hat{\xi}(\Lambda)\rangle = \xi(\Lambda)$.\footnote{To be precise, $\hat{\xi}(\Lambda)$ should be thought of as a Monte Carlo integral over the space of all data $d$ and sources $\theta$ as opposed to a Monte Carlo integral over only the space of sources $\theta$. Using the expression in the center of Eq.~\ref{eq: selection effects estimator}, the expectation can be taken directly over the sources $d_j, \theta_j$, without accounting for the data-dependent $N_{\rm found}$.}

However, while the variance in each Monte Carlo integral is important, a more appropriate quantity to track is the total variance in the log-likelihood estimator in Eq.~\ref{eq: estimated hierarchical likelihood}. The variance in each Monte Carlo ingredient propagates to the total log-likelihood variance as
\beq
\hat{\sigma}^2_{\ln\mathcal{L}}(\Lambda) = \left(\sum_{i=1}^{\Nobs}\frac{\hat{\sigma}^2_{p_i}}{\hat{p}(d_i|\Lambda)^2}\right) + \frac{\Nobs^2\hat{\sigma}^2_{\xi}}{\hat{\xi}(\Lambda)^2},
\label{eq: likelihood variance estimator}
\eeq
where $\hat{\sigma}^2_{p_i}$ is the variance calculated from Eq.~\ref{eq: variance estimator} for $\hat{p}(d_i|\Lambda)$ and $\hat{\sigma}^2_{\xi}$ is the variance for $\hat{\xi}(\Lambda)$.
\citet{Talbot:2023pex} conjecture that thresholding on this variance estimator ensures a safe hyperposterior. By using a generalization of Eq.~\ref{eq: likelihood variance estimator}---the covariance in the log-likelihood at two hyperparameters $\Lambda$ and $\Lambda'$, first introduced by \citet{Essick:2022ojx}---we show in Sec.~\ref{sec: robustness measures} that the log-likelihood variance threshold indeed ensures a safe hyperposterior.

\subsection{Nonlinear functions of a Monte Carlo integral}
\label{subsec: monte carlo bias}

In the hierarchical likelihood for a population inference (Eq.~\ref{eq: rate marginalized likelihood}), we must multiply by a selection efficiency integral, raised to the negative power of the number of events in the catalog $\Nobs$. That is, we run the estimator $\hat{\xi}(\Lambda)$ through the function $g(x) = x^{-\Nobs}$. We may write the expectation of this nonlinear function of the Monte Carlo integral
\beq
\expect{g(\hat{I})} = \expectlarge{\hat{I}^{-\Nobs}} = \int \left(\frac{1}{M}\sum_{i=1}^M \frac{f(\theta_i)}{p(\theta_i)}\right)^{-\Nobs} p(\theta_1) \dd\theta_1 p(\theta_2) \dd\theta_2 ... p(\theta_M) \dd\theta_M \neq I^{-\Nobs}
\eeq
which cannot be simplified using the techniques from before. In fact, this is generically a \textit{biased} estimator for $I^{-\Nobs}$, and since $g(x) = x^{-\Nobs}$ is a convex function for $x > 0$, then, by Jensen's inequality, the estimator is biased \textit{too large}: $\expect{\hat{I}^{-\Nobs}} > I^{-\Nobs}$ \citep{kendall1977advanced, MacKay2003}. 

We can approximate the bias by a Taylor expansion to leading order. Expanding each weight $f(\theta_i) / p(\theta_i)$ around the expectation of that weight, $\expect{f(\theta_i) / p(\theta_i)} = I$, 
\beq
\expect{g(\hat{I})} \approx \int \left[g(I) + \frac{g'(I)}{M} \sum_{i=1}^M \left(\frac{f(\theta_i)}{p(\theta_i)} - I\right)  + \frac{g''(I)}{2M^2}\sum_{i=1}^M \left(\frac{f(\theta_i)}{p(\theta_i)} - I\right)^2 + \mathcal{O}\left(\frac{1}{M^2}\right)\right] p(\theta_1) \dd\theta_1 p(\theta_2) \dd\theta_2 ... p(\theta_M) \dd\theta_M,
\eeq
where $g'(I)$ and $g''(I)$ are the first and second derivatives of $g$, evaluated at $I$. Using the same techniques as in Eq.~\ref{eq: expectation of Monte Carlo integral} and Eq.~\ref{eq: variance of Monte Carlo integral}, we can simplify to
\beq
\expect{g(\hat{I})} \approx g(I) + \frac{g''(I)}{2}\variance{\hat{I}} + \mathcal{O}\left(\frac{1}{M^2}\right) = I^{-\Nobs}\left[1 + \frac{\Nobs(\Nobs + 1)}{2}\frac{\variance{\hat{I}}}{I^2} + \mathcal{O}\left(\frac{1}{M^2}\right)\right].
\label{eq: bias for mean raised to power}
\eeq

\begin{figure}
    \centering
    \includegraphics[width=\linewidth]{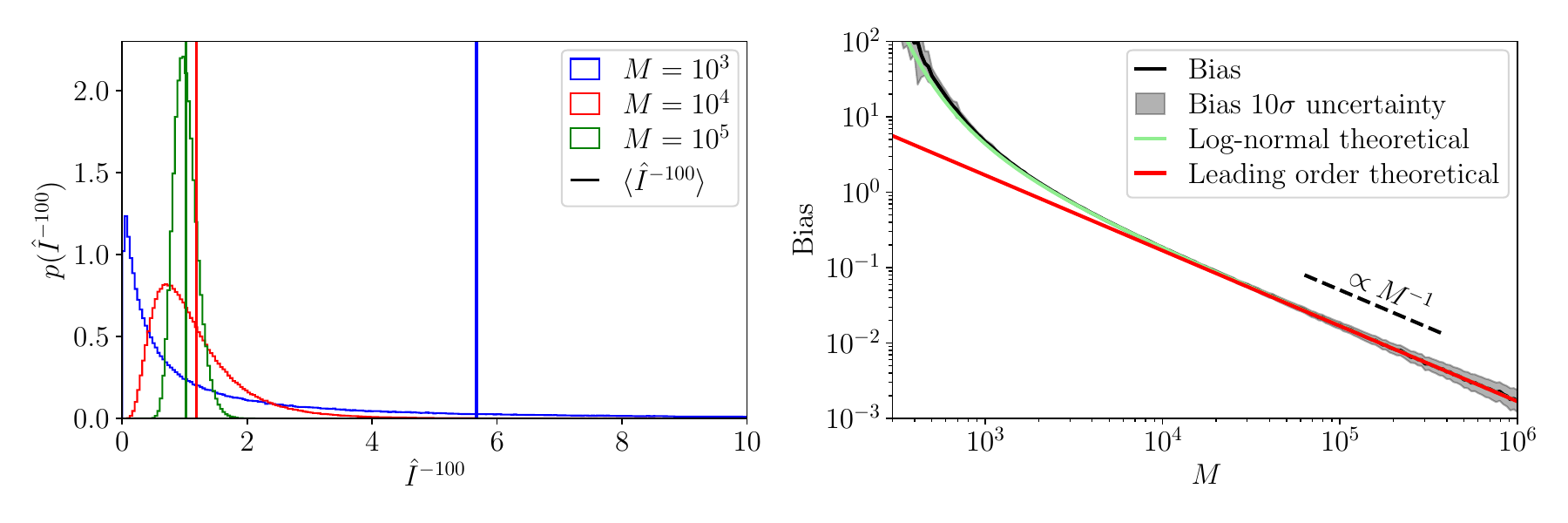}
    \caption{
    \textit{Left}: Histograms of $10^6$ draws of a $\hat{I}^{-100}$ estimated from reweighting a sample of $M$ values from a broader exponential distribution, for various values of $M$. The solid vertical lines show the mean of these draws, $\expect{\hat{I}^{-100}}$. The mean becomes severely biased for small $M$, as the distribution of $\hat{I}^{-100}$ becomes strongly skewed. 
    \textit{Right}: The bias $\expect{\hat{I}^{-100}} - I^{-100}$ as a function of $M$. In black is the empirical bias computed from an average of $10^6$ estimates with sample size $M$, with the $10\sigma$ (so the envelope is visible) shaded uncertainty envelope. The uncertainty is due to averaging with finite number statistics ($10^6$). In green is the theoretical bias assuming the $\hat{I}$ is distributed according to a log-normal, which closely matches the empirically calculated bias, and in red is the leading order theoretical bias, which does not assume an underlying distribution for $\hat{I}$ but deviates at small values of $M$ (see Eq.~\ref{eq: expectation of exponential biased estimator}).
    }
    \label{fig: exponential powers histograms}
\end{figure}

The approach of Eq.~\ref{eq: bias for mean raised to power} does not assume anything about the distribution for $\hat{I}$ beyond that the mean and variance exist. However, we can make a more reasonable assumption about the distribution of $\hat{I}$, namely, that it is distributed as a normal distribution by the central limit theorem. If we assume $M$ is sufficiently large, we can assume that the normal distribution is narrow, and thus can also be approximated as a narrow log-normal distribution. It is easy to raise a log-normal distribution to an arbitrary power, and under this assumption one obtains that (also shown in \citet{Essick:2022ojx})\footnote{Another advantage of a log-normal distribution is that it only has support for positive values, and we know the estimator must be positive.}
\beq
\expect{\hat{I}^{-\Nobs}} \approx I^{-\Nobs}\exp\left(\frac{\Nobs(\Nobs + 1)}{2}\frac{\variance{\hat{I}}}{I^2}\right).
\label{eq: bias for mean raised to power assuming lognormal}
\eeq
For an illustrative example, consider the simple case where we want to estimate the integral of an exponential function raised to the $\Nobs=-100$ power
\beq
I_{\rm expo}^{-100} = \left(\int_0^\infty e^{-x} \dd x\right)^{-100} = 1.
\eeq
Pretending that we don't already know the analytic value, let's use a Monte Carlo estimator with draws from a \textit{broader} exponential $x_i \sim p(x) = \frac{1}{2}e^{-x/2}$,
\beq
\hat{I}_{\rm expo}^{-100} = \left(\frac{1}{M}\sum_{i=1}^M \frac{2e^{-x_i}}{e^{-x_i/2}}\right)^{-100} = \left(\frac{1}{M}\sum_{i=1}^M 2e^{-x_i/2}\right)^{-100}.
\label{eq: exponential reweighting scheme}
\eeq
We can compute the variance in closed form, which gives us the expectation
\beq
\variance{\hat{I}_{\rm expo}} = \frac{1}{3M} \implies \expect{\hat{I}_{\rm expo}^{-100}} = \begin{dcases}
1 + \dfrac{5050}{3M} + \mathcal{O}\left(\dfrac{1}{M^2}\right) & \text{to leading order} \\
\exp\left(\dfrac{5050}{3M}\right) & \text{assuming log-normal}
\end{dcases}
\label{eq: expectation of exponential biased estimator}
\eeq
To verify these approximations, we show the histograms of many random realizations using different fixed values of $M$ in Fig.~\ref{fig: exponential powers histograms} and a comparison of the empirical expectation with the theoretical expectation of this biased estimator (Eq.~\ref{eq: expectation of exponential biased estimator}) as a function of $M$. 

\subsection{Correction Factor}

We can correct for the bias due to the nonlinearity using a correction factor, which we derive in general in Appendix~\ref{app: correction in general}, under the restrictions that the improved estimator should remain positive everywhere, and remove the leading order bias. For the function $g(x) = x^{-\Nobs}$, the corrected estimator for $I^{-\Nobs}$ is
\beq
\hat{I}_{\rm corr}^{-\Nobs} \equiv \hat{I}^{-\Nobs}\exp(-\frac{\Nobs(\Nobs+1)}{2}\frac{\hat{\sigma}^2_{I}}{\hat{I}^2}).
\label{eq: corrected integral power estimator}
\eeq

\begin{figure}
    \centering
    \includegraphics[width=\linewidth]{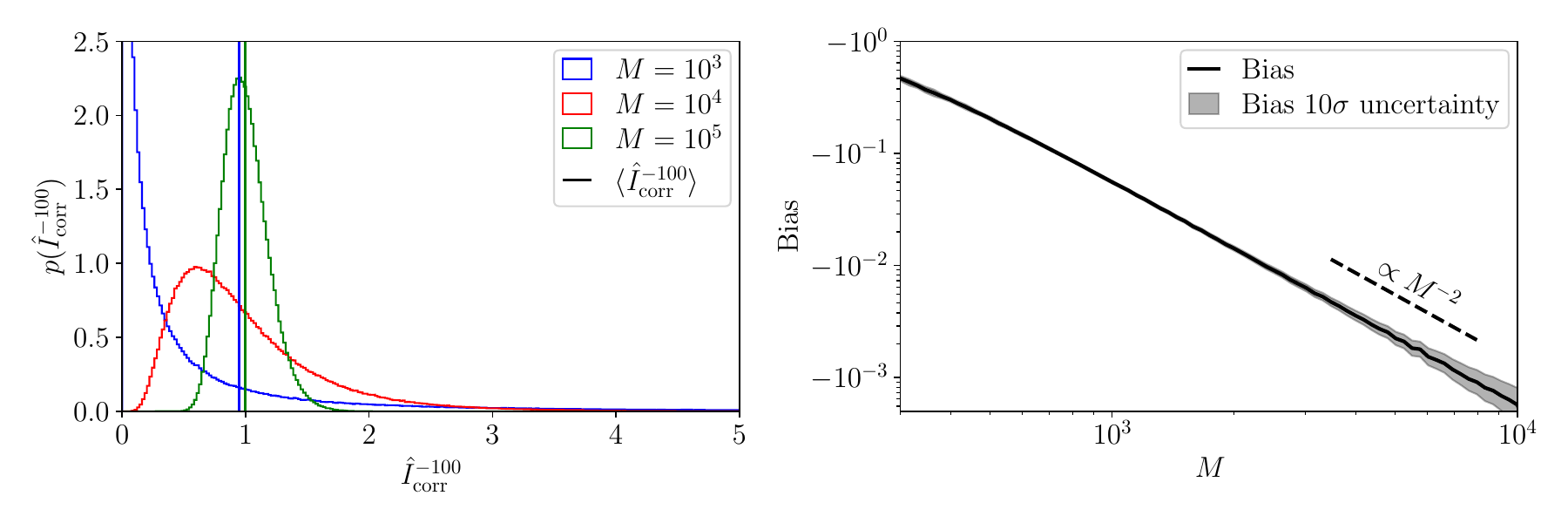}
    \caption{
    Analogous to Fig.~\ref{fig: exponential powers histograms} except using the corrected estimator $\hat{I}_{\rm corr}$. Note the bias is now negative and the reduced axes relative to Fig.~\ref{fig: exponential powers histograms}. We show the bias now scales as $\mathcal{O}(M^{-2})$ as expected.
    }
    \label{fig: corrected exponential powers histograms}
\end{figure}

Having removed the leading order bias at $\mathcal{O}(M^{-1})$, the remaining bias occurs at $\mathcal{O}(M^{-2})$. We show the analogous figure to Fig.~\ref{fig: exponential powers histograms} in Fig.~\ref{fig: corrected exponential powers histograms} and verify the remaining bias scales as expected. 

\section{Bias in population inference}
\label{sec: bias in population inference}


\subsection{Likelihood bias and correction}

We saw in Section~\ref{subsec: monte carlo bias} that running an \textit{unbiased} Monte Carlo estimator through a nonlinear function produces a biased estimator. In \ac{GW} population analyses, the likelihood is estimated by raising the selection effects estimator to the $-\Nobs$ power: $\hat{\xi}(\Lambda)^{-\Nobs}$.
Without any correction, this means the likelihood estimator is biased as in Eqs.~\ref{eq: bias for mean raised to power} and \ref{eq: bias for mean raised to power assuming lognormal}. 

For an unbiased likelihood estimator, we need to include the following modifying term.\footnote{Note the similarity but opposite sign to the modifying term proposed \citet{Farr:2019rap}. Both approaches compute a bias term $\sim \exp(\Nobs^2/2\Neff)$, though \citet{Farr:2019rap} suggests including this term explicitly as a marginalization over the uncertainty. That is, the correction in \citet{Farr:2019rap} is analogous to the bias term in Eq.~\ref{eq: bias for mean raised to power assuming lognormal}. We suggest that the expectation of the likelihood should be unbiased, and so we \textit{undo} this term, hence it obtains a minus sign in Eq.~\ref{eq: unbiased likelihood estimator}.}
\beq
\hat{\mathcal{L}}_{\rm corr}(\{d_i\} | \Lambda) \equiv \hat{\xi}(\Lambda)^{-\Nobs}\exp\left(-\frac{1}{2}\frac{\Nobs(\Nobs+1)\hat{\sigma}_\xi^2(\Lambda)}{\hat{\xi}(\Lambda)^2}\right)\prod_{i=1}^{\Nobs}\hat{p}(d_i|\Lambda)
\label{eq: unbiased likelihood estimator}
\eeq
where $\hat{\sigma}^2_{\xi}$ is the estimated variance in the Monte Carlo integral for $\hat{\xi}$
\beq
\hat{\sigma}^2_\xi = \frac{1}{\Ninj-1}\left(\frac{1}{\Ninj}\sum_{j=1}^{\Ninj}\frac{p(\theta_j|\Lambda)^2}{p(\theta_j|{\rm draw})^2} - \hat{\xi}(\Lambda)^2\right).
\eeq
With the inclusion of the modifying term, the bias in the likelihood estimator is removed to leading order (that is, it is now at order $\Ninj^{-2}$ rather than $\Ninj^{-1}$). It does not reduce the precision of the estimator meaningfully, and so represent a simple improvement to the likelihood estimator. For the mathematical details on this modifying term, see the Appendices \ref{app: bias in general} and \ref{app: correction in general}.

Now since the estimators $\hat{\xi}(\Lambda)$ and $\hat{p}(d_i|\Lambda)$ are all uncorrelated---the Monte Carlo samples are from unrelated distributions---the expectation over Monte Carlo samples of the likelihood estimator in Eq.~\ref{eq: unbiased likelihood estimator} factors and is now unbiased to $\mathcal{O}(\Ninj^{-2})$
\beq
\expectlarge{\hat{\mathcal{L}}_{\rm corr}(\{d_i\} | \Lambda)} = \expectlarge{\hat{\xi}(\Lambda)^{-\Nobs}\exp\left(-\frac{1}{2}\frac{\Nobs(\Nobs+1)\hat{\sigma}_\xi^2(\Lambda)}{\hat{\xi}(\Lambda)^2}\right)}\prod_{i=1}^{\Nobs}\expect{\hat{p}(d_i|\Lambda)} = \mathcal{L}(\{d_i\}|\Lambda)[1 + \mathcal{O}(\Ninj^{-2})].
\eeq

\subsection{Posterior bias}

However, the issue with the bias extends beyond the likelihood. Even if the likelihood estimator is unbiased, the \textit{posterior} can be biased. This is for a similar reason as the bias in the likelihood: the posterior is a nonlinear function of the likelihood, and so marginalizing the posterior over the uncertainty in the likelihood will leave some residual bias. Mathematically, this is due to the nonlinearity in the normalization and covariance in the likelihood estimator between multiple locations in parameter space $\Lambda, \Lambda'$. Denoting $\mathcal{L}(\{d_i\}|\Lambda) = \mathcal{L}(\Lambda)$ and the scatter $\hat{\mathcal{L}}(\Lambda)/\mathcal{L}(\Lambda) -1 \equiv \hat{\delta}(\Lambda)$,
\begin{align}
\expect{\hat{p}(\Lambda|\{d_i\})} 
&= \expectlarge{\frac{\pi(\Lambda)\mathcal{L}(\Lambda)[1 + \hat{\delta}(\Lambda)]}{\int \dd\Lambda' \pi(\Lambda')\mathcal{L}(\Lambda')[1 + \hat{\delta}(\Lambda')]}} \label{eq: bias in posterior heuristic} \\
&\approx p(\Lambda)\left(1 + \expect{\hat{\delta}(\Lambda)} - \int \dd\Lambda' p(\Lambda')\expect{\hat{\delta}(\Lambda')} - \int \dd\Lambda' p(\Lambda')\expect{\hat{\delta}(\Lambda)\hat{\delta}(\Lambda')} + \int \dd\Lambda'\dd\Lambda''p(\Lambda')p(\Lambda'')\expect{\hat{\delta}(\Lambda')\hat{\delta}(\Lambda'')}\right) \nonumber
\end{align}
where we have defined $p(\Lambda) = \pi(\Lambda)\mathcal{L}(\Lambda) / \int \dd\Lambda' \pi(\Lambda')\mathcal{L}(\Lambda')$ as the true posterior, and we have used the binomial expansion to second order for the denominator. 
$\expect{\hat{\delta}(\Lambda)}$ is proportional to the bias in the likelihood, and is zero for an unbiased likelihood (or $\mathcal{O}(\Ninj^{-2})$ for the corrected likelihood in Eq.~\ref{eq: unbiased likelihood estimator}). Notice there are additional biasing terms in the posterior estimator, due to $\expect{\hat{\delta}(\Lambda)\hat{\delta}(\Lambda')}$ terms or two-point \textit{covariance} terms between different locations in parameter space. 

It may seem surprising that we should worry about nonlinearity in a normalization term that only appears implicitly in most applications. We discuss this in Appendix~\ref{app: normalization discussion}.

We derive the general bias to leading order in the Appendix~\ref{app: bias in general}. For the \ac{GW} population inference problem approximated with Monte Carlo integration (Eq.~\ref{eq: estimated hierarchical likelihood}), the bias $b$ is such that
\beq
\expect{\hat{p}(\Lambda | \{d_i\})} = p(\Lambda | \{d_i\})\left(1 +b(\Lambda) - \int b(\Lambda')\dd\Lambda' \right)
\eeq
where $b(\Lambda)$ varies across parameter space and the second term simply ensures the posterior remains normalized. To leading order, 
\beq
b(\Lambda) = \frac{1}{2}\frac{\Nobs(\Nobs+1)\mathbb{V}[\hat{\xi}(\Lambda)]}{\xi(\Lambda)^2} - \int C_{\ln\mathcal{L}}(\Lambda, \Lambda')p(\Lambda'|\{d_i\})\dd\Lambda'.
\label{eq: posterior bias}
\eeq
The first term is the bias from the nonlinearity at which the selection effects estimator enters the likelihood; $\mathbb{V}[\hat{\xi}(\Lambda)]$ is the variance in the estimator $\hat{\xi}(\Lambda)$. As shown in detail in Appendix~\ref{app: bias in general}, the second term is the \textit{posterior} bias where $C_{\ln\mathcal{L}}(\Lambda, \Lambda')$ is the covariance in the log-likelihood estimator between points $\Lambda$ and $\Lambda'$, and for the \ac{GW} population inference problem, may be estimated with
\beq
\hat{C}_{\ln\mathcal{L}}(\Lambda, \Lambda') = \sum_{i=1}^{\Nobs}\frac{\hat{C}_{p_i}(\Lambda, \Lambda')}{\hat{p}(d_i|\Lambda)\hat{p}(d_i|\Lambda')}
+ \frac{\Nobs^2\hat{C}_{\xi}(\Lambda, \Lambda')}{\hat{\xi}(\Lambda)\hat{\xi}(\Lambda')}
\label{eq: population covariance with selection}    
\eeq
where 
\beq
\hat{C}_{p_i}(\Lambda, \Lambda') = \frac{1}{\NPE - 1}\left(\left[\frac{1}{\NPE}\sum_{j=1}^{\NPE}\frac{p(\theta_{ij}|\Lambda)}{\pi(\theta_{ij}|{\rm PE})}\frac{p(\theta_{ij}|\Lambda')}{\pi(\theta_{ij}|{\rm PE})}\right] - \hat{p}(d_i|\Lambda)\hat{p}(d_i|\Lambda')\right)
\label{eq: single event covariance estimator}
\eeq
is the covariance between the estimators $\hat{p}(d_i|\Lambda)$ and $\hat{p}(d_i|\Lambda')$ and
\beq
\hat{C}_{\xi}(\Lambda, \Lambda') = \frac{1}{\Ninj - 1}\left(\left[\frac{1}{\Ninj}\sum_{j=1}^{\Nfound}\frac{p(\theta_j|\Lambda)}{p(\theta_j|{\rm draw})}\frac{p(\theta_j|\Lambda')}{p(\theta_j|{\rm draw})}\right] - \hat{\xi}(\Lambda)\hat{\xi}(\Lambda')\right)
\label{eq: selection covariance estimator}
\eeq
is the covariance between the selection efficiency estimators $\hat{\xi}(\Lambda)$ and $\hat{\xi}(\Lambda')$.
These covariances come from the fact that each estimator is using the \textit{same} sets of samples to reweight across $\Lambda$ parameter space (also see \citet{Essick:2022ojx} and \citet{Talbot:2023pex}).

\begin{figure}
    \centering
    \includegraphics[width=0.8\linewidth]{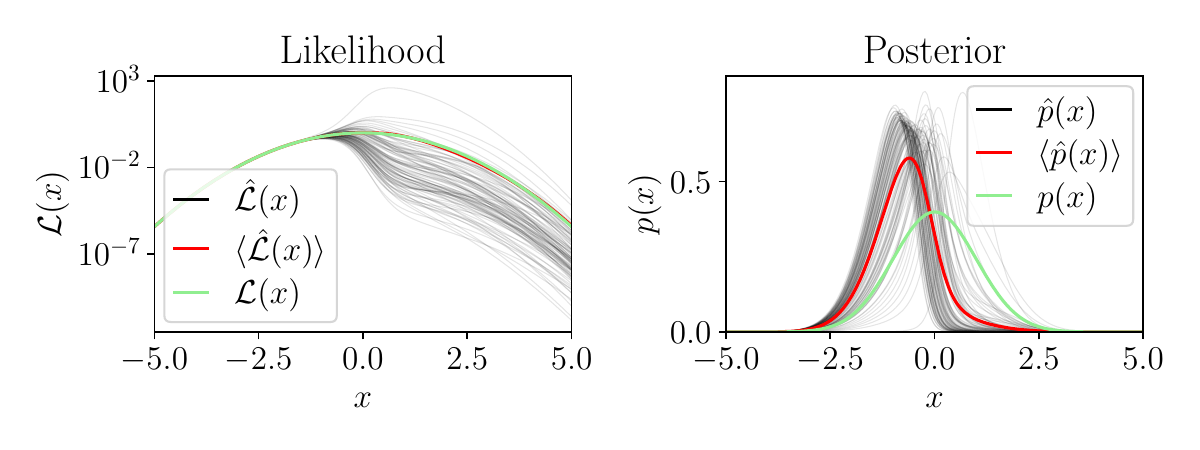}
    \caption{A schematic for how an unbiased likelihood estimator leads to a biased posterior estimator. We draw likelihood scatter from a zero mean log-Gaussian process where the uncertainty is small for negative $x$ and large for positive $x$. Each black curve represents a likelihood (left) or normalized posterior (right) estimator, the red curve is the mean of the estimators and the green curve is the true likelihood and posterior. The likelihood estimator is unbiased (the mean of the likelihood estimators lies on top of the true likelihood estimator) but the corresponding posterior estimators are biased. We drop the implicit conditioning on data.}
    \label{fig: schematic of posterior bias}
\end{figure}

We illustrate the intuition behind the posterior bias in Fig.~\ref{fig: schematic of posterior bias}. Suppose the true likelihood is a standard normal, and we have an unbiased likelihood estimator scattered by a zero-mean log-Gaussian process. Each draw from the scattering process represents a likelihood estimator $\hat{\mathcal{L}}$. We show several draws of the likelihood estimator in the left-hand panel of Fig.~\ref{fig: schematic of posterior bias}. Notice, due to the form of the log-Gaussian process, the uncertainty is small towards the negative end of the parameter space, and grows larger towards the positive end. We also show the true likelihood ($\mathcal{L}$, green) and the average of the likelihood estimators ($\langle\hat{\mathcal{L}}\rangle$, red) to show the likelihood estimator is unbiased, as designed.

However, the resulting posterior estimator is biased. As shown mathematically in Eq.~\ref{eq: bias in posterior heuristic}, this is because of the nonlinearity of the normalization, and we can understand this intuitively from the schematic in Fig.~\ref{fig: schematic of posterior bias}. Consider two cases: one where the likelihood estimator for $x>0$ is scattered upwards, and one where the likelihood estimator is scattered downwards. The likelihood estimator is unbiased, so these cases should average out over an ensemble of estimators to the correct likelihood. That is, the amount of upwards scatter balances the amount of downwards scatter in the likelihood estimator. However, when we consider the corresponding \textit{normalized} posterior estimators, this balance is broken. The normalization is larger for the estimator which scattered upwards and smaller for the estimator which scattered downwards. The normalization then down-weights the upwardly scattered estimator and up-weights the downwardly scattered estimators. In general, this means that regions with larger uncertainty in the likelihood will be negatively biased, and regions with smaller uncertainty will be positively biased. 
Fig.~\ref{fig: schematic of posterior bias} confirms our intuition: where the likelihood estimator uncertainty is small ($x<0$), the posterior is positively biased. We quantify this bias using the \textit{average} of the posterior estimators, denoted $\langle\hat{p}(x)\rangle$ (and dropping the implicit conditioning on data). 

For completeness, we include the discussion of general posterior correction in the Appendix~\ref{app: correction in general} and as it applies to some examples in Appendix~\ref{app: examples of posterior correction}. That said, in practice it is expensive to include the full posterior correction factor because it requires an integral over the uncorrected posterior. That typically means performing an initial biased inference followed by a second corrected inference. Another drawback is that in the corrected posterior, we lose the ability to estimate the remaining posterior bias and there is often still significant \textit{uncertainty} which actually dominates the total error. For this reason, it is usually enough to use the corrected likelihood but neglect the posterior correction term. In Section~\ref{sec: robustness measures}, we introduce an estimator to quantify the total error (bias plus uncertainty) in the uncorrected posterior, and the analyst should compute this error statistic to assess the validity of their approximate inference.


%% file: robustness_measures.tex
\section{Robustness measures and uncertainty thresholding}
\label{sec: robustness measures}

In this section, we introduce a unified statistic $\hat{E}$ for evaluating the total error in an approximated posterior. From this discussion, the concepts of accuracy and precision for posterior estimators naturally appear. We also define the accuracy statistic $\hat{A}$ and precision statistic $\hat{\Pi}$ for estimating these quantities, where the error statistic is the sum $\hat{E} = \hat{A} + \hat{\Pi}$. We note that the precision statistic is proportional to the doubly-integrated Eq. A10 from \citet{Essick:2022ojx}.

To quantify the error between the posterior estimator and the true posterior, we use the \KL divergence. Given a reference probability distribution $p$ and a test distribution $q$ over a space $\Lambda$, the \KL divergence quantifies the information lost from approximating $p$ with $q$. In units of bits \citep{MacKay2003}
\beq
\KLD(p|q) = \int p(\Lambda) \log_2\left(\frac{p(\Lambda)}{q(\Lambda)}\right)\dd\Lambda.
\label{eq: kl divergence}
\eeq
Suppose $q(\Lambda) \propto p(\Lambda)(1+\delta(\Lambda))$ for a small perturbation function $\delta(\Lambda)$. Expanding the logarithm and ensuring $q(\Lambda)$ is normalized gives\footnote{We can also think of Eq.~\ref{eq: kl divergence at leading order} as a different quantification for the distance between distributions: the $L^2$ norm of the centralized perturbation function $||\delta(\Lambda) - \int p(\Lambda')\delta(\Lambda')\dd\Lambda'||^2$ on the probability space defined by $p(\Lambda)$. This metric (which approximates the \KL divergence) is zero if and only if $\delta(\Lambda)-\int p(\Lambda')\delta(\Lambda')\dd\Lambda' = 0$ over the entire support of $p$.}
\beq
\KLD(p|q) \approx \frac{1}{2\ln(2)} \left[\int p(\Lambda)\delta(\Lambda)^2\dd\Lambda - \left(\int p(\Lambda)\delta(\Lambda)\dd\Lambda\right)^2\right] + \mathcal{O}(\delta^3).
\label{eq: kl divergence at leading order}
\eeq

Here, $p$ represents the true posterior and $q=\hat{p}$ the estimator, and so $\delta(\Lambda) = b(\Lambda) + \varepsilon(\Lambda)$ is the sum of a constant bias term $b(\Lambda)$ (defined in Eq.~\ref{eq: posterior bias}) and a (zero mean) scatter term $\varepsilon(\Lambda)$. Taking the expectation of both sides and using that $\expect{\varepsilon(\Lambda)} = 0$, $\expect{\varepsilon(\Lambda)\varepsilon(\Lambda')} = C_{\ln\mathcal{L}}(\Lambda, \Lambda')$ and $C_{\ln\mathcal{L}}(\Lambda, \Lambda) \equiv \sigma^2_{\ln\mathcal{L}}(\Lambda)$
\beq
\expect{\KLD(p|\hat{p})} \approx \frac{1}{2\ln(2)} \left[\int p(\Lambda)[b(\Lambda)^2 + \sigma^2_{\ln\mathcal{L}}(\Lambda)]\dd\Lambda - \int p(\Lambda)p(\Lambda')[b(\Lambda)b(\Lambda') + C_{\ln\mathcal{L}}(\Lambda,  \Lambda')]\dd\Lambda\dd\Lambda'\right] + \mathcal{O}(\delta^3).
\label{eq: expected kl divergence to estimator}
\eeq
This motivates the definition of the error statistic $\hat{E}$, which estimates Eq.~\ref{eq: expected kl divergence to estimator}. Given $N_{\rm samp}$ samples $\{\Lambda_n\}$ from the hyperposterior estimator $\hat{p}$, we define the error statistic as 
\beq
\hat{E}[\hat{p}] = \hat{A}[\hat{p}] + \hat{\Pi}[\hat{p}]
\label{eq: error statistic}
\eeq
where $\hat{A}$ estimates the terms containing the bias $b(\Lambda)$, and so the overall effect of the bias in the posterior, and $\hat{\Pi}$ estimates the uncertainty, 
\begin{align}
    \hat{\Pi}[\hat{p}] &= \frac{1}{2\ln(2)}\left[\frac{1}{N_{\rm samp}}\sum_{n=1}^{N_{\rm samp}}\hat{\sigma}^2_{\ln\mathcal{L}}(\Lambda_n) - \frac{1}{N_{\rm samp}^2}\sum_{n=1}^{N_{\rm samp}}\sum_{m=1}^{N_{\rm samp}}\hat{C}_{\ln\mathcal{L}}(\Lambda_n, \Lambda_m)\right]
    \label{eq: precision statistic} \\
    \hat{A}[\hat{p}] &= \frac{1}{2\ln(2)}\frac{1}{N_{\rm samp}-1}\left[\sum_{n=1}^{N_{\rm samp}}w_n^2 - \frac{1}{N_{\rm samp}}\left(\sum_{n=1}^{N_{\rm samp}}w_n\right)^2\right] = \frac{1}{2\ln(2)}{\rm var}[\{w_n\}],
    \label{eq: accuracy statistic}
\end{align}
and $\hat{\sigma}^2_{\ln\mathcal{L}}(\Lambda_n)\equiv\hat{C}_{\ln\mathcal{L}}(\Lambda_n, \Lambda_n)$ is the estimator of the variance of the log-likelihood defined in Eq.~\ref{eq: likelihood variance estimator}, $\hat{C}_{\ln\mathcal{L}}(\Lambda_n, \Lambda_m)$ is the estimator for the covariance of the log-likelihood between points $\Lambda_n$ and $\Lambda_m$ defined in Eq.~\ref{eq: population covariance with selection}, and the weights are given by
\beq
w_n = 
\begin{cases}
\dfrac{1}{2}\dfrac{\Nobs(\Nobs+1)\hat{\sigma}^2_{\xi}(\Lambda_n)}{\hat{\xi}(\Lambda_n)^2} - \dfrac{1}{N_{\rm samp}}\displaystyle\sum_{m=1}^{N_{\rm samp}} \hat{C}_{\ln\mathcal{L}}(\Lambda_n, \Lambda_m) & \text{if using uncorrected likelihood in Eq.~\ref{eq: estimated hierarchical likelihood}} \\ 
\dfrac{1}{N_{\rm samp}}\displaystyle\sum_{m=1}^{N_{\rm samp}} \hat{C}_{\ln\mathcal{L}}(\Lambda_n, \Lambda_m)  & \text{if using corrected likelihood in Eq.~\ref{eq: unbiased likelihood estimator}}
\end{cases}
\eeq
where the sum estimates the average covariance between the fixed point $\Lambda_n$ and another random sample in the posterior $\hat{p}$. 

The accuracy statistic $\hat{A}[\hat{p}]$ approximates the amount of bias in the posterior estimate $\hat{p}$ from the true posterior $p$. Given that we do not have access to an ensemble of posterior estimators nor the true posterior, we can only estimate the bias to leading order. We neglect the higher order terms that should also contribute to the accuracy. A large accuracy statistic indicates that higher order terms would also contribute significantly to the bias, and the posterior estimator is not trustworthy. 

Similarly, the precision statistic $\hat{\Pi}[\hat{p}]$ measures the average amount of scatter in a theoretical ensemble of posterior estimators. The scatter is the distance between the average posterior estimator (which in general is biased from the true posterior) and the particular realization of the posterior estimator. Again, we do not have access to this theoretical ensemble, and so we can only estimate the precision to leading order based on the estimator we have. 
We verify the efficacy of the accuracy and precision statistics for a variety of hierarchical inference problems in Sections~\ref{sec: gaussian example} and \ref{sec: gw example}. 
Additionally, note that the precision statistic is also proportional to the average variance in the difference of the log-likelihoods for two draws from the hyperposterior, discussed in \citet{Essick:2022ojx} and shown in Eq. A11 therein for the single-event terms.

\begin{figure}
    \centering
    \includegraphics[width=0.8\linewidth]{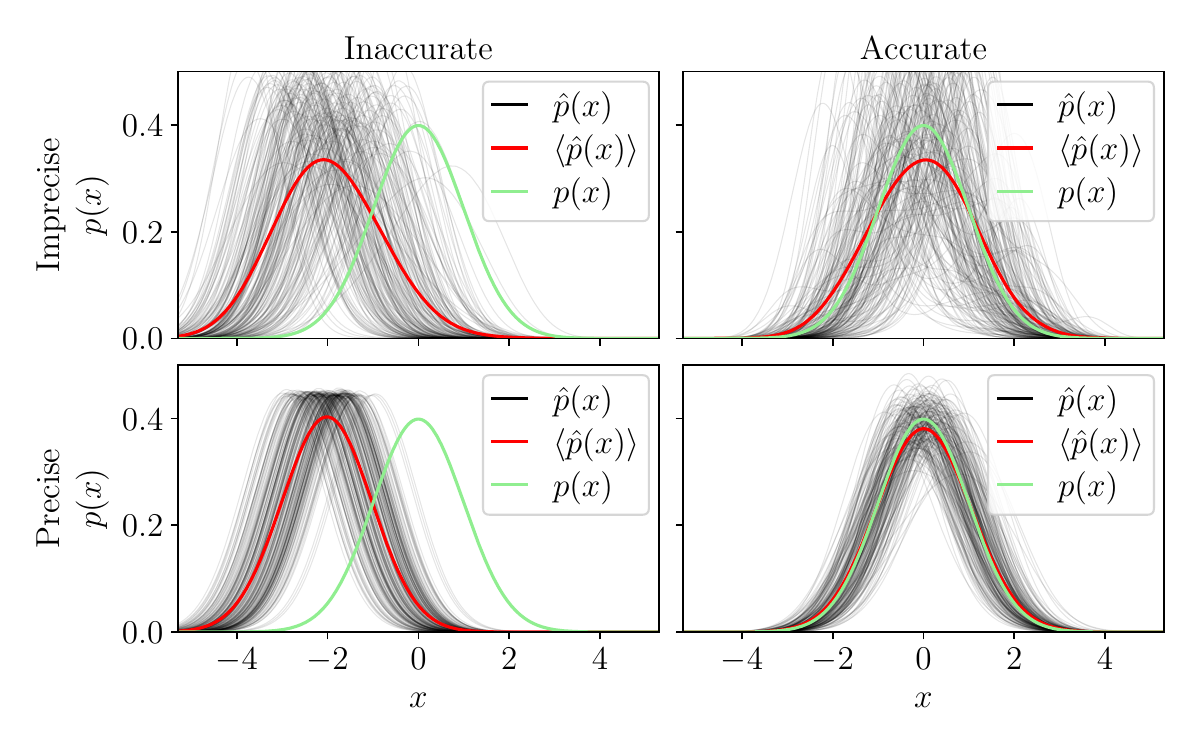}
    \caption{Schematic showing the concept of precise and accurate posterior estimators. In all panels the black traces are draws from the posterior estimator $\hat{p}(x)$, the red curve is the mean of the posterior estimators, $\langle \hat{p}(x)\rangle$, and the green is the true posterior. \textit{Top left}: draws from an inaccurate, imprecise posterior estimator. The posterior estimators are inaccurate because the mean $\langle \hat{p}(x)\rangle$ posterior is significantly different from the true posterior $p(x)$, and the estimator is imprecise due to the large scatter around the true posterior. \textit{Top right}: the same as the top left, but with an accurate but imprecise posterior estimator. \textit{Bottom left}: draws from an inaccurate yet precise posterior estimator. \textit{Bottom right}: draws from an accurate and precise posterior estimator.}
    \label{fig: accuracy vs precision}
\end{figure}

For a variety of problems we consider, we find that when $\hat{E}[\hat{p}] \lesssim 0.2$ bits, the posterior estimator is reliable, and the smaller the error statistic, the more reliable the posterior estimator is. See Fig.~\ref{fig: accuracy vs precision} for a schematic to illustrate the concepts of accuracy and precision in a posterior estimator.

Notice that the accuracy statistic is on the order of the \textit{square} of the covariance estimator and the precision is on the order of covariance estimator. As the uncertainty in the log-likelihood estimator decreases, the error is dominated by the uncertainty rather than the bias. In practice, though, there are scenarios where the bias can dominate the overall error. Bias and uncertainty in the estimator are due to different aspects of the ``noise process'' in the likelihood estimator. The bias in the posterior happens due to a changing variance/covariance across the support of the posterior, while the uncertainty in the posterior estimator comes from a variance in the log-likelihood estimator which is not matched by equal \textit{covariance} across the posterior support.

\subsection{Thresholding on the uncertainty}

Another perspective on the error in the posterior estimator is as a sum of contributions which occur at higher and higher order in the uncertainty of the log-likelihood. The leading order contribution to the error is the precision statistic, which occurs at first order in the covariance of the log-likelihood $C$. The bias occurs at order $C^2$, and in principle, there are contributions at third, fourth, etc. order in the covariance.\footnote{At higher order, there are also higher order moments which contribute, e.g. $\expect{\varepsilon(\Lambda)\varepsilon(\Lambda')\varepsilon(\Lambda'')}$, which must be accounted for.} 
However we neglect these to describe only the leading order contribution of the precision and the accuracy. If we bound the covariance to be below 1, that bounds the leading order sources of error and also ensures that the contributions at high order in the covariance are suppressed.

Fortunately, the covariance in the log-likelihood must satisfy
\beq
|C_{\ln\mathcal{L}}(\Lambda, \Lambda')| \le \sqrt{|C_{\ln\mathcal{L}}(\Lambda, \Lambda)||C_{\ln\mathcal{L}}(\Lambda', \Lambda')|} \equiv \sigma_{\ln\mathcal{L}}(\Lambda)\sigma_{\ln\mathcal{L}}(\Lambda'),
\label{eq: covariance Schwarz inequality}
\eeq
due to the Cauchy-Schwarz inequality, where $\sigma_{\ln\mathcal{L}}(\Lambda)$ is the square root of the variance of the log-likelihood estimator. The same inequality holds for the covariance and variance \textit{estimators}
\beq
|\hat{C}_{\ln\mathcal{L}}(\Lambda, \Lambda')| \le \hat{\sigma}_{\ln\mathcal{L}}(\Lambda)\hat{\sigma}_{\ln\mathcal{L}}(\Lambda').
\label{eq: covariance Schwarz inequality estimator}
\eeq
Showing these inequalities does require some careful application of the Cauchy-Schwarz inequality, which we do in Appendix~\ref{app: covariance cauchy-Schwarz}.

If we require that the variance in the log-likelihood estimator be bounded below some value (e.g. below 1, as recommended by \citet{Talbot:2023pex}), then this shows that the leading order contributions to error in the posterior estimators are bounded. Now, we do not have access to the true variance of the log-likelihood estimator, but we do have access to an estimator of the variance of the log-likelihood. We can cut out regions of $\Lambda$ parameter space wherever our estimator of the variance in the log-likelihood exceeds 1, and we show here that this bounds the error in the posterior estimator.\footnote{The uncertainty in the estimator for the variance in the log-likelihood is typically negligible, especially when the variance estimator itself is below 1.} This verifies the conjectured threshold on the log-likelihood put forth by \citet{Talbot:2023pex}.\footnote{That being said, there are pathological examples where this argument breaks down: for example, the arguments here fail if the likelihood does not admit a Taylor expansion.}

Using the above inequalities of Eqs.~\ref{eq: covariance Schwarz inequality} and \ref{eq: covariance Schwarz inequality estimator}, and a variance threshold $\sigma^2_{\rm MAX}$, the precision, accuracy and error statistics are bounded by 
\beq
\hat{\Pi}[\hat{p}] \le \frac{\sigma^2_{\rm MAX}}{\ln(2)} \qquad \qquad \hat{A}[\hat{p}] \le \frac{\sigma^4_{\rm MAX}}{2\ln(2)} \qquad \qquad \hat{E}[\hat{p}] \le \frac{\sigma^2_{\rm MAX}}{\ln(2)} + \frac{\sigma^4_{\rm MAX}}{2\ln(2)} 
\label{eq: error statistics bounds}
\eeq
but in practice tend to be much smaller than these bounds (typically by an order of magnitude or so).

While we show the log-likelihood variance threshold suggested in \citet{Talbot:2023pex} guarantees a trustworthy posterior estimator, it only does so within the domain of $\Lambda$ which is not below the variance threshold. Furthermore, \citet{Essick:2022ojx} point out that a threshold on the log-likelihood variance might be sufficient for accurate inference, but also that there are cases where thresholding on the log-likelihood variance is much stronger than necessary; these are the cases where we believe the error statistic is useful (\citet{Essick:2022ojx} and \citet{Talbot:2023pex} come to similar conclusions and define a statistic similar to our precision statistic). We discuss examples of the variance threshold being stronger than necessary in Section~\ref{sec: gw example}.

Another common threshold previously used in the literature is based on the effective sample size of a Monte Carlo integral. Somewhat confusingly, there are two related but subtly different definitions of the effective sample size, both of which are used. One definition, defined in \citet{Farr:2019rap}, is commonly applied to the selection efficiency Monte Carlo integral
\beq
N_{{\rm eff}, \xi} \equiv \frac{\hat{\xi}^2}{\hat{\sigma}_{\xi}^2}.
\eeq
\citet{Farr:2019rap} point out that, when $N_{{\rm eff}, \xi} < 4\Nobs$, then the likelihood estimator is extremely biased and so is almost certainly unreliable. However, the converse is not necessarily true: if $N_{{\rm eff},\xi} > 4\Nobs$ it is still possible that the likelihood estimator is too uncertain to be trusted. Indeed, the variance in the log-likelihood due to the selection efficiency Monte Carlo integral is $\Nobs^2 / N_{{\rm eff},\xi}$, and so for $\Nobs \gg 4$, we can have a large effective sample size, but a variance which is much larger than one, with no \textit{guarantee} that the posterior estimator is reliable (only the lack of a guarantee that the posterior estimator is \textit{not} reliable).

Historically, some of the first problems due to single-event Monte Carlo integrals (eq.~\ref{eq: single event estimator}) were encountered by \citet{KAGRA:2021duu}. To handle such issues, another definition of effective sample size due to \citet{kish1965survey} was used to ensure convergence, where
\beq
N_{{\rm eff}, p} \equiv \frac{\left(\sum_{j=1}^{\NPE} w_j\right)^2}{\sum_{j=1}^{\NPE} w_j^2} = \frac{\hat{p}^2}{\hat{\sigma}^2_{p} + \hat{p}^2/(\NPE-1)} \qquad\qquad \mathrm{where} \qquad \qquad w_j = \frac{p(\theta_j | \Lambda)}{\pi(\theta_j|{\rm PE})}.
\eeq
The threshold used by \citet{KAGRA:2021duu} was to require that \textit{all} single-event Monte Carlo integrals satisfy $N_{{\rm eff}, p} > \Nobs$. Other analyses have used different thresholds, requiring instead that $N_{{\rm eff}, p} > 10$ or some other fixed number. 
In particular, the thresholds we refer to as ``effective sample size'' or ``$\Neff$'' thresholds are
\begin{enumerate}
    \item $N_{{\rm eff},\xi} > 4\Nobs$
    \item $N_{{\rm eff},p} > 10, 30, \Nobs$ depending on the analysis threshold.
\end{enumerate}
Our claim is that these thresholds are not enough to ensure a reliable posterior estimator. Though these may produce reasonable results in some applications, we do not have any theoretical guarantee of reliable results, and we show examples in Section~\ref{sec: gw example} where these thresholds can be insufficient.

%% file: gaussian.tex
\section{Example with Gaussian hierarchical inference}
\label{sec: gaussian example}
\subsection{Without selection effects}
\label{subsec: example without selection effects}

We now consider a more concrete example with a simple hierarchical inference. To eliminate the effects of bias in the likelihood, we consider an example where there are no selection effects; all signals are detected. However, there is some noise associated with each signal, and this uncertainty must be marginalized over. In particular, our population draws events $\theta \sim \mathcal{N}(\mu, \sigma)$ from an underlying Gaussian population with mean $\mu$ and standard deviation $\sigma$. We note these simulation studies are very similar to those in Sec. III in \citet{Essick:2022ojx}.

Suppose there is some noise $n\sim\mathcal{N}(0, \sigma_n)$ where we observe $d=\theta+n$. Then, we have a Gaussian likelihood for each event
\beq
p(d|\theta) = \frac{1}{\sqrt{2\pi\sigma_n^2}}\exp\left(-\frac{(d-\theta)^2}{2\sigma_n^2}\right)
\eeq
For a collection of events $\{d_i\}$, we can compute the hierarchical likelihood of Eq.~\ref{eq: rate marginalized likelihood} in closed form (ie. we have a ``ground truth'' for comparison), and also estimate using the traditional Monte Carlo approach of Eq.~\ref{eq: estimated hierarchical likelihood}
\begin{align}
\mathcal{L}(\{d_i\}|\Lambda) &= \prod_{i=1}^{\Nobs} \int p(d_i|\theta)p(\theta|\Lambda)\dd\theta = [2\pi(\sigma^2 + \sigma_n^2)]^{-N_{\rm obs} / 2}\exp\left[-\sum_{j=1}^{N_{\rm obs}}\frac{(d_j - \mu)^2}{2(\sigma^2 + \sigma_n^2)}\right] 
\label{eq: analytic gaussian likelihood} \\
\hat{\mathcal{L}}(\{d_i\}|\Lambda) &= \prod_{i=1}^{\Nobs}\hat{p}(d_i|\Lambda),
\qquad\qquad{\rm where}\qquad\qquad
\hat{p}(d_i|\Lambda)=\left[\frac{1}{\NPE}\sum_{j=1}^{\NPE}\frac{p(\theta_{ij}|\Lambda)}{\pi(\theta_{ij}|{\rm PE})}\right]_{\theta_{ij} \sim p(\theta|d_i)}.
\label{eq: estimated gaussian likelihood}
\end{align}
For all the single event \ac{PE}, we use an improper uniform prior $\pi(\theta|{\rm PE})\propto 1$, so drawing $\theta_j$ from the posterior $p(\theta|d_i)$ is drawing from a Gaussian with mean $d_i$ and standard deviation $\sigma_n$. 

We calculate the posterior on the population hyperparameters $\mu, \sigma$ with a uniform hyperprior $\pi(\mu)=U(0,2)$ and $\pi(\log_{10}\sigma) = U(-1,1)$.
Though the likelihood estimator is unbiased, we know from Section~\ref{sec: bias in population inference} that the resulting posteriors will be biased. In addition, there is uncertainty in the posterior estimators, and we quantify the uncertainty and bias with the precision and accuracy statistics of Eqs.~\ref{eq: precision statistic} and \ref{eq: accuracy statistic}. To compute these statistics we need
\beq
\hat{C}_{\ln\mathcal{L}}(\Lambda, \Lambda') =  \sum_{i=1}^{\Nobs}\frac{\hat{C}_{p_i}(\Lambda, \Lambda')}{\hat{p}(d_i|\Lambda)\hat{p}(d_i|\Lambda')},
\label{eq: gaussian example covariance}
\eeq
which is the estimator for the covariance in the log-likelihood of two hyperparameter points $\Lambda$ and $\Lambda'$. 

\begin{figure}
    \centering
    \includegraphics[width=\linewidth]{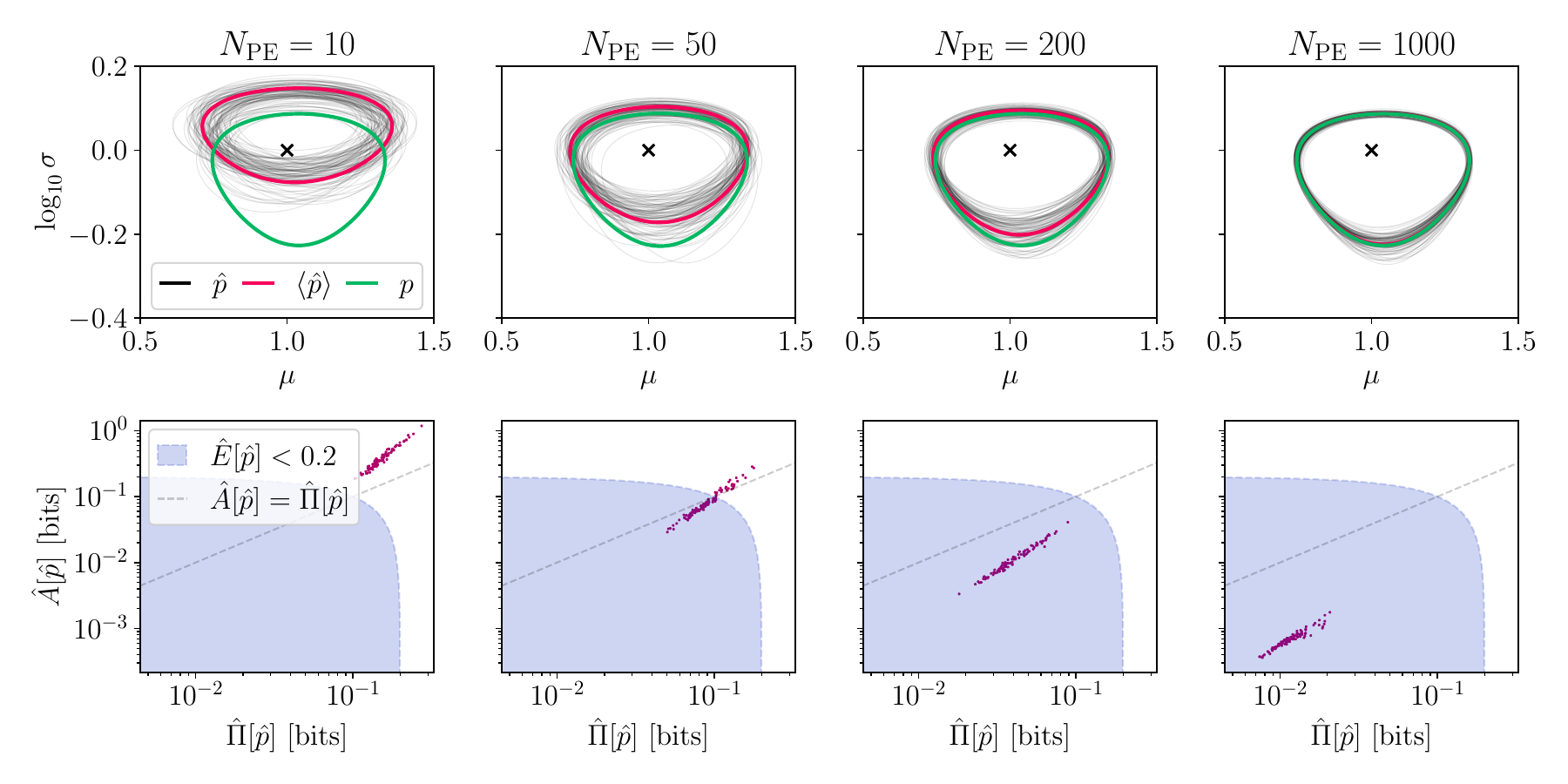}
    \caption{Inference in a Gaussian hierarchical inference with 100 observations and a variable number of \ac{PE} samples. Here, the scale of the noise is comparable to the size of the population ($\sigma_n=\sigma=1$). In the upper panels, we show the 90\% \HPD contours of the Monte Carlo posterior estimators $\hat{p}$ (black) compared to the true posterior $p$ (green) and the mean of the posterior estimators $\langle\hat{p}\rangle$. We show the true value of the population hyperparameters with a black $\times$ symbol ($\mu=1$ and $\sigma = 1$). In the lower panels, we show the accuracy and precision statistics for each posterior estimator and the threshold for a reliable posterior estimator $\hat{E}[\hat{p}] < 0.2$ bits.}
    \label{fig: nobs 100 gaussian example}
\end{figure}

In this simplified setting, where the hyperposterior is only two dimensional, we can directly compute the analytic posterior density (with Eq.~\ref{eq: analytic gaussian likelihood}) over a grid of appropriate $\mu$, $\sigma$ values, the Monte Carlo estimator of the posterior density, and the accuracy and precision statistics (using Eqs.~\ref{eq: precision statistic},~\ref{eq: accuracy statistic} and \ref{eq: gaussian example covariance}). In Fig.~\ref{fig: nobs 100 gaussian example}, we show a comparison between the Monte Carlo posterior estimates (in black), the average of the Monte Carlo posteriors (red), and the true analytic posterior (green). Each Monte Carlo posterior estimate has a different randomized set of samples drawn from the single-event posteriors. In the upper panels, the Monte Carlo posterior estimators compared to the true posterior and the mean of the posterior estimators. The lower panels show the precision and accuracy statistics for each posterior estimator. As the number $\NPE$ of \PE samples increases for each Monte Carlo integral, the bias begins to disappear below noise between the different posterior estimators -- the accuracy statistic decreases below the precision statistic. Fig.~\ref{fig: nobs 100 gaussian example} is for an example with $\Nobs=100$ observations.

\begin{figure}
    \centering
    \includegraphics[width=\linewidth]{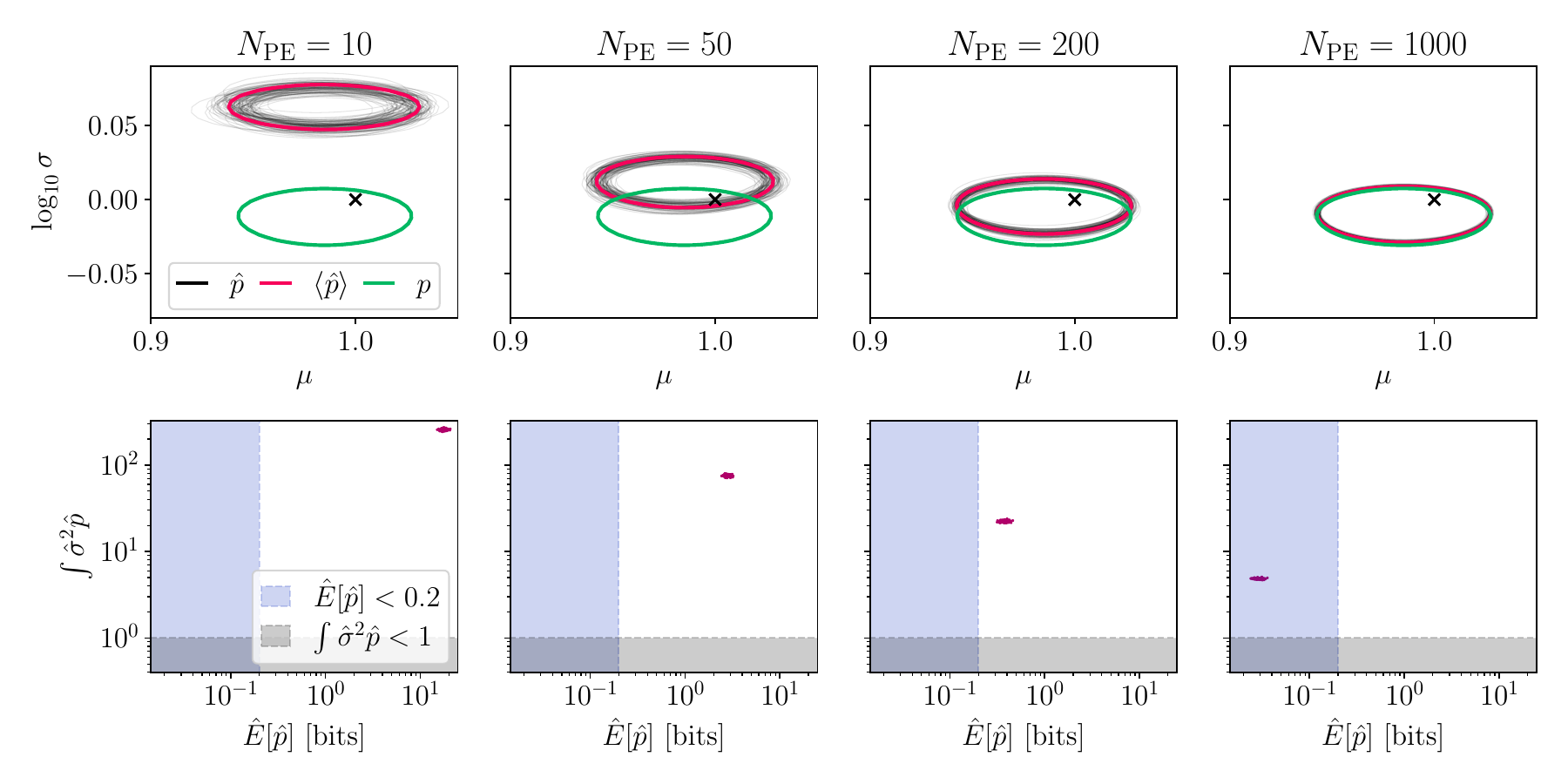}
    \caption{Analogous to Fig.~\ref{fig: nobs 100 gaussian example} with $\Nobs = 5000$. In the bottom row, we compare the mean variance estimator across the posterior ($\int \hat{\sigma}^2\hat{p}$) to the error statistic for each posterior estimator and demonstrate that the error in a posterior estimator is more accurately described by the error statistic than the variance in the log-likelihood estimator.}
    \label{fig: nobs 5000 ehat vs var}
\end{figure}

We showed in Section~\ref{sec: robustness measures} that thresholding on the variance in the log-likelihood estimator is enough to bound the total error statistic. However sometimes it is much stronger than necessary, especially for simple population models (like the Gaussian population model we use here) with a large number of observations. In the bottom row of Fig.~\ref{fig: nobs 5000 ehat vs var}, we compare the error statistics to the mean log-likelihood variance across the posterior when using $\Nobs = 5000$ observations. We show that the error statistic can be very small even when the variance in the log-likelihood is large --- see the furthest right panels, using $\NPE = 1000$. The posterior estimators are clearly strongly clustered on the correct posterior (upper panel) and the error statistics correctly reflect this: $\hat{E}[\hat{p}] \sim 0.03$ bits, while the variance is quite large, $\hat{\sigma}^2 \sim 5$.

We show further examples with $\Nobs = 1000$ and $\Nobs = 5000$ and a comparison where the population structure $\sigma$ is smaller than the noise in each single event $\sigma_n$ in the Appendix~\ref{app: gaussian pe uncertainty only}. 

\begin{figure}[h!]
    \centering
    \includegraphics[width=\linewidth]{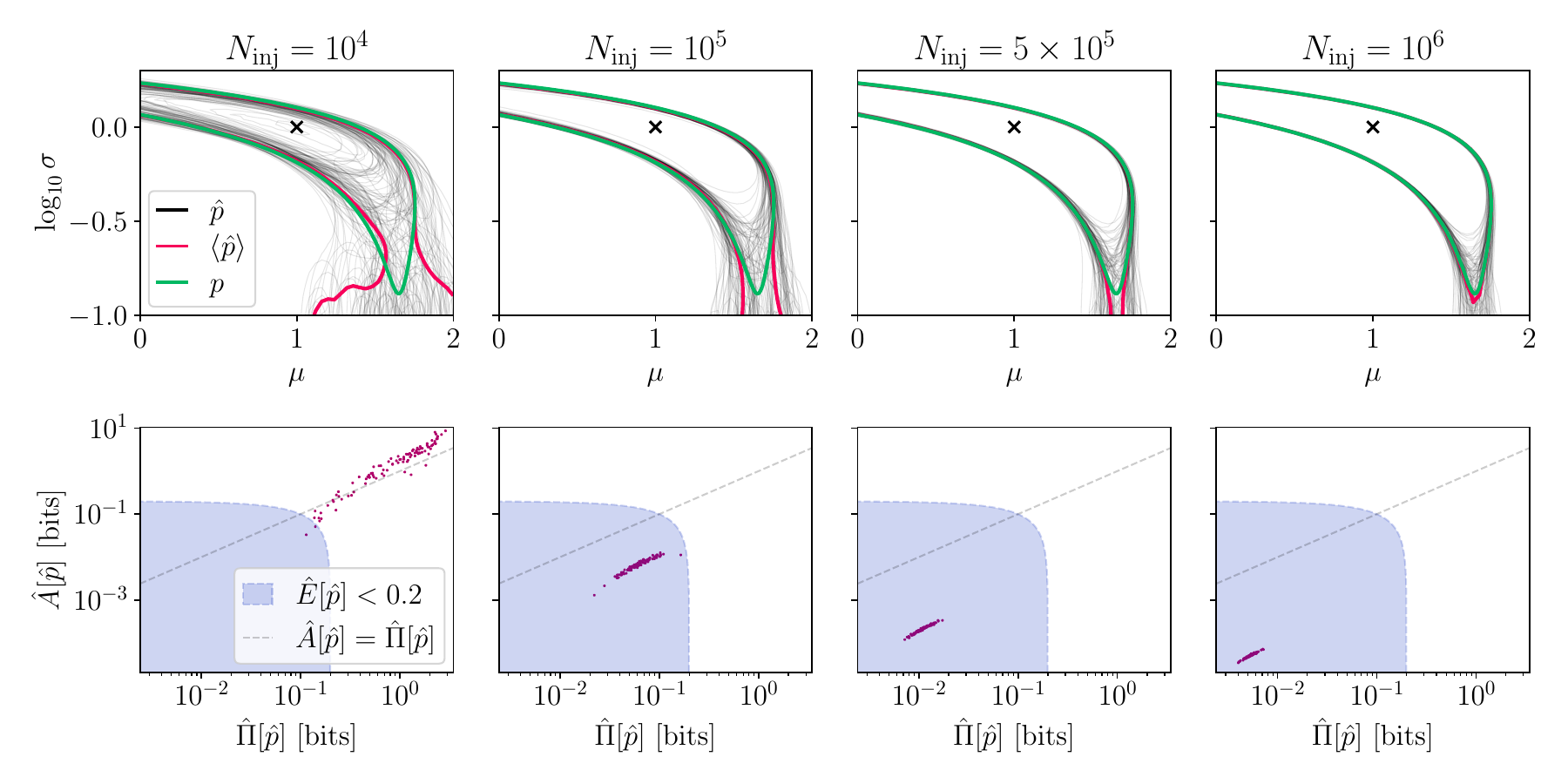}
    \caption{Analogous to Fig.~\ref{fig: nobs 100 gaussian example} where we select events for detection using $d=\theta+n > \thresh$ where $\thresh = 1$. We estimate the likelihood using Eq.~\ref{eq: estimated gaussian likelihood with vt}. We use $\Nobs=100$ observations and compare $\Ninj = 10^4,10^5,5\times10^5$ and $10^6$.}
    \label{fig: nobs 100 vt 1 gaussian example}
\end{figure}

\subsection{Example with selection effects}
\label{subsec: example with selection effects}

As another example, we consider a similar example as the above, where events are drawn from an underlying Gaussian population $\theta \sim\mathcal{N}(\mu, \sigma)$ and we observe data $d=\theta+n$ obscured by noise $n\sim\mathcal{N}(0,\sigma_n)$. This time, however, we include a selection effect where we only detect events such that $d > \thresh$ exceeds the threshold $\thresh$. We again note these simulation studies are very similar to those in Sec. III in \citet{Essick:2022ojx}.

We may still compute the hierarchical likelihood of Eq.~\ref{eq: rate marginalized likelihood} in closed form (ground truth)
\beq
\mathcal{L}(\{d_i\}|\Lambda) = \left[\frac{1}{2}{\rm erfc}\left(\frac{\thresh-\mu}{\sqrt{2(\sigma^2+\sigma_n^2)}}\right)\right]^{-\Nobs}[2\pi(\sigma^2 + \sigma_n^2)]^{-N_{\rm obs} / 2}\exp\left[-\sum_{j=1}^{N_{\rm obs}}\frac{(d_j - \mu)^2}{2(\sigma^2 + \sigma_n^2)}\right],
\label{eq: analytic gaussian likelihood with vt}
\eeq
which is identical to Eq.~\ref{eq: analytic gaussian likelihood} but now includes an analytic selection effects term $\alpha(\Lambda)^{-\Nobs}$ using the complementary error function erfc.

We also compute a variant of the hierarchical likelihood where we calculate the single-event integrals analytically but we estimate the selection effects term with a Monte Carlo integral
\beq
\hat{\mathcal{L}}(\{d_i\}|\Lambda) = \left[\frac{1}{\Ninj}\sum_{j=1}^{\Nfound}\frac{p(\theta_{i}|\Lambda)}{p(\theta_{i}|{\rm draw})}\right]_{\theta_{i} \sim p(\theta|{\rm draw})}^{-\Nobs}[2\pi(\sigma^2 + \sigma_n^2)]^{-N_{\rm obs} / 2}\exp\left[-\sum_{j=1}^{N_{\rm obs}}\frac{(d_j - \mu)^2}{2(\sigma^2 + \sigma_n^2)}\right].
\label{eq: estimated gaussian likelihood with vt}
\eeq
We can also think of this as the traditional estimator of Eq.~\ref{eq: estimated hierarchical likelihood} when we take the limit $\NPE \to \infty$. We take the draw distribution for estimating selection effects to be $p(\theta|{\rm draw}) = \mathcal{N}(\mu_{\rm vt}=1,\sigma_{\rm vt}=1.5)$.\footnote{Using Eq.~\ref{eq: variance of Monte Carlo integral}, one can show that the Monte Carlo estimator for the selection effects has a infinite variance when $\sigma^2 > 2\sigma_{\rm vt}^2$, and so the estimator is always unreliable for $\sigma \gtrsim 2.12$. We restrict the hyperprior on $\log_{10}\sigma$ to avoid this domain. In practice, the correction term in Eq.~\ref{eq: unbiased likelihood estimator} should also strongly downweight this region.}

We assume uniform hyperpriors $\pi(\mu)=U(0,2)$ and $\pi(\log_{10}\sigma) = U(-1,0.3)$.
We show a comparison between the posterior estimators, the mean of the posterior estimators, and the true posterior in Fig.~\ref{fig: nobs 100 vt 1 gaussian example} where $\Nobs=100$ observations and the selection given by $\thresh = 1$. In the lower panels we show the accuracy and precision statistics for each posterior estimator compared to the region $\hat{E}[\hat{p}] < 0.2$ bits.
The accuracy and precision statistics (Eqs.~\ref{eq: precision statistic} and \ref{eq: accuracy statistic}) rely on the covariance estimator, which for this problem is given as the \textit{second} term in Eq.~\ref{eq: population covariance with selection},
\beq
\hat{C}_{\ln\mathcal{L}}(\Lambda, \Lambda') = \frac{\Nobs^2\hat{C}_{\xi}(\Lambda, \Lambda')}{\hat{\xi}(\Lambda)\hat{\xi}(\Lambda')}.
\label{eq: vt gaussian example covariance}    
\eeq

Moving to the right in Fig.~\ref{fig: nobs 100 vt 1 gaussian example}, we increase the size $\Ninj$ of the Monte Carlo integrals. Notice the error statistics drop below the recommended threshold, and the error becomes dominated by the precision statistic (the points in the lower panel move below the diagonal dashed line).

\begin{figure}
    \centering
    \includegraphics[width=\linewidth]{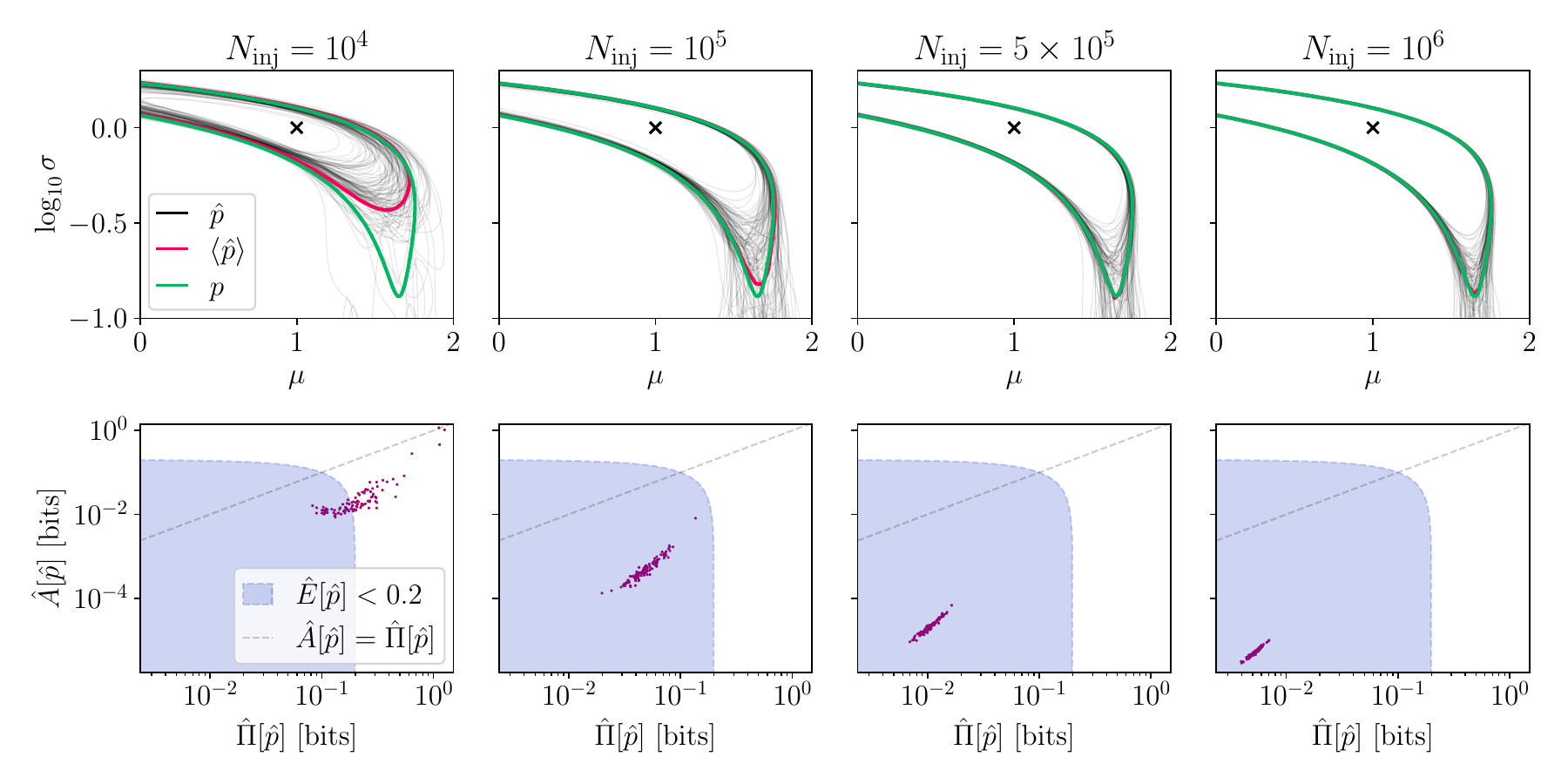}
    \caption{Analogous to Fig.~\ref{fig: nobs 100 vt 1 gaussian example} using the corrected likelihood estimator of Eq.~\ref{eq: unbiased likelihood estimator}. Notice, particularly for small $\Ninj$, the improved behavior of the mean posterior estimator $\langle\hat{p}\rangle$.}
    \label{fig: nobs 100 corrected vt 1 gaussian example}
\end{figure}

For small $\Ninj$, the likelihood estimator in Eq.~\ref{eq: estimated gaussian likelihood with vt} is significantly biased. When we include the likelihood correction term of Eq.~\ref{eq: unbiased likelihood estimator}, this alters the posterior estimators and shifts them away from the biased regime (small $\log_{10}\sigma$); see Fig.~\ref{fig: nobs 100 corrected vt 1 gaussian example}. There is still bias in the posterior, which we discuss in greater detail in Appendix~\ref{app: additional gaussians}.

\subsection{Asymptotic scaling and verification}

\begin{figure}[h!]
    \centering
    \includegraphics[width=0.6\linewidth]{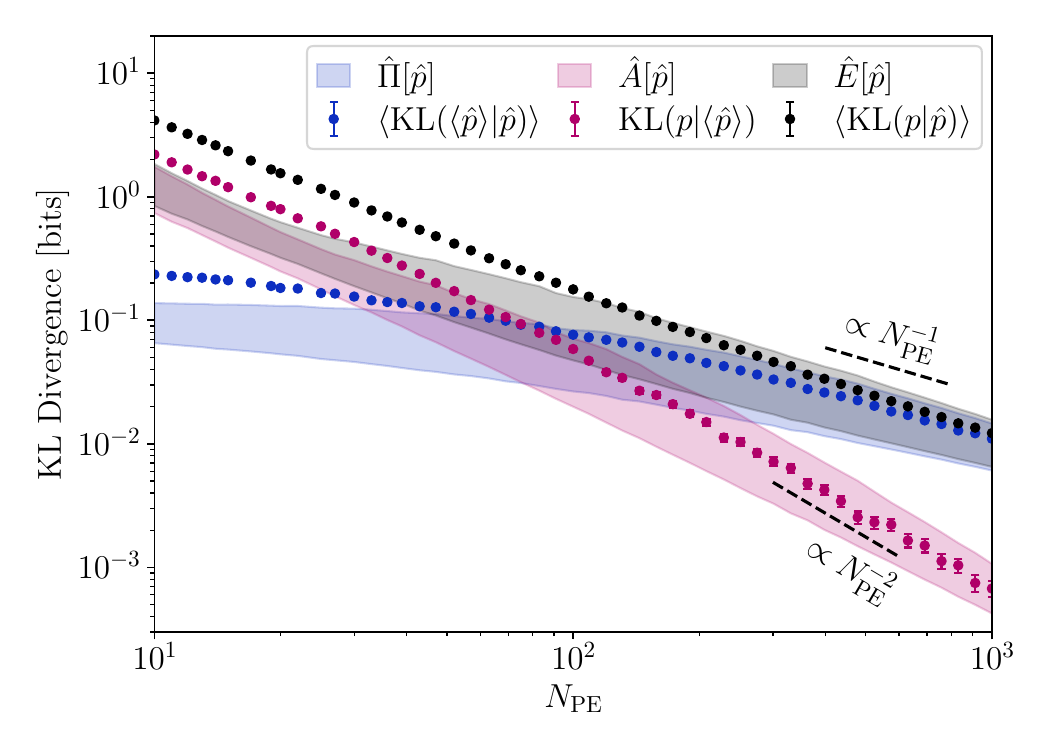}
    \caption{A comparison between the empirically measured \KL divergences (using $10^4$ posterior estimators) and their corresponding estimators, as a function of $\NPE$. We use the simple Gaussian hierarchical inference problem of Section~\ref{subsec: example without selection effects} with $\Nobs = 100$ where the prior for the mean is fixed to the true value $\pi(\mu) = \delta(\mu-1)$. In blue, we show the empirically measured average \KL divergence from the mean of the posterior estimators $\langle\KLD(\langle\hat{p}\rangle|\hat{p})\rangle$. For each posterior estimator, this \KL divergence is itself estimated by the precision statistic $\hat{\Pi}[\hat{p}]$. We show the central 90\% interval for the precision statistic in the blue shaded region, and show that $\langle\KLD(\langle\hat{p}\rangle|\hat{p})\rangle$ and $\hat{\Pi}[\hat{p}]$ asymptotically scale as $\NPE^{-1}$. Similarly, in red we show the empirically measured \KL divergence from the true posterior to the mean of the posterior estimators $\KLD(p|\langle\hat{p}\rangle)$. This is estimated by the accuracy statistic $\hat{A}[\hat{p}]$; the 90\% interval of $\hat{A}[\hat{p}]$ is shaded in red. These quantities, $\KLD(p|\langle\hat{p}\rangle)$ and $\hat{A}[\hat{p}]$ asymptotically scale as $\NPE^{-2}$. Finally, the black points show the empirically measured mean \KL divergence from the true posterior to a posterior estimator $\langle\KLD(p|\hat{p})\rangle$. To leading order, we estimate this with the error statistic $\hat{E}[\hat{p}] = \hat{\Pi}[\hat{p}] + \hat{A}[\hat{p}]$, shaded in black. At large values of $\NPE$, the precision statistic is the dominant contribution to the error and so the error scales as $\NPE^{-1}$. At small values of $\NPE$, higher order terms in the \KL divergence contribute and so the statistics tend to underestimate to the \KL divergences. As $\NPE$ increases, the precision, accuracy and error statistics begin to center on their correct, empirically measured values.} 
    \label{fig: random KL fn of npe}
\end{figure}

We show here that the precision, accuracy and error statistics are valid estimators for the information-loss they are meant to quantify. The precision statistic $\hat{\Pi}[\hat{p}]$ estimates the average information loss in the posterior estimator due to random scatter. Formally, it estimates the average \KL divergence from the \textit{mean} of the posterior estimators $\langle\KLD(\langle\hat{p}\rangle|\hat{p})\rangle$. The accuracy statistic $\hat{A}[\hat{p}]$ estimates the information lost due to bias in the posterior estimators; the \KL divergence from the true posterior to the mean of the posterior estimators $\KLD(p|\langle\hat{p}\rangle)$. The error statistic, which we calculate as the sum of the accuracy and precision statistics $\hat{E}[\hat{p}] = \hat{\Pi}[\hat{p}] + \hat{A}[\hat{p}]$, estimates the leading order average \KL divergence from the true posterior $\langle\KLD(p|\hat{p})\rangle$. 

We use the simple Gaussian hierarchical inference model introduced above in Section~\ref{subsec: example without selection effects} to empirically calculate the \KL divergences $\langle\KLD(\langle\hat{p}\rangle|\hat{p})\rangle$, $\KLD(p|\langle\hat{p}\rangle)$ and $\langle\KLD(p|\hat{p})\rangle$ from an ensemble of $10^4$ posterior estimators. 
We show that the precision statistic, accuracy statistic and error statistics all reliably estimate the associated \KL divergences at larger values of $\NPE$. 
At small values of $\NPE$ (large uncertainties in the log-likelihood estimator), the statistics \textit{under}-estimate the true \KL divergences, as higher order terms in the expansion begin to contribute meaningfully. 

We also verify that the precision statistic scales as $\NPE^{-1}$ (it scales with the variance in the log-likelihood) at large values of $\NPE$, and the accuracy statistic asymptotically scales with $\NPE^{-2}$ (the \textit{square} of the uncertainty in the log-likelihood).

\begin{table*}[h]
\normalsize
\renewcommand{\arraystretch}{1.5}
\centering
\begin{tabular}{lllll}
\hline\hline
& \multicolumn{2}{l}{PE uncertainty dominates} & \multicolumn{2}{l}{Selection uncertainty} dominates \\ 
Statistic \quad\, & Asymptotic \quad\, & Worst case \quad\, & Asymptotic \quad\, & Worst case \quad\, \\ \hline 
$\hat{\Pi}$ & \(\displaystyle \propto \NPE^{-1}\) & \(\displaystyle \propto \Nobs\NPE^{-1}\) & \(\displaystyle \propto \Nobs\Ninj^{-1}\) & \(\displaystyle \propto \Nobs^2\Ninj^{-1}\) \\ 
$\hat{A}$ & \(\displaystyle \propto \Nobs\NPE^{-2}\) & \(\displaystyle \propto \Nobs^2\NPE^{-2}\) & \(\displaystyle \propto \Nobs^3\Ninj^{-2}\) & \(\displaystyle \propto \Nobs^4\Ninj^{-2}\) \\
\hline\hline
\end{tabular}
\caption{Scaling of the accuracy and precision statistics as a function of $\Nobs$, $\NPE$ and $\Ninj$, assuming limiting behavior for the hyperposteriors (Gaussian, with support scaling as $1 / \sqrt{\Nobs}$) and a leading order quadratic expansion of the covariance $C_{\ln \mathcal{L}}(\Lambda, \Lambda')$ in the log-likelihood estimator. In the worst case where these assumptions break down, the scalings are more appropriately bounded without the $\Nobs^{-1}$ term \citep{Essick:2022ojx, Talbot:2023pex}.
} 
\label{tab: scaling}
\end{table*}

The scaling of the accuracy and precision statistics with $\Nobs$ are, in general, very complex. However, if we assume limiting behavior (large $\Nobs$), then the posteriors $p(\Lambda | \{d_i\})$ should be approximately Gaussian. If we approximate the covariance in the log-likelihood estimator to second order in $\Lambda$,\footnote{That is, we approximate $C(\Lambda, \Lambda') \approx a_0 + b_i\sum_i\Lambda_{i} + b'_i\sum_i\Lambda'_{i} + \sum_{ij}c_{ij}\Lambda_i\Lambda'_j + \sum_{ij}d_{ij}\Lambda_i\Lambda_j + \sum_{ij}d'_{ij}\Lambda'_i\Lambda'_j$, and $i,j$ index the components of the $\Lambda$ or $\Lambda'$ vectors. We assume that each of the $a_0$, $b_i$, $b_i'$, $c_{ij}$, $d_{ij}$ and $d_{ij}'$ scale with $\Nobs$, $\NPE$ and $\Ninj$ as Eq.~\ref{eq: population covariance with selection} does. 
We may justify this by noting that the coefficients of the expansion can be calculated in a usual Taylor series formalism.} 
then we may show that the accuracy and precision statistics scale as in Table~\ref{tab: scaling}.
We can justify this approximation over the support of the posterior $p(\Lambda | \{d_i\})$ in the limit of large $\Nobs$, as the support asymptotically becomes smaller as $1/\sqrt{\Nobs}$. The scalings come from the fact that the precision statistic scales linearly with the covariance estimator $\hat{\Pi} \sim \hat{C}$ and the accuracy statistic scales quadratically $\hat{A} \sim \hat{C}^2$. Both are suppressed by an additional factor $\Nobs^{-1}$ to reflect the narrowing support of the posterior \cite{Essick:2022ojx, Gair:2022fsj}. We verify that these scalings approximately hold in the additional examples shown in Appendix~\ref{app: additional gaussians}. The asymptotic scalings of the precision statistic are exactly analogous to the Eq. 31 of \citet{Essick:2022ojx}. Additionally, note the Figs. 2 and 3 in \citet{Essick:2022ojx} empirically verify the asymptotic scalings of the precision statistic $\hat{\Pi}$.

However, we emphasize that this asymptotic scaling behavior only holds under the following important assumptions \cite{Gair:2022fsj}:
\begin{enumerate}
    \item The posterior is in the asymptotic regime (approximately Gaussian, with support scaling as $1/\sqrt{\Nobs}$. See the Bernstein--von Mises theorem \citep{Kleijn:2012bvm, Gair:2022fsj}).
    \item The population model is fixed as $\Nobs$ increases, and has a fixed and finite dimension (e.g., this argument fails for trans-dimensional or infinite dimensional models \citep{Freedman:1999bvm, Miller:2013ipy, ghosal2017fundamentals}).
    \item The support of the posterior is much smaller than the structure in the log-likelihood covariance, such that the log-likelihood covariance is adequately approximated to second order in the hyperparameters $\Lambda$.
\end{enumerate}
In the worst case, if one of these assumptions does not hold, the scalings shown in Table~\ref{tab: scaling} may not include the additional $\Nobs^{-1}$ term. See a more extensive discussion about the scaling for realistic models in \citet{Talbot:2023pex}. 

We include these scalings as a theoretical statement: we stress that they should not be used to evaluate the reliability of a posterior estimator. These scalings hold for a \textit{fixed population model}; the dependence on the population model is nontrivial and so even if an inference with one model is stable to Monte Carlo uncertainty, inference with another different model using the same set of events and injections (same $\Nobs, \NPE, \Ninj$) may have vastly different error statistics and may not be reliable.

%% file: gw_example.tex
\section{Gravitational Wave Example}
\label{sec: gw example}

\begin{table*}
\renewcommand{\arraystretch}{1.25}
\centering
\begin{tabular}{lllll}
\hline\hline
Parameter & Description & True Population & VT Population & Strong Model Prior \\
\hline $\alpha$ & $m_1$ power-law index & $3.4$ & $3$ & 3.4 \\
$\beta_q$ & $q$ power-law index & $1.1$ & 1 & $1.1$ \\
$m_\mathrm{min}$ & minimum BH mass & $5M_\odot$ & $5M_\odot$ & $5M_\odot$ \\
$m_\mathrm{max}$ & maximum BH mass & $87M_\odot$ & $100M_\odot$ & $87M_\odot$\\
$\lambda_\mathrm{peak}$ & fraction of BBHs in Gaussian component & $0.04$ & $0.05$ & $0.04$ \\
$\mu_m$ & $m_1$ Gaussian component mean & $34M_\odot$ & $35 M_\odot$ & $34M_\odot$ \\
$\sigma_m$ & $m_1$ Gaussian component standard deviation  & $3.6M_\odot$ & $5M_\odot$ & $3.6M_\odot$  \\
$\delta_m$ & low-mass smoothing parameter & $4.8M_\odot$ & $4M_\odot$ & $4.8M_\odot$ \\
\hline
$\kappa$ & $z$ power-law index & $2.73$ & 2 & $2.73$ \\ 
\hline
$\alpha_\chi$ & spin magnitude $\alpha$ in Beta distribution & 1.67 & 1 & --- \\
$\beta_\chi$ & spin magnitude $\beta$ in Beta distribution & 4.43 & 1 & --- \\ 
$\mu_\chi$ & implied mean of spin magnitude Beta distribution & $0.274$ & $0.5$ & $U(0, 1)$ \\
$\sigma^2_\chi$ & implied variance of spin magnitude Beta distribution & $0.028$ & $0.083$ & $U(0.005, 0.25)$\\ \hline
$f_{q=1}$ & fraction in cosine-tilt Gaussian at equal mass ratio & 1 & 0 & 1 \\
$n$ & index in mass ratio, cosine-tilt correlation & 2 & --- & 2 \\
$\mu_\tau$ & mean of cosine-tilt Gaussian & 1 & --- & 1 \\
$\sigma_\tau$ & width of cosine-tilt Gaussian & 1.15 & --- & 1.15 \\
\hline\hline
\end{tabular}
\caption{Parameters for the simulated population. In the ``True Population'' column, we show the true numerical values from \citet{Vitale:2025lms}. The ``VT Population'' column shows the parameter values for the source population for estimating selection effects. The ``Strong Model Prior'' column shows the model priors in our strong model (Beta spin distribution). Wherever we list a numerical value for a prior, we use a delta function prior at the true population value.}
\label{tab:true_hyper}
\end{table*}

To show how this works in a realistic \ac{GW} case, we show example hierarchical inferences with synthetic \ac{GW} source populations. We implement the hierarchical likelihood using \texttt{gwpopulation} \citep{Talbot:2019okv, Talbot:2024yqw}, and sample the posterior estimator using the \texttt{Bilby} \citep{Ashton:2018jfp} implementation of \texttt{dynesty} \citep{Speagle_2020}.
As our \ac{GW} source population, we use the large catalog described in \citet{Vitale:2025lms}. For estimating the selection efficiency, we created a injection set with $\Ninj = 4\times 10^9$ total injections, using a network of the Hanford and Livingston \ac{GW} interferometers with sensitivity defined in \citet{O4_psds}. We used a slightly stricter detection threshold than that used in \citet{Vitale:2025lms}.\footnote{We defined the network matched filter SNR as the \textit{real} parts of the matched filter SNRs summed in quadrature, where \citet{Vitale:2025lms} used the magnitude. \citet{Vitale:2025lms} also used a Virgo interferometer in addition to Hanford and Livingston. Therefore, the network SNRs computed in \citet{Vitale:2025lms} are always larger than the SNRs we compute here, so it is safe to down-select: a detection for our catalog is always a detection in \citet{Vitale:2025lms}.} Therefore, we down-selected the 1599 detections described therein to 1066 detections in order to be consistent with our selection criteria. We consider a catalog of 200 observations and a catalog of 1000 observations, and we show the following.

\begin{enumerate}
    \item We show further examples of the uncertainty in a posterior estimator and the quantification with the error statistic. We show how the inferred \ac{GW} populations can be subject to large errors if adequate error statistics are not met.
    \item For simple, strong models with well-constrained hyperparameters, thresholding on estimator convergence may be unnecessary (such as a log-likelihood variance threshold). The analyst can instead show the error statistic $\hat{E}[\hat{p}] \lesssim 0.2$ bits to demonstrate a safe analysis. 
    \item For weaker models, such as nonparametric models, it is necessary to adopt some sort of threshold to ensure a reliable posterior estimator. Thresholding on the log-likelihood variance is the only way to \textit{guarantee} that an analysis is safe, but other heuristic thresholds (e.g. based on Monte Carlo effective samples sizes $\Neff$) may be adopted, so long as the error statistic is shown to be small.
\end{enumerate}

\begin{figure}[h!]
    \centering
    \includegraphics[width=0.78\linewidth]{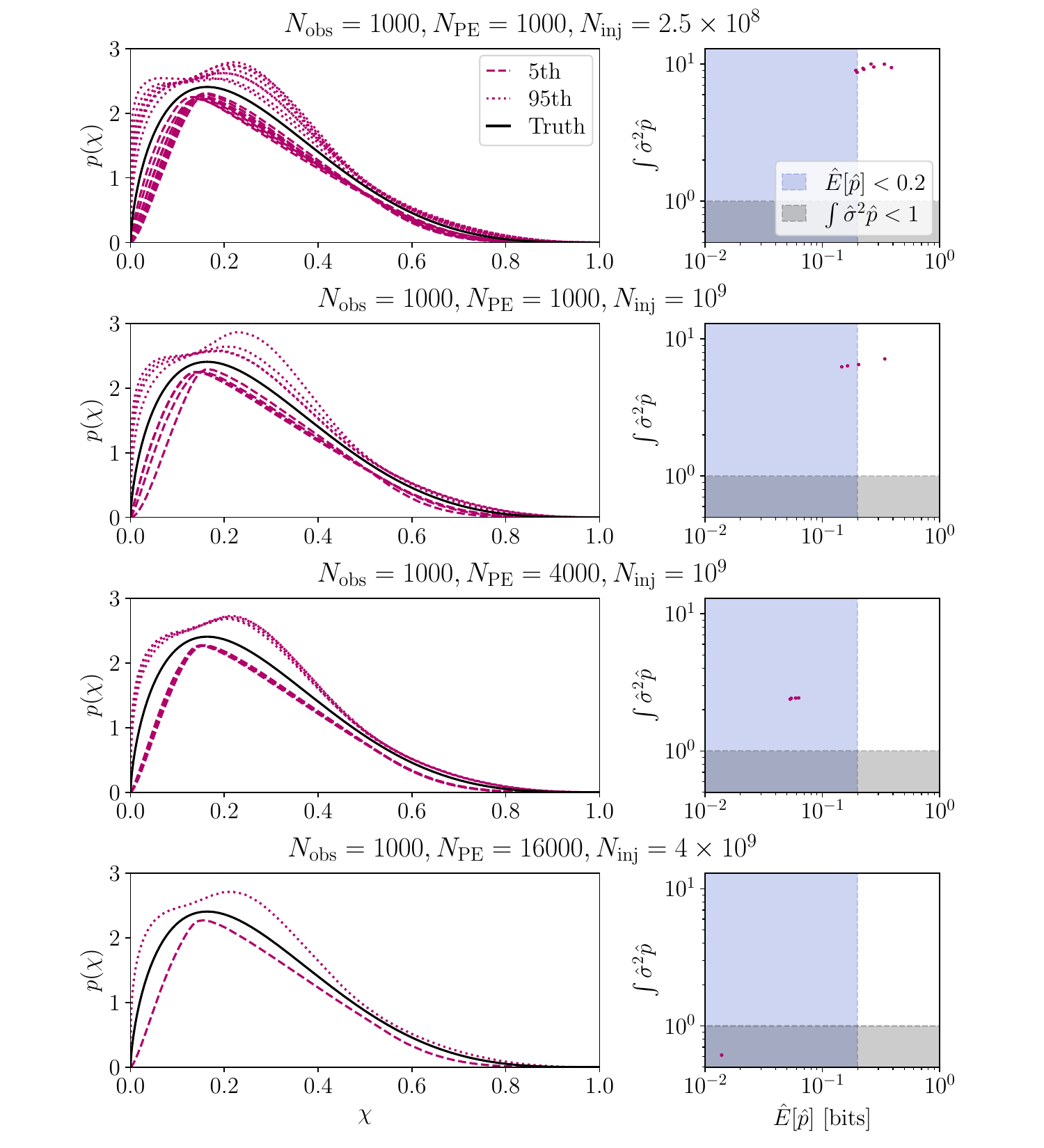}
    \caption{The inferred spin distributions for different independent posterior estimators using a strong model. We have $\Nobs = 1000$ observations, and we vary the number of independent \ac{PE} samples $\NPE$ and the number of injections $\Ninj$ for estimating the selection efficiency. In the top row, we have eight independent posterior estimators with $\NPE = 1000$ and $\Ninj = 2.5\times 10^8$, the the second row we have four independent posterior estimators with $\NPE = 1000$ and $\Ninj = 10^9$, the third row has four independent posterior estimators with $\NPE = 4000$ and $\Ninj = 10^9$, and in the last row, we only have 1 posterior estimator with $\NPE = 16000$ and $\Ninj = 4\times 10^9$. In the left column we show the 90\% credible interval on the inferred spin distributions for each independent posterior estimator, where the dashed line is the $5^{\rm th}$ percentile and the dotted line is the $95^{\rm th}$ percentile. The black line shows the true spin distribution. In the right column we compare the error statistic $\hat{E}[\hat{p}]$ with the mean log-likelihood variance, showing that some inferences have log-likelihood variances greater than 1 but the error statistics are adequate and the inference is stable.}
    \label{fig:spins comparison strong}
\end{figure}

\begin{figure}[h!]
    \centering
    \includegraphics[width=0.78\linewidth]{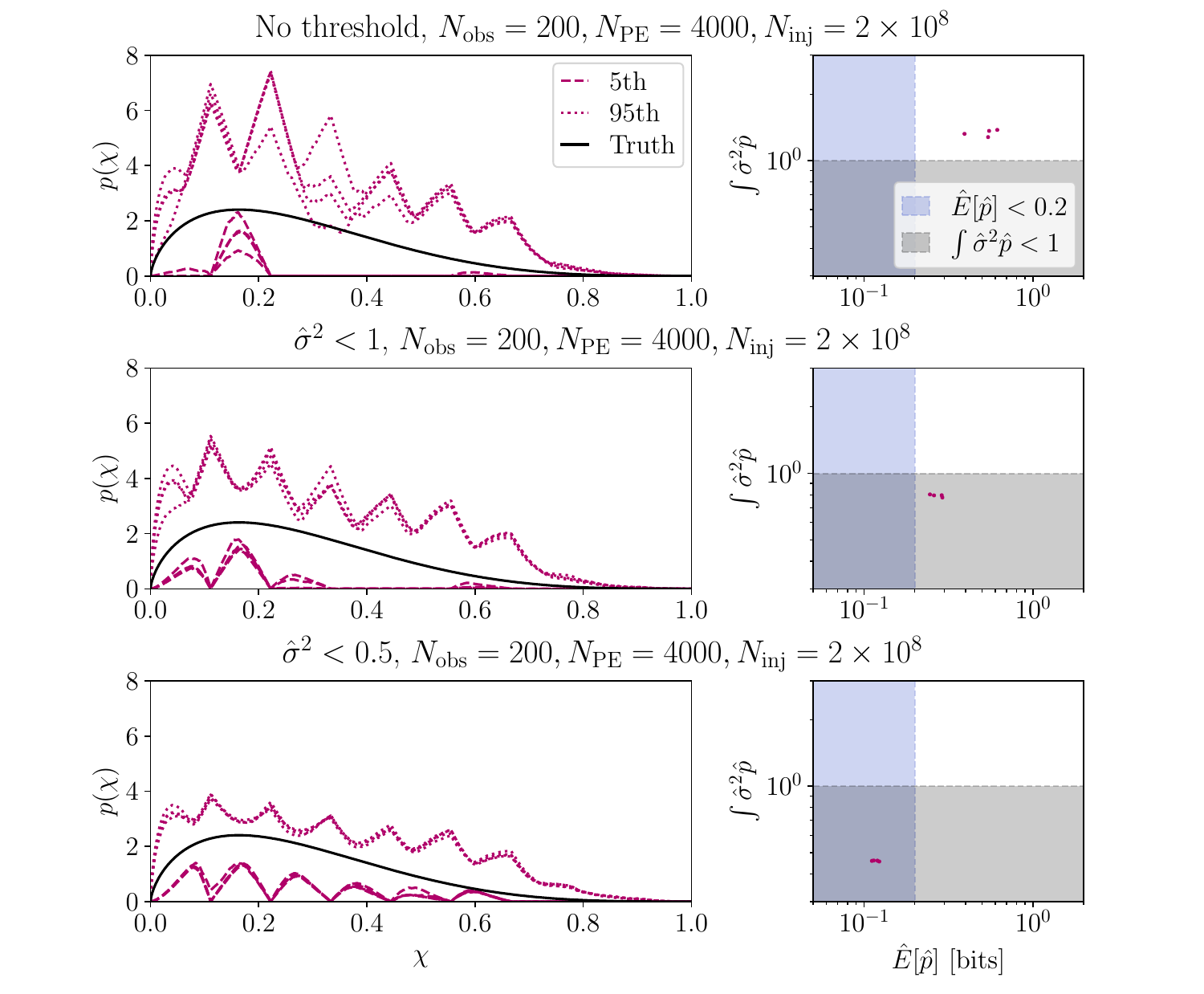}
    \caption{Analogous to Fig.~\ref{fig:spins comparison strong} for a set of independent posterior estimators using a weak model, using $\Nobs = 200$ observations. Here we compare the impact of adopting different variance thresholds. In the top row, we use no threshold, and there is a significant amount of scatter between the independent posterior estimators (some even exclude the truth from the 90\% credible region). Adopting the fiducial variance threshold $\hat{\sigma}^2 < 1$ ensures significantly better behaved posterior estimators and an error statistic $\hat{E}[\hat{p}] \approx 0.25$ bits, which is acceptable although strains our recommendation $\hat{E}[\hat{p}] \lesssim 0.2$ bits. We show a further variance threshold for $\hat{\sigma}^2 < 0.5$ which places the error statistic well within the acceptable range, but at the cost of dramatically reducing the space of spin distributions explored.}
    \label{fig:spins comparison weak varcut}
\end{figure}

As discussed in Sec.~\ref{sec: gaussian example}, thresholding on the uncertainty can be more restrictive than is absolutely necessary. We use a strong spin model in Fig.~\ref{fig:spins comparison strong}, assuming a Beta distribution for the spins, where we infer the mean and variance of the Beta spin distribution as described in \citet{Wysocki:2018mpo} and \citet{KAGRA:2021duu}. In each row, we show the results of a number of inferences using unique realizations of $\NPE$ posterior samples for each observed event and unique realizations of $\Ninj$ draws for estimating the selection efficiency. For example, in the second row we show the results of 4 independent posterior estimators (we produced $4\times10^9$ samples from the VT population and so can down select to $4$ copies of $10^9$ unique sets for estimating the selection efficiency). In the left column, we show the $5^{\rm th}$ and $95^{\rm th}$ percentiles for the predicted spin distribution using the Beta spin model, as compared to the truth. We see some variation between different independent posterior estimators, and to estimate the degree of variation we compute the error statistic for each posterior estimator (the x-axis in the right column) compared to the average variance in the log-likelihood estimator (the y-axis in the right column).

Notice in Fig.~\ref{fig:spins comparison strong} that the variance in the log-likelihood estimator can be as high as $\hat{\sigma}^2 \sim 10$, but the error statistic is in a tolerable range, $\hat{E}[\hat{p}] \lesssim 0.2$ bits. Furthermore, notice that the error statistic qualitatively tracks the degree of variation within the ensemble of posterior estimators (and verified quantitatively in Sections~\ref{sec: robustness measures} and \ref{sec: gaussian example}). For instance, with $\NPE = 4000$ and $\Ninj = 10^9$, the variance in the log-likelihood estimator is $\sim 2-3$. If the analyst was requiring that the log-likelihood variance was less than 1, this would require many more \ac{PE} samples and injections which can become computationally intractable. However, the error statistics are $\hat{E}[\hat{p}] \sim 0.05$ bits, well within a reliable range and indeed, the ensemble of 4 independent posterior estimators in the left column shows that the results are highly stable. 

Since the results with a strong model are stable with no convergence thresholds, they will also be reliable when using the effective sample size threshold. However, analysts should still be careful if they use an effective sample size threshold and check the error statistic is small in order to ensure the systematic uncertainty is acceptable.

\begin{figure}
    \centering
    \includegraphics[width=0.78\linewidth]{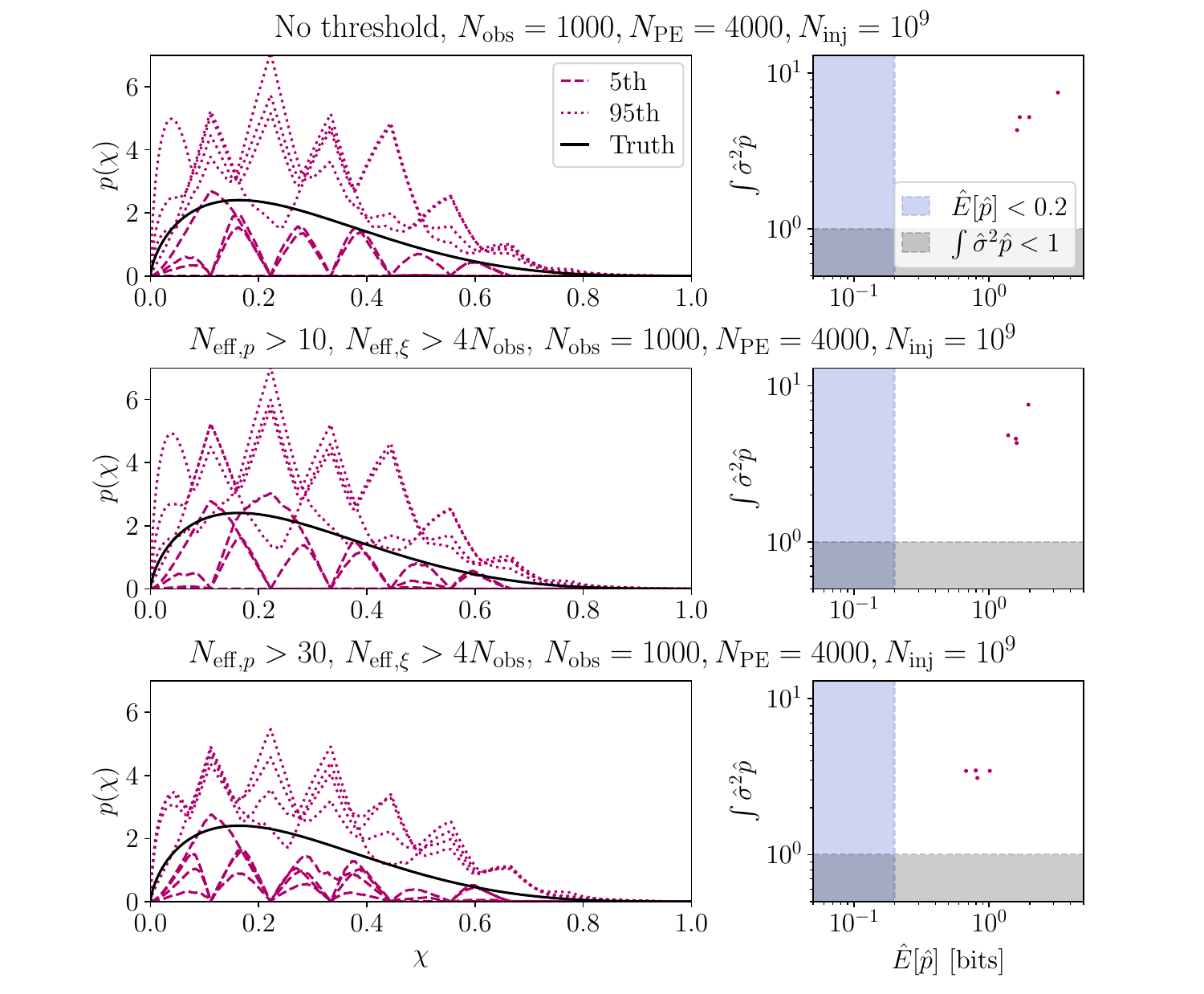}
    \caption{Analogous to Fig.~\ref{fig:spins comparison weak varcut} with $\Nobs = 1000$ detections. Here we compare the impact of adopting effective sample size thresholds. Note that none of the inferences shown here are reliable: the independent posterior estimators are quite different, reflected by the large error statistics computed in each inference. Note the log-likelihood variances are large in each of these inferences as well. These represent an example of the insufficiency of effective sample size thresholds \textit{on their own} to produce a reliable posterior estimator. In each of these, we also check that the effective sample size for the selection function estimator $N_{{\rm eff},\xi} \gtrsim 10^5 > 4\Nobs=4000$ to satisfy the threshold proposed in \citet{Farr:2019rap}.}
    \label{fig:spins comparison weak}
\end{figure}

We also show results using a weaker model (sometimes referred to as a data-driven or nonparametric model in the \ac{GW} community) for the spin distribution
\begin{equation}
    p(\chi|\{s_n\}) \propto p(\chi|\alpha_{\chi} = 1.67, \beta_{\chi} = 4.43)L(\chi| \exp[\{s_n\}])
    \label{eq: weak spin model}
\end{equation}
where $L(\chi|\exp[\{s_n\}])$ is a linear interpolation for $\chi$ between the nodes $(\chi_n, e^{s_n})$. In log space, this represents adding a linear interpolation scattering term to a base-Beta distribution (using the true population values), similar to the Power Law Spline models in \citet{Edelman:2021zkw, KAGRA:2021duu, Edelman:2022ydv, Godfrey:2023oxb}. We use 10 linear interpolation nodes and the x-axis positions of the nodes are separated linearly between 0 and 1. We use a uniform prior on $s_n \sim U(-10, 10)$ between $-10$ and $10$. Note if all $s_n=0$, then the spin population model reduces to the true spin distribution.

When using a weaker model as in Eq.~\ref{eq: weak spin model}, the resulting hyperposterior will be broader. The posterior estimator is well converged if, for each point in the support, the log-likelihood variance in that point is equally balanced by that point's average covariance with another point in the posterior. Over a larger hyperposterior support, we do not expect the covariance in the log-likelihood estimator surface to balance the variance. In such cases, the analyst may be forced to threshold on the uncertainty in the likelihood estimator, \textit{a posteriori} rejecting samples where some convergence statistic exceeds some threshold. Common thresholds considered in the literature are based on the effective sample size of individual Monte Carlo integrals \citep[e.g.][]{KAGRA:2021duu, Farr:2019rap} or on the total variance in the log-likelihood estimator \citep{Talbot:2023pex}. 

We showed in Section~\ref{sec: robustness measures} that a threshold on the variance of the log-likelihood estimator is \textit{guaranteed} to bound the overall error in the posterior estimator, and we suggested that thresholds on the effective sample size may not be sufficient. Here, we show two examples to demonstrate that (Fig.~\ref{fig:spins comparison weak varcut}) thresholding on the variance of the log-likelihood bounds the error in the posterior estimator and (Fig.~\ref{fig:spins comparison weak}) effective sample size thresholds are not always sufficient. 

In Fig.~\ref{fig:spins comparison weak varcut}, we compare the results between no threshold and a variance cut threshold, where we use $\Nobs = 200, \NPE = 4000, \Ninj = 10^8$. Without using any convergence threshold, there is a large amount of scatter between each posterior estimator and some exclude the true population from the 90\% credible region. Correspondingly, the error statistics are $\hat{E}[\hat{p}]\sim 0.6-0.8$ bits, too large to fall within the tolerable range. Once we turn on a variance threshold, the posterior estimators become much more reliable, and the error statistics drop towards a more acceptable $\hat{E}[\hat{p}] \sim 0.25$ bits, though they are still straining the recommended range $\hat{E}[\hat{p}] \lesssim0.2$ bits. However, the space of populations which can be explored is more limited. 
We also consider an even stricter threshold, where the variance in the log-likelihood is required to be less than $0.5$. The error statistic of these inferences are now well within the acceptable range $\hat{E}[\hat{p}] \sim 0.1$ bits, but at the cost of a severe reduction in the range of hyperparameters allowed by the variance threshold.

In Fig.~\ref{fig:spins comparison weak}, we show how effective sample size thresholds can sometimes be insufficient. We expect this to be particularly bad for weak models with a large number of observations, where the uncertainty in the likelihood is large and there may not be much covariance across the support of the posterior. 
We compare thresholds where $N_{{\rm eff}, p} > 10$, $N_{{\rm eff}, p} > 30$, and an example without any single-event Monte Carlo integral convergence threshold. In each example we additionally ensure that the effective sample size for the selection function estimator $N_{{\rm eff},\xi} > 4\Nobs$ to satisfy the threshold proposed in \citet{Farr:2019rap}. In practice, the effective sample size of the selection function estimator is $N_{{\rm eff},\xi} \gtrsim 10^5$, well above the suggested threshold and so it does not come into play. Notice the large scatter between independent posterior estimators (the 5th and 95th percentiles on the inferred spin distribution can be significantly different across independent sets of $\NPE$ and $\Ninj$ parameter estimation samples and injection set samples), and that some even exclude the truth.

%% file: conclusion.tex
\section{Conclusion}
\label{sec: conclusion}

In this paper, we focus on the estimation of a hierarchical posterior with Monte Carlo integrals. We reconcile two apparently competing ideas for tracking the accuracy of the posterior estimator with a unified statistic $\hat{E}[\hat{p}]$ for calculating the average information loss due to the approximation. We show that the likelihood estimator suffers from a bias which we argue can be easily corrected, but that the posterior estimator is also biased and, though it may be corrected in principle (examples in Appendix~\ref{app: examples of posterior correction}), it is expensive. In practice, the posterior bias is smaller than the uncertainty in the posterior estimator, and to this end, we derive a precision statistic to quantify this uncertainty (similar to metrics introduced in \citet{Essick:2022ojx}; \citet{Talbot:2023pex}). We suggest that analysts should ensure their posterior estimator does not suffer from significant Monte Carlo error by calculating the error statistic in post-processing and verifying $\hat{E}[\hat{p}] \lesssim 0.2$ bits. 

To enable analysts to compute the error statistics, we also publish a \texttt{pip}-installable python package called \texttt{population-error} \citep{population_error2025github} powered by \texttt{jax} \citep{jax2018github}.  \texttt{population-error} takes as input the set of $\Nsamp$ samples from the posterior estimator $\hat{p}(\Lambda | \{d_i\})$, the population model $p(\theta|\Lambda)$ and the $\Nobs\times\NPE$ posterior samples and $\Nfound$ found injections for computing the accuracy, precision and error statistics.


Even with current catalog sizes of $\mathcal{O}(100)$ events, posterior uncertainty is a major systematic source of error. We show in Sec.~\ref{sec: gw example} that the standard Monte Carlo approach may remain reliable in certain simple models as the number of observations increases, but only under somewhat idealized assumptions (e.g. that simple models will remain appropriate for informative data). It quickly becomes computationally intractable for weak models and large catalogs. Ultimately, a new approach to avoid Monte Carlo uncertainties entirely may become necessary. One idea jointly samples the hierarchical posterior by fitting density estimates to each single event, promising to avoid the single-event Monte Carlo integrals \citep{Rinaldi:2021bhm,Callister:2022qwb,Hussain:2024qzl,Mancarella:2025uat}. However, it will also be important to remove the Monte Carlo approximation to the selection efficiency, which may be tackled by analytic or machine learning approaches \citep{Lorenzo-Medina:2024opt, Talbot:2020oeu, Gerosa:2020pgy, Callister:2024qyq}. Another promising technique proposes avoiding approximate likelihoods entirely via likelihood-free, simulation-based inference \citep{Leyde:2023iof}.
Though the impact of uncertainty due to the approximations in these proposals are typically neglected, we argue that our approach (though based on the common Monte Carlo method for likelihood estimation) can be extended to these in the future as long as one can construct an estimate for the covariance in the log-likelihood surface. 

\section{Acknowledgments}

We gratefully acknowledge helpful conversations with Reed Essick, Will Farr, Jacob Golomb, Matthew Mould and Colm Talbot. 
Additionally, we thank Reed Essick for agreeing to perform the LIGO internal document review.
J.H. is supported by the NSF Graduate Research Fellowship under Grant No. DGE1122374.
J.H. and S.V. are partially supported by the NSF grant PHY-2045740.
This material is based upon work supported by NSF's LIGO Laboratory which is a major facility fully funded by the National Science Foundation.
The authors are grateful for computational resources provided by the LIGO Laboratory and supported by National Science Foundation Grants PHY-0757058 and PHY-0823459.

%% file: general_case.tex
\section{Why does normalization matter?}
\label{app: normalization discussion}

Probabilistic samplers (MCMC, nested sampling, etc.) do not require a normalized estimate of the posterior uncertainty, and so it may seem surprising that sampling routines can be biased by a normalization factor that does not appear explicitly. Briefly, we argue that bias appears in the following manner: consider an ensemble of 1000 independent unnormalized likelihood times prior estimators and samples of 1000 points from each. If the samples are combined uniformly across the ensemble of estimators, then this is equivalent to sampling from the mean of the 1000 \textit{normalized posteriors}. The actual overall normalization itself doesn’t matter, but combining the samples this way means the \textit{relative} normalizations between the likelihood times prior estimators do matter. If the samples are combined by instead weighting by their corresponding evidence, this would be equivalent to sampling from the mean of the unnormalized likelihoods. 

We used the mean posterior as one method for quantifying a more general phenomenon: with an uncertain likelihood estimator---even if we know that the likelihood estimator is unbiased---one finds that the ``typical'' likelihood estimator is systematically scattered. Likelihood uncertainty tends to be very skewed (a lognormal distribution is often a good approximation; the hierarchical likelihood is the product of many uncertain quantities and hence the multiplicative central limit theorem applies).
 
To say it in another way, if we want to quantify the bias from using an uncertain likelihood estimator, we need to decide how to average over the samples from the ensemble of uncertain likelihood estimators. If we just want an unbiased likelihood estimator and we therefore weight each set of samples by their evidences, then we will tend to have an extremely inefficient weighting: we have to wait a long time for some very rare likelihood estimator realizations from within the high-skew tail. If we instead demand an unbiased posterior estimator and we combine samples uniformly, then a typical realization from the unbiased posterior estimator should be closer to the average. 

\section{Derivation of bias in a general posterior estimator}
\label{app: bias in general}

Consider the general case where we have a Bayesian posterior, 
\beq
p(\Lambda) = \frac{\pi(\Lambda)\mathcal{L}(\Lambda)}{\int {\rm d}\Lambda' \pi(\Lambda')\mathcal{L}(\Lambda')}
\eeq
where the likelihood times prior is not tractable analytically. Suppose we estimate $\pi(\Lambda)\mathcal{L}(\Lambda) \approx g(\hat{\bm X}(\Lambda))$ with a random variable $\hat{\bm X}(\Lambda)$ which varies across parameter space. For instance, $\hat{\bm X}(\Lambda)$ may be a Monte Carlo integral reweighted across $\Lambda$ parameter space, as we examine in detail in this paper. The posterior estimator is then
\beq
\hat{p}(\Lambda) = \frac{g(\hat{\bm X}(\Lambda))}{\int {\rm d}\Lambda' g(\hat{\bm X}(\Lambda'))}.
\label{eq: estimated posterior}
\eeq
In general, $\hat{\bm X}(\Lambda)$ is a vector of random variables; in the Monte Carlo example for hierarchical inference, $\hat{\bm X}(\Lambda)$ is the list of the $\Nobs$ single-event Monte Carlo integrals and the selection efficiency Monte Carlo estimator.\footnote{Formally $\hat{\bm X}(\Lambda)$ is a vector of random \textit{processes}. A single component $\hat{X}_i(\Lambda)$ describes a random \textit{function} across $\Lambda$ parameter space, which are known as a random processes.}

We assume that $\hat{\bm X}$ is designed so that $g(\expect{\hat{\bm X}}) = \pi(\Lambda)\mathcal{L}(\Lambda)$ is exactly equal to the likelihood times prior we are trying to approximate. However, because $g$ may be a nonlinear function of $\hat{\bm X}$, then (as we discuss in Sec.~\ref{subsec: monte carlo bias}) $\expect{g(\hat{\bm X})} \neq g(\expect{\hat{\bm X}})$. We refer to this phenomenon as \textit{likelihood bias}, and we compute a recipe for removing the leading order biasing term. However, because of the nonlinearity in the normalization by the evidence, posterior estimators acquire an additional bias. We derive an additional correction that removes this bias, although it is typically expensive to calculate and in many practical cases it tends to be small.

To compute the bias, we use the same approach as in Sec.~\ref{subsec: monte carlo bias} to take the expectation of Eq.~\ref{eq: estimated posterior} over the random variable $\hat{\bm X}(\Lambda)$. We write the expectation as $\expect{\hat{\bm X}(\Lambda)}$, and the deviation from the expectation as $\hat{\bm \delta}(\Lambda) = \hat{\bm X}(\Lambda) - \expect{\hat{\bm X}(\Lambda)}$. The deviation has components $\hat{\delta}(\Lambda)_i$. Partial derivatives of $g$ with respect to the $i^{\rm th}$ component of $\hat{\bm X}$ will be written $\partial_i$, and repeated indices are implicitly summed over. 

Approximating Eq.~\ref{eq: estimated posterior} to second order in the numerator and denominator using a Taylor expansion around the expectation $\expect{\hat{\bm X}(\Lambda)}$,
\beq
\hat{p}(\Lambda) = \frac{g(\expect{\hat{\bm X}(\Lambda)}) + \partial_ig(\expect{\hat{\bm X}(\Lambda)}) \hat{\delta}_i(\Lambda) + \frac{1}{2}\partial_i\partial_jg(\expect{\hat{\bm X}(\Lambda)})\hat{\delta}_i(\Lambda)\hat{\delta}_j(\Lambda) + \mathcal{O}(\hat{\bm \delta}^3)}
{\int {\rm d}\Lambda' g(\expect{\hat{\bm X}(\Lambda')}) + \partial_kg(\expect{\hat{\bm X}(\Lambda')}) \hat{\delta}_k(\Lambda') + \frac{1}{2}\partial_k\partial_lg(\expect{\hat{\bm X}(\Lambda')})\hat{\delta}_k(\Lambda')\hat{\delta}_l(\Lambda') + \mathcal{O}(\hat{\bm \delta}^3)} 
\eeq
Using the binomial expansion for the denominator and keeping terms only to second order, we are left with
\begin{align}
\hat{p}(\Lambda) = \frac{
g(\expect{\hat{\bm X}(\Lambda)})
}{
\mathcal{Z}
}
&\Bigg(
1 
+ 
\frac{\partial_ig(\expect{\hat{\bm X}(\Lambda)})}{g(\expect{\hat{\bm X}(\Lambda)})}\hat{\delta}_i(\Lambda)
+ 
\frac{1}{2}\frac{\partial_i\partial_jg(\expect{\hat{\bm X}(\Lambda)})}{g(\expect{\hat{\bm X}(\Lambda)})}\hat{\delta}_i(\Lambda)\hat{\delta}_j(\Lambda)
-
\frac{1}{\mathcal{Z}}\int {\rm d}\Lambda'\partial_ig(\expect{\hat{\bm X}(\Lambda')})\hat{\delta}_i(\Lambda') \nonumber \\
&- \frac{1}{2\mathcal{Z}}\int {\rm d}\Lambda' \partial_i\partial_jg(\expect{\hat{\bm X}(\Lambda')})\hat{\delta}_i(\Lambda')\hat{\delta}_j(\Lambda')
-
\frac{\partial_ig(\expect{\hat{\bm X}(\Lambda)})}{g(\expect{\hat{\bm X}(\Lambda)})}\hat{\delta}_i(\Lambda) \frac{1}{\mathcal{Z}}\int {\rm d}\Lambda'\partial_jg(\expect{\hat{\bm X}(\Lambda')})\hat{\delta}_j(\Lambda')
\nonumber \\ 
&+
\frac{1}{\mathcal{Z}^2}\int {\rm d}\Lambda' {\rm d}\Lambda''\partial_ig(\expect{\hat{\bm X}(\Lambda')})\partial_jg(\expect{\hat{\bm X}(\Lambda'')})\hat{\delta}_i(\Lambda')\hat{\delta}_j(\Lambda'') + \mathcal{O}(\hat{\bm \delta}^3)\Bigg)
\end{align}
To compute the bias, we take the expectation with respect to $\hat{\bm X}$. Each term linear in $\hat{\bm \delta}$ vanishes, since the expectation of $\hat{\bm \delta}$ is zero, and only the quadratic terms remain. The represents the expected posterior probability density over many draws of the random variable $\hat{\bm X}$
\begin{align}
\expect{\hat{p}(\Lambda&)} = \frac{
g(\expect{\hat{\bm X}(\Lambda)})
}{
\mathcal{Z}
}
\Bigg(
1 
+ 
\frac{1}{2}\frac{\partial_i\partial_jg(\expect{\hat{\bm X}(\Lambda)})}{g(\expect{\hat{\bm X}(\Lambda)})}C_{i,j}(\Lambda, \Lambda)
- 
\frac{1}{\mathcal{Z}}\frac{\partial_ig(\expect{\hat{\bm X}(\Lambda)})}{g(\expect{\hat{\bm X}(\Lambda)})} \int {\rm d}\Lambda'\partial_jg(\expect{\hat{\bm X}(\Lambda')})C_{i,j}(\Lambda, \Lambda') \nonumber \\
- &\frac{1}{2\mathcal{Z}}\int {\rm d}\Lambda' \partial_i\partial_jg(\expect{\hat{\bm X}(\Lambda')})C_{i,j}(\Lambda', \Lambda')
+
\frac{1}{\mathcal{Z}^2}\int {\rm d}\Lambda' {\rm d}\Lambda'' \partial_i g(\expect{\hat{\bm X}(\Lambda')})\partial_j g(\expect{\hat{\bm X}(\Lambda'')})C_{i,j}(\Lambda', \Lambda'')+ \mathcal{O}(\hat{\bm \delta}^3)
\Bigg)
\label{eq: full bias}
\end{align}
where $C_{i,j}(\Lambda_0, \Lambda_1) = \expect{(\hat{X}_i(\Lambda_0) - \expect{\hat{X}_i(\Lambda_0)})(\hat{X}_j(\Lambda_1) - \expect{\hat{X}_j(\Lambda_1)})} =  \expect{\hat{\delta}_i(\Lambda_0)\hat{\delta_j}(\Lambda_1)}$ is the covariance between $\hat{X}_i$ and $\hat{X}_j$ at points in parameter space $\Lambda_0$ and $\Lambda_1$ respectively (for independent Monte Carlo integrals, this vanishes if $i\neq j$). Explicitly, 
\beq
C_{i,j}(\Lambda_0, \Lambda_1) = \int {\rm d}\hat{\bm X} p(\hat{\bm X}) \left(\hat{X}_i(\Lambda_0) - \expect{\hat{X}_i(\Lambda_0)}\right)\left(\hat{X}_j(\Lambda_1) - \expect{\hat{X}_j(\Lambda_1)}\right)
\label{eq: covariance of single MC integral}
\eeq
where $p(\hat{\bm X})$ is the probability density over the random variable $\hat{\bm X}$. 

In Eq.~\ref{eq: full bias} the first two terms vary across $\Lambda$ parameter space, and so contribute nontrivially to the posterior. We refer to these kinds of biasing terms as \textit{parameteric bias}, as they vary across parameter space. The second two terms are simply constants which ensure the posterior is normalized over $\Lambda$. Writing the derivatives out explicitly and noticing that $\partial g /g = \partial \ln g$, then the parameteric bias is given by
\begin{gather}
b(\Lambda)= \frac{1}{2}\frac{1}{g(\expect{\hat{\bm X}(\Lambda)})}\frac{\partial^2g(\expect{\hat{\bm X}(\Lambda)})}{[\partial\hat{X}_i(\Lambda)][\partial\hat{X}_j(\Lambda)]}C_{i,j}(\Lambda, \Lambda) - \int {\rm d}\Lambda'p(\Lambda')\frac{\partial\ln g(\expect{\hat{\bm X}(\Lambda)})}{\partial\hat{X}_i(\Lambda)} \frac{\partial\ln g(\expect{\hat{\bm X}(\Lambda')})}{\partial\hat{X}_j(\Lambda')}C_{i,j}(\Lambda, \Lambda') 
\label{eq: the parametric bias} 
\end{gather}
and so 
\beq
\expect{\hat{p}(\Lambda)} = p(\Lambda)\left[1 + b(\Lambda) - \int b(\Lambda')\dd\Lambda'\right]
\eeq
The first biasing term in Eq.~\ref{eq: the parametric bias} comes from any nonlinearity in the function $g$, or covariance in the estimator $\hat{\bm X}$. For the case of gravitational wave population inference, it becomes the likelihood biasing term in Eq.~\ref{eq: bias for mean raised to power}. If the estimators $\hat{X}_i$ are each independent (the cross terms vanish) and there is no nonlinearity of the estimators (the second derivatives vanish), then the first biasing term is zero. 
However, in essentially all cases, the second biasing term is nonzero and in some cases, may contribute non-negligibly to the posterior. Notice the integrand in the second term
\beq
C_{\ln g}(\Lambda, \Lambda') \equiv \sum_{i,j}\frac{\partial\ln g(\expect{\hat{\bm X}(\Lambda)})}{\partial\hat{X}_i(\Lambda)} \frac{\partial\ln g(\expect{\hat{\bm X}(\Lambda')})}{\partial\hat{X}_j(\Lambda')}C_{i,j}(\Lambda, \Lambda') 
\label{eq: general covariance in ln g}
\eeq
is exactly the propagation of uncertainty rules for the covariance in $\ln g$: the covariance in the log likelihood (here we write out the implicit sum over $i,j$).

\section{Correcting likelihood bias and posterior bias}
\label{app: correction in general}

We would like to handle the bias in Eq.~\ref{eq: the parametric bias}. Here, we will restrict to the case where $\hat{\bm X}(\Lambda)$ is a family of independent Monte Carlo estimators.
\beq
\hat{X}_j(\Lambda) = \frac{1}{M_j}\sum_{i=1}^{M_j}\frac{f_j(\theta_i | \Lambda)}{p_{{\rm draw},j}(\theta_i)},
\label{eq: family of monte carlo integrals}
\eeq
where $g$ is some function of these Monte Carlo integrals. However, to show the modifying term we propose is correct, we will have to consider $g$ from a more general perspective, as a function of the \textit{weights}, where the Monte Carlo integrals are just the averages of the weights along the $i$ axis\footnote{We use a slightly streamlined notation for $\hat{W}_{ij}^\Lambda$ to avoid too many repeated parentheses.}
\beq
\hat{W}_{ij}^\Lambda = \frac{f_j(\theta_i | \Lambda)}{p_{{\rm draw},j}(\theta_i)} \qquad\qquad \hat{X}_j(\Lambda) = \frac{1}{M_j}\sum_{i=1}^{M_j}\hat{W}_{ij}^\Lambda.
\eeq
In Eq.~\ref{eq: expectation of Monte Carlo integral} we pointed out that the expectation of a Monte Carlo integral (even with just one sample) is simply the integral
\beq
\expectlarge{\hat{W}_{ij}^\Lambda} = \int {\rm d}\theta f_j(\theta | \Lambda)
\eeq
and so the expectation is independent of $i$.

We must consider each weight as a separate, independent random variable. Let there be an additional modifying term $m$ so that the leading order bias vanishes. 
\beq
g(\hat{\bm X}(\Lambda)) \to g\left(\left\{\frac{1}{M_j}\sum_{i=1}^{M_j}\hat{W}_{ij}^\Lambda\right\}_{j=1}^{N}\right)m\left(\left\{\hat{W}_{ij}^\Lambda\right\}_{i=1,j=1}^{M_j,N}, \left\{\hat{W}_{ij}^{\Lambda'}\right\}_{i=1,j=1}^{M_j,N}\right) \equiv g(\hat{\bm W}^\Lambda)m(\hat{\bm W}^\Lambda, \hat{\bm W}^{\Lambda'}).
\label{eq: definition of m}
\eeq
In order to retain the property that $g(\expect{\hat{\bm X}(\Lambda)}) = \pi(\Lambda)\mathcal{L}(\Lambda)$, we demand that $m(\expect{\hat{\bm W}^\Lambda}, \expect{\hat{\bm W}^{\Lambda'}}) = 1$.

So long as the modifying term is biased in exactly the opposite way to the bias in Eq.~\ref{eq: the parametric bias} at leading order, then the leading order bias will vanish. 
Based on the form of the bias in Eq.~\ref{eq: the parametric bias}, the following modifying term will work: to leading order it is exactly the opposite of the biasing terms in Eq.~\ref{eq: the parametric bias}. Furthermore, it is positive everywhere and, as we will verify, evaluates to 1 at the expectation $\expect{\hat{\bm W}^\Lambda}$
\beq
m(\hat{\bm W}^\Lambda, \hat{\bm W}^{\Lambda'}) = \exp\left(\sum_{p,q=1}^{M_q,N}\left[-\frac{1}{2}\frac{\partial^2g(\hat{\bm W}^\Lambda)}{(\partial \hat{W}^\Lambda_{pq})^2}\frac{\hat{C}^2_q(\hat{\bm W}^\Lambda, \hat{\bm W}^\Lambda)}{g(\hat{\bm W}^\Lambda)} + \int {\rm d}\Lambda' \hat{p}(\Lambda') \frac{\partial \ln g(\hat{\bm W}^\Lambda)}{\partial \hat{W}^\Lambda_{pq}}\frac{\partial \ln g(\hat{\bm W}^{\Lambda'})}{\partial \hat{W}^{\Lambda'}_{pq}}\hat{C}^2_q(\hat{\bm W}^\Lambda, \hat{\bm W}^{\Lambda'})\right]\right)
\label{eq: modifying term}
\eeq
where 
\beq
\hat{C}^2_q(\hat{\bm W}^\Lambda, \hat{\bm W}^{\Lambda'}) = \frac{1}{M_q-1}\left[\sum_{p=1}^{M_q}\hat{W}_{pq}^\Lambda\hat{W}_{pq}^{\Lambda'} - \frac{1}{M_q}\left(\sum_{i=1}^{M_q}\hat{W}_{iq}^\Lambda\right)\left(\sum_{j=1}^{M_q}\hat{W}_{jq}^{\Lambda'}\right)\right]
\label{eq: general variance estimator}
\eeq
is the minimum variance unbiased estimator for the covariance between $\hat{X}_q(\Lambda)$ and $\hat{X}_q(\Lambda')$. Note the integrand (after performing the sum) in the second term of Eq.~\ref{eq: modifying term} is covariance in the log-likelihood estimator $\ln g$ via the standard propagation of uncertainty rules
\beq
\hat{C}_{\ln g}(\Lambda, \Lambda') \equiv \sum_{p,q=1}^{M_q,N} \frac{\partial \ln g(\hat{\bm W}^\Lambda)}{\partial \hat{W}^\Lambda_{pq}}\frac{\partial \ln g(\hat{\bm W}^{\Lambda'})}{\partial \hat{W}^{\Lambda'}_{pq}}\hat{C}^2_q(\hat{\bm W}^\Lambda, \hat{\bm W}^{\Lambda'}).
\label{eq: general covariance term}
\eeq

To verify Eq.~\ref{eq: modifying term} removes the bias, we must compute the derivatives
\begin{align}
\frac{\partial \hat{C}^2_q(\hat{\bm W}^\Lambda, \hat{\bm W}^{\Lambda'})}{\partial W_{ij}^\Lambda} &= \begin{cases}
0 & {\rm if} \; j \neq q \\
\frac{1}{M_q-1}\left[\hat{W}_{ij}^{\Lambda'} - \frac{1}{M_q}\left(\sum_{p=1}^{M_q}\hat{W}_{pq}^{\Lambda'}\right)\right] & {\rm if}\; j = q 
\end{cases} \nn
\\
\frac{\partial \hat{C}^2_q(\hat{\bm W}^\Lambda, \hat{\bm W}^{\Lambda})}{\partial W_{ij}^\Lambda} &= \begin{cases}
0 & {\rm if} \; j \neq q \\
\frac{2}{M_q-1}\left[\hat{W}_{ij}^{\Lambda} - \frac{1}{M_q}\left(\sum_{p=1}^{M_q}\hat{W}_{pq}^{\Lambda}\right)\right] & {\rm if}\; j = q 
\end{cases} \nn
\\
\frac{\partial \hat{C}^2_q(\hat{\bm W}^\Lambda, \hat{\bm W}^{\Lambda'})}{(\partial W_{ij}^\Lambda)^2} &= 0 \nn
\\
\frac{\partial^2 \hat{C}^2_q(\hat{\bm W}^\Lambda, \hat{\bm W}^{\Lambda'})}{(\partial W_{ij}^\Lambda)(\partial W_{ij}^{\Lambda'})} &= \begin{cases}
0 & {\rm if} \; j \neq q \\
\frac{1}{M_q-1}\left[1 - \frac{1}{M_q}\right] & {\rm if}\; j = q 
\end{cases} = \begin{cases}
0 & {\rm if} \; j \neq q \\
\frac{1}{M_q} & {\rm if}\; j = q \end{cases} \nn
\\
\frac{\partial^2 \hat{C}^2_q(\hat{\bm W}^\Lambda, \hat{\bm W}^{\Lambda})}{(\partial W_{ij}^\Lambda)^2} &= \begin{cases}
0 & {\rm if} \; j \neq q \\
\frac{2}{M_q-1}\left[1 - \frac{1}{M_q}\right] & {\rm if}\; j = q 
\end{cases} = \begin{cases}
0 & {\rm if} \; j \neq q \\
\frac{2}{M_q} & {\rm if}\; j = q 
\end{cases}
\label{eq: variance estimator derivatives}
\end{align}
Notice, however, when we evaluate Eq.~\ref{eq: general variance estimator} and Eq.~\ref{eq: variance estimator derivatives} on the expectation of $\hat{\bm W}$, we obtain zero for the estimator, zero for the first derivative, and the second derivative does not change
\begin{gather}
\hat{C}^2_q(\expect{\hat{\bm W}^\Lambda}, \expect{\hat{\bm W}^{\Lambda'}}) = 0 \qquad\quad \;\qquad \frac{\partial \hat{C}^2_q(\expect{\hat{\bm W}^\Lambda}, \expect{\hat{\bm W}^\Lambda}) }{\partial W_{ij}^\Lambda} = 0
\qquad \qquad \;\;  \frac{\partial^2 \hat{C}^2_q(\expect{\hat{\bm W}^\Lambda}, \expect{\hat{\bm W}^\Lambda})}{(\partial W_{ij}^\Lambda)^2} = \begin{cases}
0 & {\rm if} \; j \neq q \\
\frac{2}{M_q} & {\rm if}\; j = q
\end{cases} \quad\quad \nn \\ 
\frac{\partial \hat{C}^2_q(\expect{\hat{\bm W}^\Lambda}, \expect{\hat{\bm W}^{\Lambda'}}) }{\partial W_{ij}^\Lambda} = 0 \qquad \qquad  \frac{\partial^2 \hat{C}^2_q(\expect{\hat{\bm W}^\Lambda}, \expect{\hat{\bm W}^{\Lambda'}})}{(\partial W_{ij}^\Lambda)^2} = 0 \qquad \qquad 
\frac{\partial^2 \hat{C}^2_q(\expect{\hat{\bm W}^\Lambda}, \expect{\hat{\bm W}^{\Lambda'}})}{(\partial W_{ij}^\Lambda)(\partial W_{ij}^{\Lambda'})} = \begin{cases}
0 & {\rm if} \; j \neq q \\
\frac{1}{M_q} & {\rm if}\; j = q.
\end{cases} 
\label{eq: variance estimator derivatives expectation}
\end{gather}
We want to show that the modifying term corrects the bias.
To see this, Taylor expand
\beq
g(\hat{\bm W}^{\Lambda})\exp\left(\sum_{p,q=1}^{M_q,N}\left[-\frac{1}{2}\frac{\partial^2g(\hat{\bm W}^\Lambda)}{(\partial \hat{W}^\Lambda_{pq})^2}\frac{\hat{C}^2_q(\hat{\bm W}^\Lambda, \hat{\bm W}^\Lambda)}{g(\hat{\bm W}^\Lambda)} + \int {\rm d}\Lambda' \hat{p}(\Lambda') \frac{\partial \ln g(\hat{\bm W}^\Lambda)}{\partial \hat{W}^\Lambda_{pq}}\frac{\partial \ln g(\hat{\bm W}^{\Lambda'})}{\partial \hat{W}^{\Lambda'}_{pq}}\hat{C}^2_q(\hat{\bm W}^\Lambda, \hat{\bm W}^{\Lambda'})\right]\right)
\eeq
around the expectation $\expect{\hat{\bm W}^\Lambda}$. Many terms vanish immediately as only the second derivatives in Eq.~\ref{eq: variance estimator derivatives expectation} are nonzero. We are left with all the same terms as Eq.~\ref{eq: full bias}, plus the biasing terms from the modifying function, of which only two terms remain
\begin{align}
\hat{p}(\Lambda) &= \frac{
g(\expect{\hat{\bm W}^\Lambda})
}{
\mathcal{Z}
}
\Bigg(
1 
+ 
\sum_{i,j=1}^{M_j,N}\left(\frac{\partial g(\expect{\hat{\bm W}^\Lambda})}{\partial \hat{W}_{ij}^\Lambda}\frac{\hat{\delta}_{ij}^\Lambda}{g(\expect{\hat{\bm W}^\Lambda})}
-
\frac{1}{\mathcal{Z}}\int {\rm d}\Lambda'\frac{\partial g(\expect{\hat{\bm W}^{\Lambda'}})}{\partial \hat{W}_{ij}^{\Lambda'}}\hat{\delta}_{ij}^{\Lambda'}\right) \nonumber \\
&+ 
\frac{1}{2}\sum_{i,j=1}^{M_j,N}\sum_{k,l=1}^{M_l,N}
\left(\frac{\partial^2 g(\expect{\hat{\bm W}^\Lambda})}{(\partial \hat{W}_{ij}^\Lambda)^2}\frac{\hat{\delta}_{ij}^\Lambda\hat{\delta}_{kl}^\Lambda}{g(\expect{\hat{\bm W}^\Lambda})} -
\frac{\partial g(\expect{\hat{\bm X}(\Lambda)})}{\partial \hat{W}_{ij}^\Lambda}\frac{\hat{\delta}_{ij}^\Lambda}{g(\expect{\hat{\bm X}(\Lambda)})} \frac{1}{\mathcal{Z}}\int {\rm d}\Lambda'\frac{\partial g(\expect{\hat{\bm W}^{\Lambda'}})}{\partial \hat{W}_{kl}^{\Lambda'}}\hat{\delta}_{kl}^{\Lambda'} \right.
\nonumber \\ 
&- \left. \frac{1}{2\mathcal{Z}}\int {\rm d}\Lambda' \frac{\partial^2 g(\expect{\hat{\bm W}^{\Lambda'}})}{(\partial \hat{W}_{ij}^{\Lambda'})(\partial \hat{W}_{kl}^{\Lambda'})}\hat{\delta}_{ij}^{\Lambda'}\hat{\delta}_{kl}^{\Lambda'}
+
\frac{1}{\mathcal{Z}^2}\int {\rm d}\Lambda' {\rm d}\Lambda''\frac{\partial g(\expect{\hat{\bm W}^{\Lambda'}})}{\partial \hat{W}_{ij}^{\Lambda'}}\frac{\partial g(\expect{\hat{\bm W}^{\Lambda''}})}{\partial \hat{W}_{kl}^{\Lambda'}}\hat{\delta}_{ij}^{\Lambda'}\hat{\delta}_{kl}^{\Lambda''} \right. \nonumber \\
&+ 
\left. \sum_{p,q=1}^{M_q,N}\left[-\frac{1}{2} \frac{\partial^2g(\expect{\hat{\bm W}^\Lambda})}{(\partial \hat{W}^\Lambda_{pq})^2}\frac{\partial^2\hat{C}^2_q(\expect{\hat{\bm W}^\Lambda}, \expect{\hat{\bm W}^\Lambda})}{(\partial \hat{W}_{ij}^\Lambda)^2}\frac{\hat{\delta}_{ij}^\Lambda\hat{\delta}_{kl}^\Lambda}{g(\expect{\hat{\bm W}^\Lambda})} \nonumber \right.\right. \\ 
&+ 
\left.\left. \frac{\partial \ln g(\expect{\hat{\bm W}^\Lambda})}{\partial \hat{W}^\Lambda_{pq}}\int {\rm d}\Lambda' p(\Lambda') \frac{\partial \ln g(\expect{\hat{\bm W}^{\Lambda'}})}{\partial \hat{W}^{\Lambda'}_{pq}}\frac{\partial^2 \hat{C}^2_q(\expect{\hat{\bm W}^\Lambda}, \expect{\hat{\bm W}^{\Lambda'}})}{(\partial \hat{W}_{ij}^\Lambda)(\partial \hat{W}_{kl}^{\Lambda'})}\hat{\delta}_{ij}^\Lambda\hat{\delta}_{kl}^{\Lambda'} \nonumber \right.\right. \\ 
&+
\left.\left.
\frac{1}{2\mathcal{Z}} \int {\rm d}\Lambda' \frac{\partial^2g(\expect{\hat{\bm W}^{\Lambda'}})}{(\partial \hat{W}^{\Lambda'}_{pq})^2}\frac{\partial^2\hat{C}^2_q(\expect{\hat{\bm W}^{\Lambda'}}, \expect{\hat{\bm W}^{\Lambda'}})}{(\partial \hat{W}_{ij}^{\Lambda'})(\partial \hat{W}_{kl}^{\Lambda'})}\hat{\delta}_{ij}^{\Lambda'}\hat{\delta}_{kl}^{\Lambda'} \nonumber \right.\right. \\ 
&- 
\left.\left. \frac{1}{\mathcal{Z}}\int {\rm d}\Lambda' \frac{\partial g(\expect{\hat{\bm W}^{\Lambda'}})}{\partial \hat{W}^{\Lambda'}_{pq}}\int {\rm d}{\Lambda''} p(\Lambda'') \frac{\partial \ln g(\expect{\hat{\bm W}^{\Lambda''}})}{\partial \hat{W}^{\Lambda''}_{pq}}\frac{\partial^2 \hat{C}^2_q(\expect{\hat{\bm W}^{\Lambda'}}, \expect{\hat{\bm W}^{\Lambda''}})}{(\partial \hat{W}_{ij}^{\Lambda'})(\partial \hat{W}_{kl}^{\Lambda''})}\hat{\delta}_{ij}^{\Lambda'}\hat{\delta}_{kl}^{\Lambda''}
\right]\right)
+
\mathcal{O}(\hat{\bm \delta}^3)\Bigg).
\end{align}
As before, we can recognize that $\frac{1}{\mathcal{Z}} = \frac{p(\Lambda)}{g(\Lambda)}$ for any arbitrary $\Lambda$, and pull this inside integrals wherever relevant. Second, we notice that the nonzero branches of Eq.~\ref{eq: variance estimator derivatives expectation} picks out a small subset of the summation. Third and finally, we take the expectation over all $\hat{\delta}$, where the linear terms vanish and the quadratic terms have a nonzero expectation \textit{only} where $i=k, j=l$

\begin{align}
\expect{\hat{p}&(\Lambda)} = \frac{
g(\expect{\hat{\bm W}^\Lambda})
}{
\mathcal{Z}
}
\Bigg(
1 
+ 
\frac{1}{2}\sum_{i,j=1}^{M_j,N}
\left(\frac{\partial^2 g(\expect{\hat{\bm W}^\Lambda})}{(\partial \hat{W}_{ij}^\Lambda)^2}\frac{C_{ij,ij}(\Lambda, \Lambda)}{g(\expect{\hat{\bm W}^\Lambda})} 
-
\frac{\partial \ln g(\expect{\hat{\bm X}(\Lambda)})}{\partial \hat{W}_{ij}^\Lambda} \int {\rm d}\Lambda'p(\Lambda')\frac{\partial \ln g(\expect{\hat{\bm W}^{\Lambda'}})}{\partial \hat{W}_{ij}^{\Lambda'}}C_{ij,ij}(\Lambda, \Lambda') \right.
\nonumber \\ 
&- \left. \frac{1}{2}\int {\rm d}\Lambda' p(\Lambda')\frac{\partial^2 g(\expect{\hat{\bm W}^{\Lambda'}})}{(\partial \hat{W}_{ij}^{\Lambda'})(\partial \hat{W}_{ij}^{\Lambda'})}\frac{C_{ij,ij}(\Lambda', \Lambda')}{g(\expect{\hat{\bm W}^\Lambda})}
+
\int {\rm d}\Lambda' {\rm d}\Lambda'' p(\Lambda')p(\Lambda'')\frac{\partial \ln g(\expect{\hat{\bm W}^{\Lambda'}})}{\partial \hat{W}_{ij}^{\Lambda'}}\frac{\partial \ln g(\expect{\hat{\bm W}^{\Lambda''}})}{\partial \hat{W}_{ij}^{\Lambda'}}C_{ij,ij}(\Lambda', \Lambda'') \right. \nonumber \\
&+ 
\left. \frac{1}{M_j}\sum_{p}^{M_j}\left[- \frac{\partial^2g(\expect{\hat{\bm W}^\Lambda})}{(\partial \hat{W}^\Lambda_{pj})^2}
\frac{C_{ij,ij}(\Lambda, \Lambda)}{g(\expect{\hat{\bm W}^\Lambda})}
+ 
\frac{\partial \ln g(\expect{\hat{\bm W}^\Lambda})}{\partial \hat{W}^\Lambda_{pj}}\int {\rm d}\Lambda' p(\Lambda') \frac{\partial \ln g(\expect{\hat{\bm W}^{\Lambda'}})}{\partial \hat{W}^{\Lambda'}_{pj}}C_{ij,ij}(\Lambda, \Lambda') \nonumber \right.\right. \\ 
&+
\left.\left.
\int {\rm d}\Lambda' p(\Lambda')\frac{\partial^2g(\expect{\hat{\bm W}^{\Lambda'}})}{(\partial \hat{W}^{\Lambda'}_{pj})^2}\frac{C_{ij,ij}(\Lambda', \Lambda')}{g(\expect{\hat{\bm W}^{\Lambda'}})}
- 
\int {\rm d}\Lambda' {\rm d}{\Lambda''} p(\Lambda')p(\Lambda'') \frac{\partial \ln g(\expect{\hat{\bm W}^{\Lambda'}})}{\partial \hat{W}^{\Lambda'}_{pq}} \frac{\partial \ln g(\expect{\hat{\bm W}^{\Lambda''}})}{\partial \hat{W}^{\Lambda''}_{pq}}C_{ij,ij}(\Lambda', \Lambda'')
\right]\right) \nonumber \\
&+
\mathcal{O}(\hat{\bm \delta}^3)\Bigg).
\end{align}
The final simplification comes in noticing that $g$ is designed as a function of Monte Carlo sums, so within each $j^{\rm th}$ Monte Carlo sum, the partial derivative of with respect to any weight $\hat{W}_{pj}^\Lambda$ is the same for all $p$. In particular, the sum over all $p$ is just $M_j$ copies of the same partial derivative. We can fix the derivatives with respect to $\hat{W}_{pj}^\Lambda$, $\hat{W}_{pj}^{\Lambda'}$ and $\hat{W}_{pj}^{\Lambda''}$ to the $ij$ weight, and multiply by $M_j$. All terms then cancel, leaving only the third order contribution to the bias.
\beq
\expect{\hat{p}(\Lambda)} = \frac{g(\expect{\hat{\bm W}^\Lambda})}{\mathcal{Z}}\left(1 + \mathcal{O}(\hat{\bm \delta}^3)\right).
\eeq
The first biasing term (the likelihood bias) is corrected by the the first term in the exponential of the modifying term (Eq.~\ref{eq: modifying term}). The second biasing term (the posterior bias) is corrected by the second term in Eq.~\ref{eq: modifying term}. 

The general likelihood bias correction term is therefore
\beq
\exp\left(\sum_{p,q=1}^{M_q,N}-\frac{1}{2}\frac{\partial^2g(\hat{\bm W}^\Lambda)}{(\partial \hat{W}^\Lambda_{pq})^2}\frac{\hat{C}^2_q(\hat{\bm W}^\Lambda, \hat{\bm W}^\Lambda)}{g(\hat{\bm W}^\Lambda)}\right)
\label{eq: general likelihood bias correction}
\eeq
which is only nontrivial when there are nonzero second derivatives of the likelihood estimator. On the other hand, the general posterior bias correction term is always nontrivial, but can be estimated with the accuracy statistic $\hat{A}[\hat{p}]$ and tends to be negligible in practice.

We show some examples where we explicitly correct the posterior bias in Appendix~\ref{app: examples of posterior correction}.

\section{Proof of covariance inequalities}
\label{app: covariance cauchy-Schwarz}

In this appendix, we prove the covariance inequalities in Eqs.~\ref{eq: covariance Schwarz inequality} and \ref{eq: covariance Schwarz inequality estimator}. We show the inequality holds for the covariance (as opposed to the estimator) but each step can be replaced with the corresponding estimators to show the analogous inequality in Eq.~\ref{eq: covariance Schwarz inequality estimator}.

We begin with the definition of the covariance of the log-likelihood from Eq.~\ref{eq: general covariance in ln g}, where we restrict ourselves to the case where the covariance between different independent Monte Carlo integrals is zero $C_{i\neq j} = 0$. We know the following holds:
\beq
|C_{\ln g}(\Lambda, \Lambda')| = \left|\sum_{i} \frac{\partial\ln g(\expect{\hat{\bm X}(\Lambda)})}{\partial\hat{X}_i(\Lambda)} \frac{\partial\ln g(\expect{\hat{\bm X}(\Lambda')})}{\partial\hat{X}_i(\Lambda')}C_{i}(\Lambda, \Lambda')\right| \le \sum_{i} \left|\frac{\partial\ln g(\expect{\hat{\bm X}(\Lambda)})}{\partial\hat{X}_i(\Lambda)} \frac{\partial\ln g(\expect{\hat{\bm X}(\Lambda')})}{\partial\hat{X}_i(\Lambda')}C_{i}(\Lambda, \Lambda')\right|,
\eeq
where $C_i(\Lambda, \Lambda')$ is the covariance from the $i^{\rm th}$ Monte Carlo integral (Eq.~\ref{eq: covariance of single MC integral}). We may apply Cauchy-Schwarz on the $L^2$ inner product in the definition of $C_i(\Lambda, \Lambda')$
\beq
|C_{i}(\Lambda, \Lambda')| \equiv \left |\int {\rm d}\hat{\bm X} p(\hat{\bm X}) \left(\hat{X}_i(\Lambda) - \expect{\hat{X}_i(\Lambda)}\right)\left(\hat{X}_i(\Lambda') - \expect{\hat{X}_i(\Lambda')}\right)\right| \le \sqrt{C_i(\Lambda, \Lambda)C_i(\Lambda', \Lambda')}.
\eeq
$C_i(\Lambda, \Lambda) \ge 0$ so we drop the absolute value symbols inside the square root.
Multiplying both sides
\beq
\left|\frac{\partial\ln g(\expect{\hat{\bm X}(\Lambda)})}{\partial\hat{X}_i(\Lambda)} \frac{\partial\ln g(\expect{\hat{\bm X}(\Lambda')})}{\partial\hat{X}_i(\Lambda')}C_{i}(\Lambda, \Lambda')\right| \le \sqrt{\left(\frac{\partial\ln g(\expect{\hat{\bm X}(\Lambda)})}{\partial\hat{X}_i(\Lambda)}\right)^2C_i(\Lambda, \Lambda)}\sqrt{\left(\frac{\partial\ln g(\expect{\hat{\bm X}(\Lambda')})}{\partial\hat{X}_i(\Lambda')}\right)^2C_i(\Lambda', \Lambda')},
\eeq
and then by the Cauchy-Schwarz inequality applied to the usual dot product $|\sum_i a_i b_i| \leq \sqrt{\sum_i a_i^2\sum_i b_i^2}$ where $a_i$ is the first square root term and $b_i$ is the second square root term, 
\begin{multline}
\left|\sum_i\sqrt{\left(\frac{\partial\ln g(\expect{\hat{\bm X}(\Lambda)})}{\partial\hat{X}_i(\Lambda)}\right)^2C_i(\Lambda, \Lambda)}\sqrt{\left(\frac{\partial\ln g(\expect{\hat{\bm X}(\Lambda')})}{\partial\hat{X}_i(\Lambda')}\right)^2C_i(\Lambda', \Lambda')} \right| \le \\
\sqrt{\sum_i\left(\frac{\partial\ln g(\expect{\hat{\bm X}(\Lambda)})}{\partial\hat{X}_i(\Lambda)}\right)^2C_i(\Lambda, \Lambda)}
\sqrt{\sum_i\left(\frac{\partial\ln g(\expect{\hat{\bm X}(\Lambda')})}{\partial\hat{X}_i(\Lambda')}\right)^2C_i(\Lambda', \Lambda')} = \sqrt{C_{\ln g}(\Lambda, \Lambda)} \sqrt{C_{\ln g}(\Lambda', \Lambda')}.
\end{multline}
Linking together the above inequalities proves the covariance inequality of Eq.~\ref{eq: covariance Schwarz inequality}. As stated above, the same reasoning shows the analogous inequality for the covariance \textit{estimator} of Eq.~\ref{eq: covariance Schwarz inequality estimator}.

%% file: gaussian_appendix.tex
\section{Additional examples for a Gaussian hierarchical inference}
\label{app: additional gaussians}

\subsection{Parameter Estimation uncertainty only}
\label{app: gaussian pe uncertainty only}

\begin{figure}[h!]
    \centering
    \includegraphics[width=\linewidth]{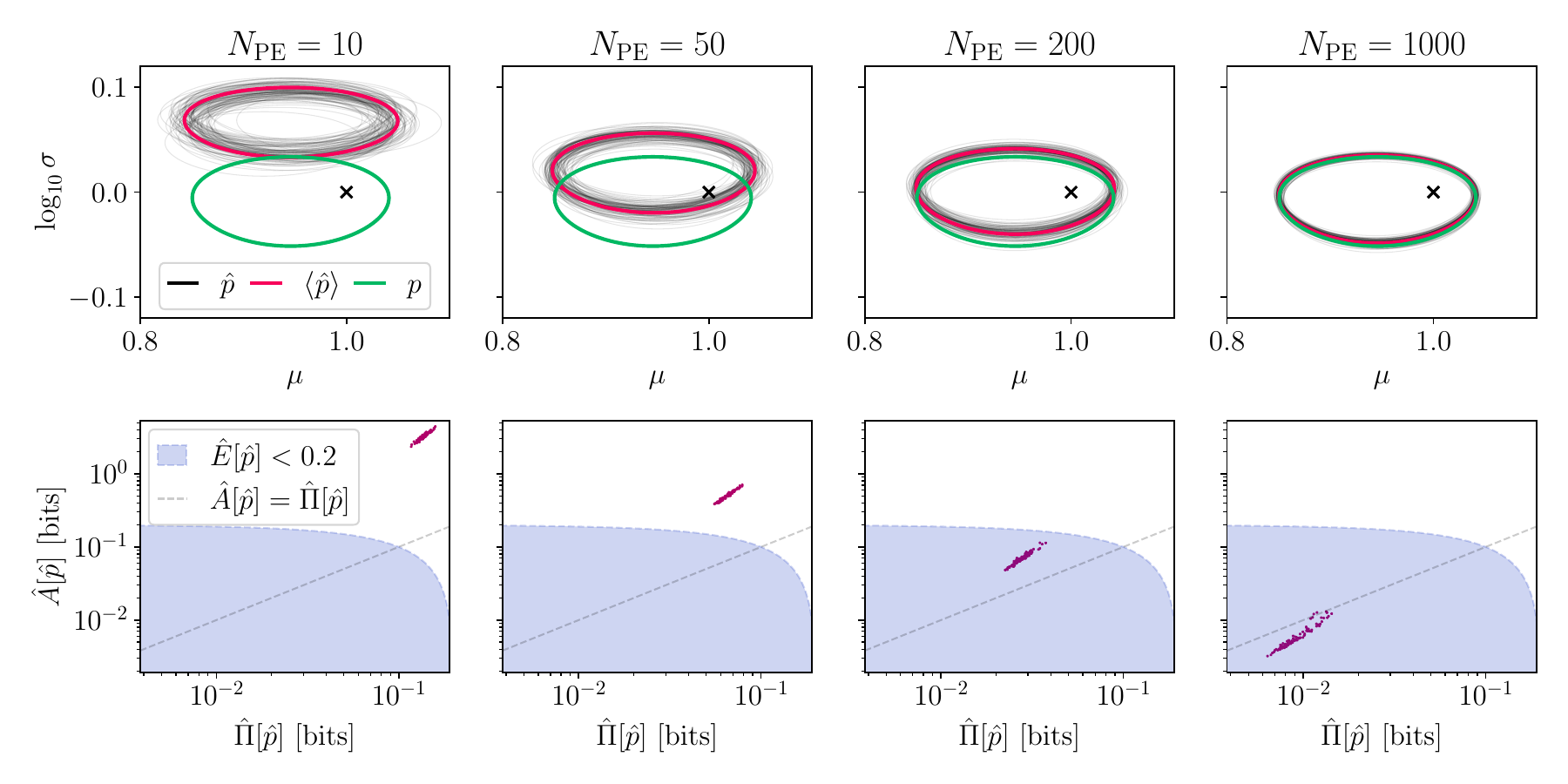}
    \caption{Same as Fig.~\ref{fig: nobs 100 gaussian example}, but with 1000 observations. Note the zoomed in x- and y-axes relative to Fig.~\ref{fig: nobs 100 gaussian example}. }
    \label{fig: nobs 1000 gaussian example}
\end{figure}

\begin{figure}[h!]
    \centering
    \includegraphics[width=\linewidth]{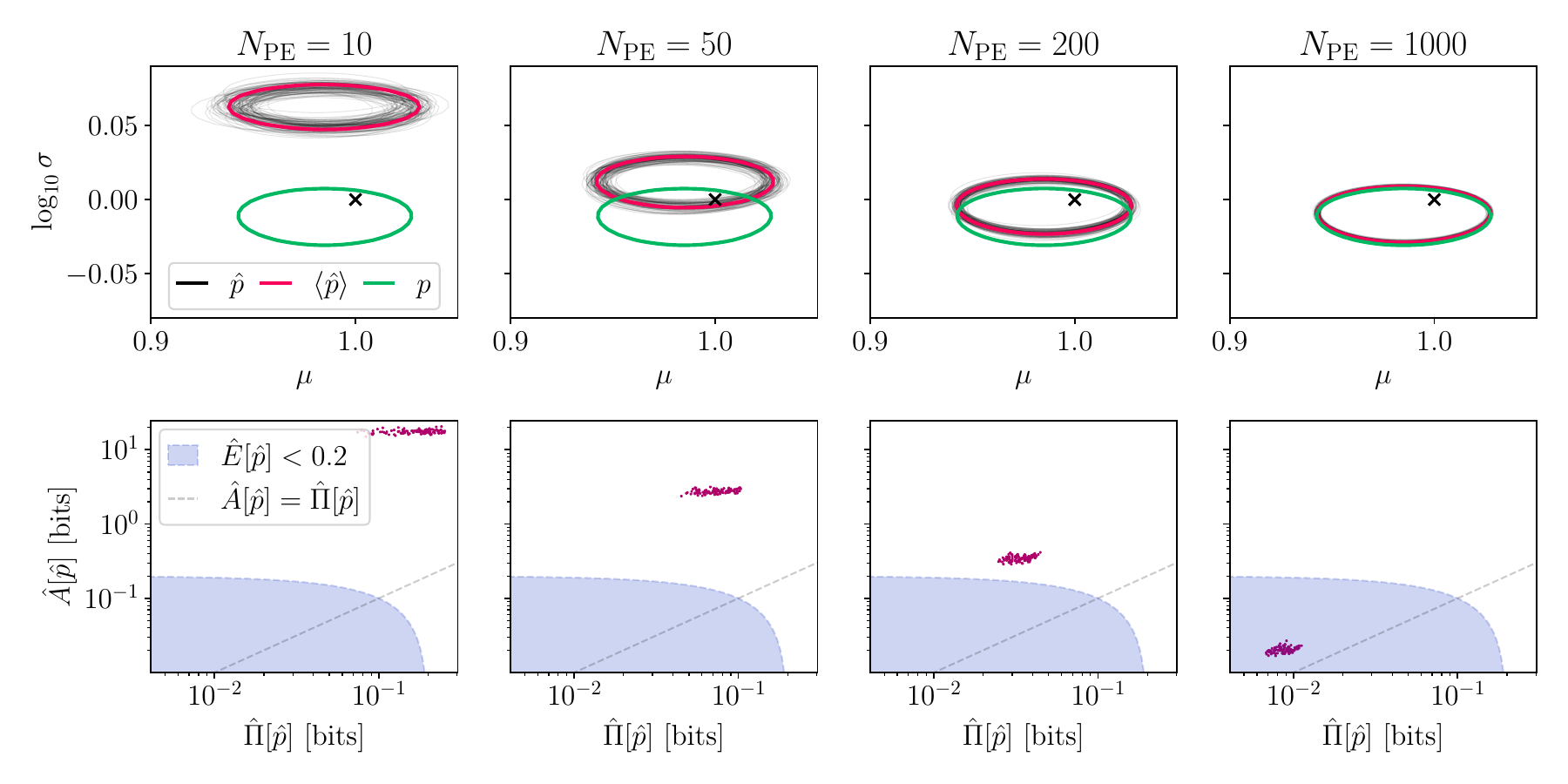}
    \caption{Same as Fig.~\ref{fig: nobs 100 gaussian example}, but with 5000 observations. Note the zoomed in x- and y-axes relative to Fig.~\ref{fig: nobs 100 gaussian example}. }
    \label{fig: nobs 5000 gaussian example}
\end{figure}

Here we show additional examples from Sec.~\ref{subsec: example without selection effects} of the Gaussian hierarchical inference with $\Nobs = 1000$ and $\Nobs = 5000$. Notice the error statistic scalings are roughly matched by the analytic predictions in Table~\ref{tab: scaling}. Restricting our attention to $\NPE = 1000$, in Figs.~\ref{fig: nobs 100 gaussian example},~\ref{fig: nobs 1000 gaussian example} and~\ref{fig: nobs 5000 gaussian example} we have $\Nobs=100$, $1000$, and $5000$ and $\hat{\Pi}[\hat{p}]\sim 10^{-2}$: roughly constant in $\Nobs$. 
The accuracy statistics scale roughly linearly: $\hat{A}[\hat{p}]\sim 5\times10^{-4}, 5\times10^{-3}, 0.02$ bits. For a more quantitative scaling analogous to our precision statistic scaling, see Fig. 2 in \citet{Essick:2022ojx}.

We also show examples where the noise is larger than the scale of the population: each single event posterior is larger than the width of the population. The hyperposteriors are correspondingly broader, and we see that we require much larger $\NPE$ in order to obtain a similar level of precision and accuracy. We include this to emphasize that the size of $\NPE$ necessary can vary significantly over different applications.

\begin{figure}[h!]
    \centering
    \includegraphics[width=\linewidth]{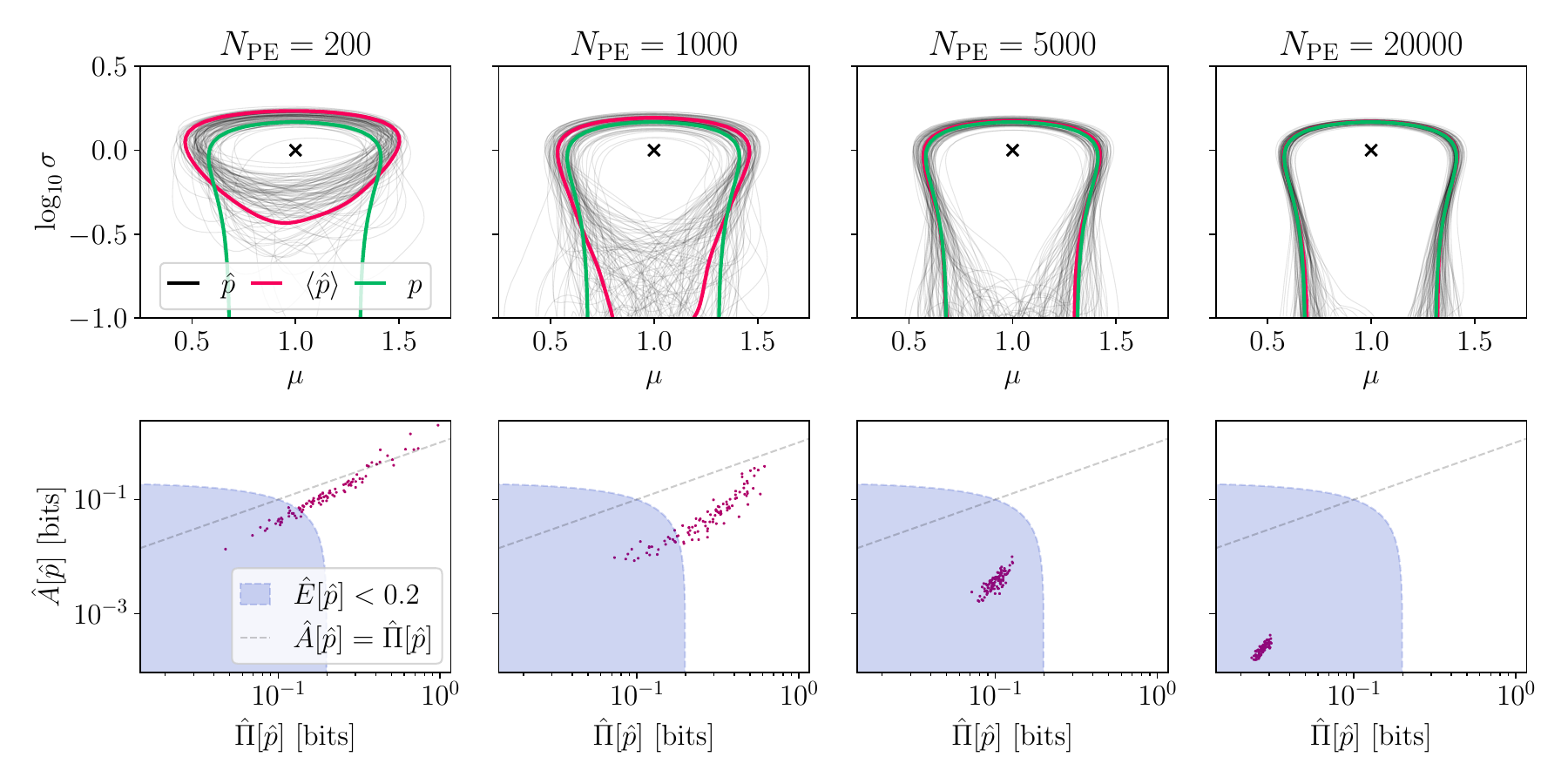}
    \caption{Same as Fig.~\ref{fig: nobs 100 gaussian example}, but with $\Nobs=100$ observations with larger noise $\sigma_n=2$ than the scale of the population $\sigma=1$. }
    \label{fig: nobs 100 larger noise}
\end{figure}

\begin{figure}[h!]
    \centering
    \includegraphics[width=\linewidth]{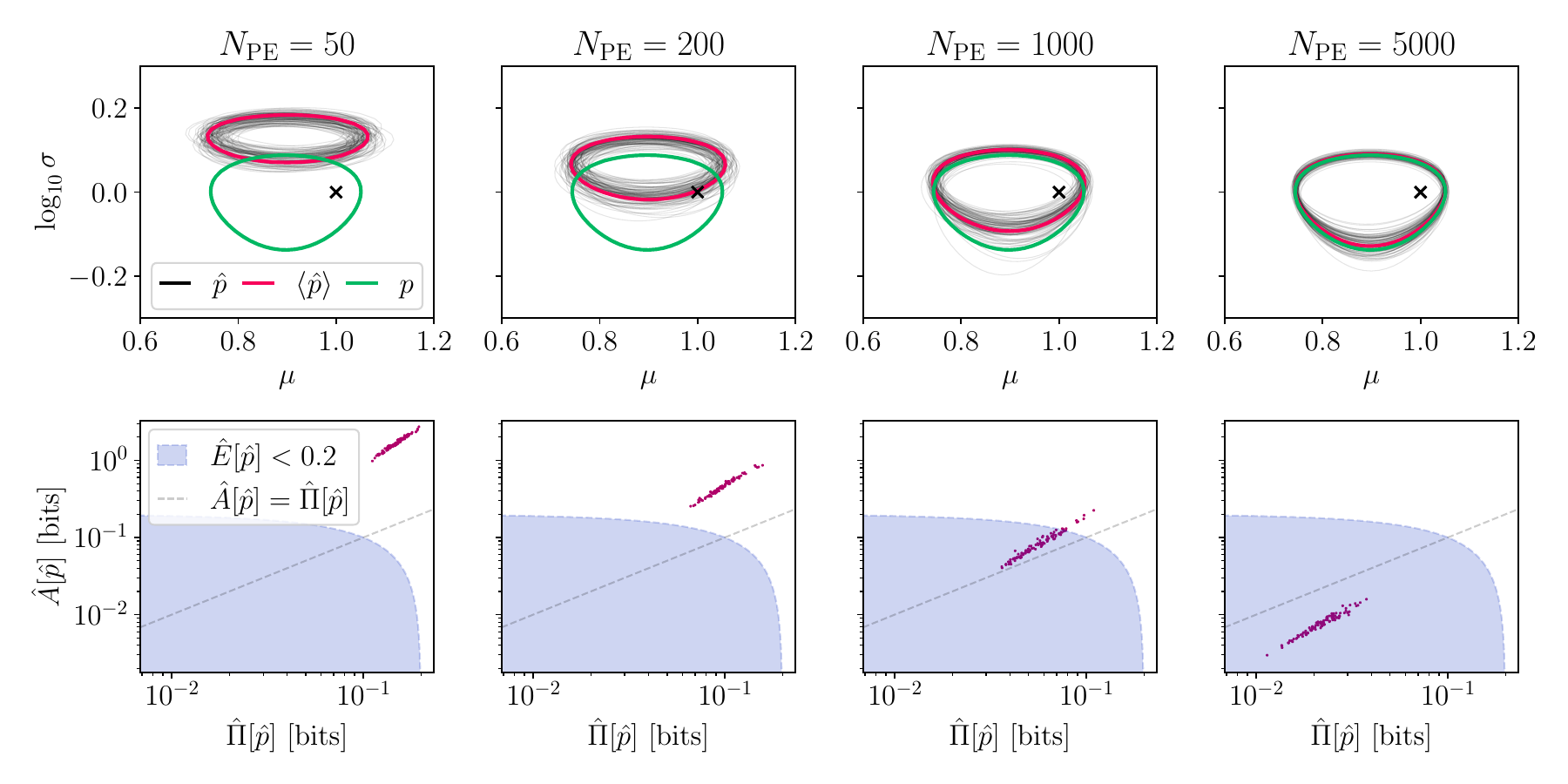}
    \caption{Same as Fig.~\ref{fig: nobs 100 gaussian example}, but with $\Nobs=1000$ observations with larger noise $\sigma_n=2$ than the scale of the population $\sigma=1$. }
    \label{fig: nobs 1000 larger noise}
\end{figure}

\begin{figure}[h!]
    \centering
    \includegraphics[width=\linewidth]{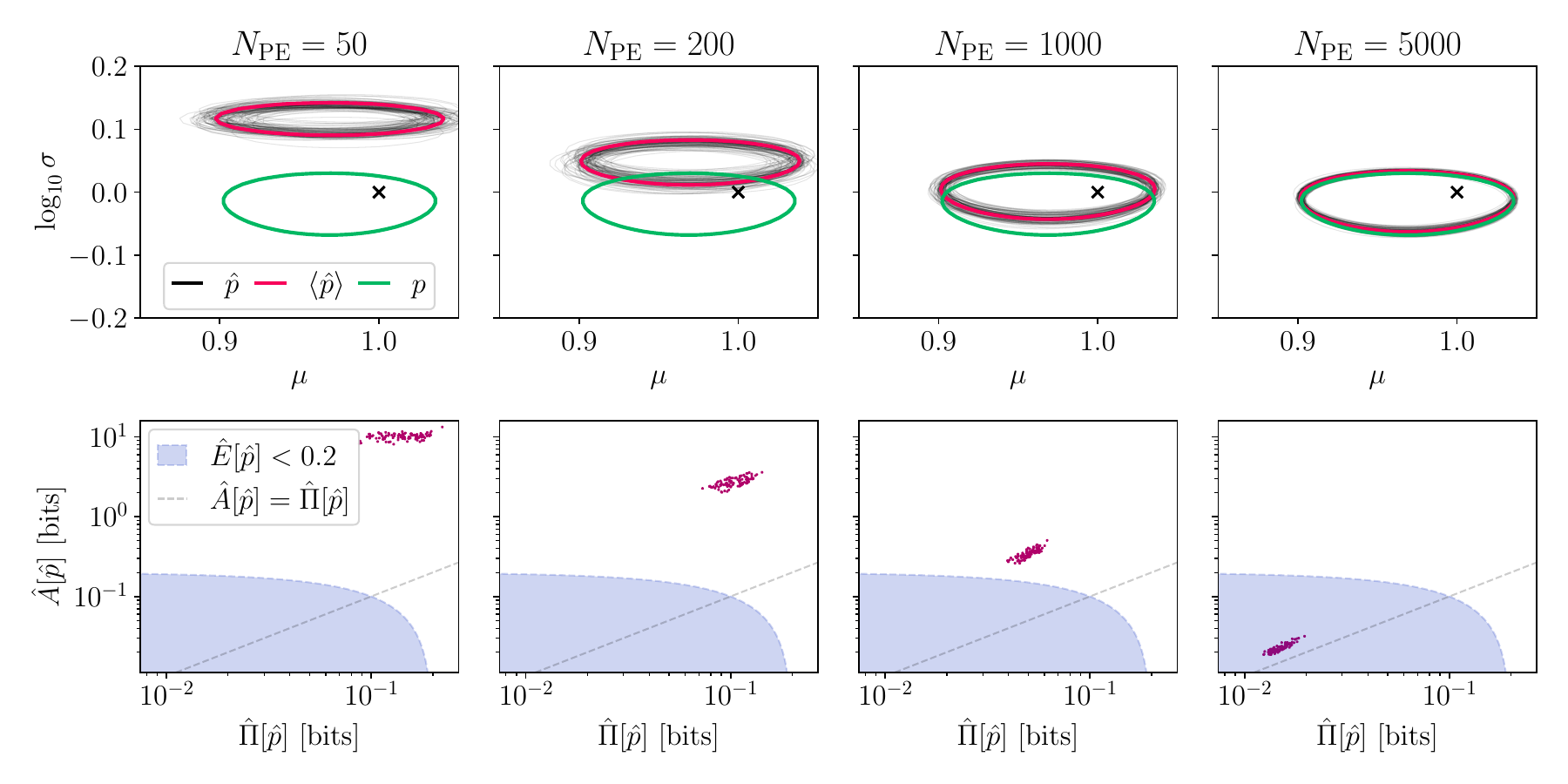}
    \caption{Same as Fig.~\ref{fig: nobs 100 gaussian example}, but with $\Nobs=5000$ observations with larger noise $\sigma_n=2$ than the scale of the population $\sigma=1$. }
    \label{fig: nobs 5000 larger noise}
\end{figure}

\clearpage

\subsection{Selection effects estimation uncertainty only}
\label{app: gaussian vt uncertainty only}

We include some additional examples to Sec.~\ref{subsec: example with selection effects} where we vary the number of injections $\Ninj$ and the strength of the selection criterion. Some of the posterior estimators shown here are limited by the resolution on the underlying grid of $\sigma$ and $\mu$. 

\begin{figure}[h!]
    \centering
    \includegraphics[width=\linewidth]{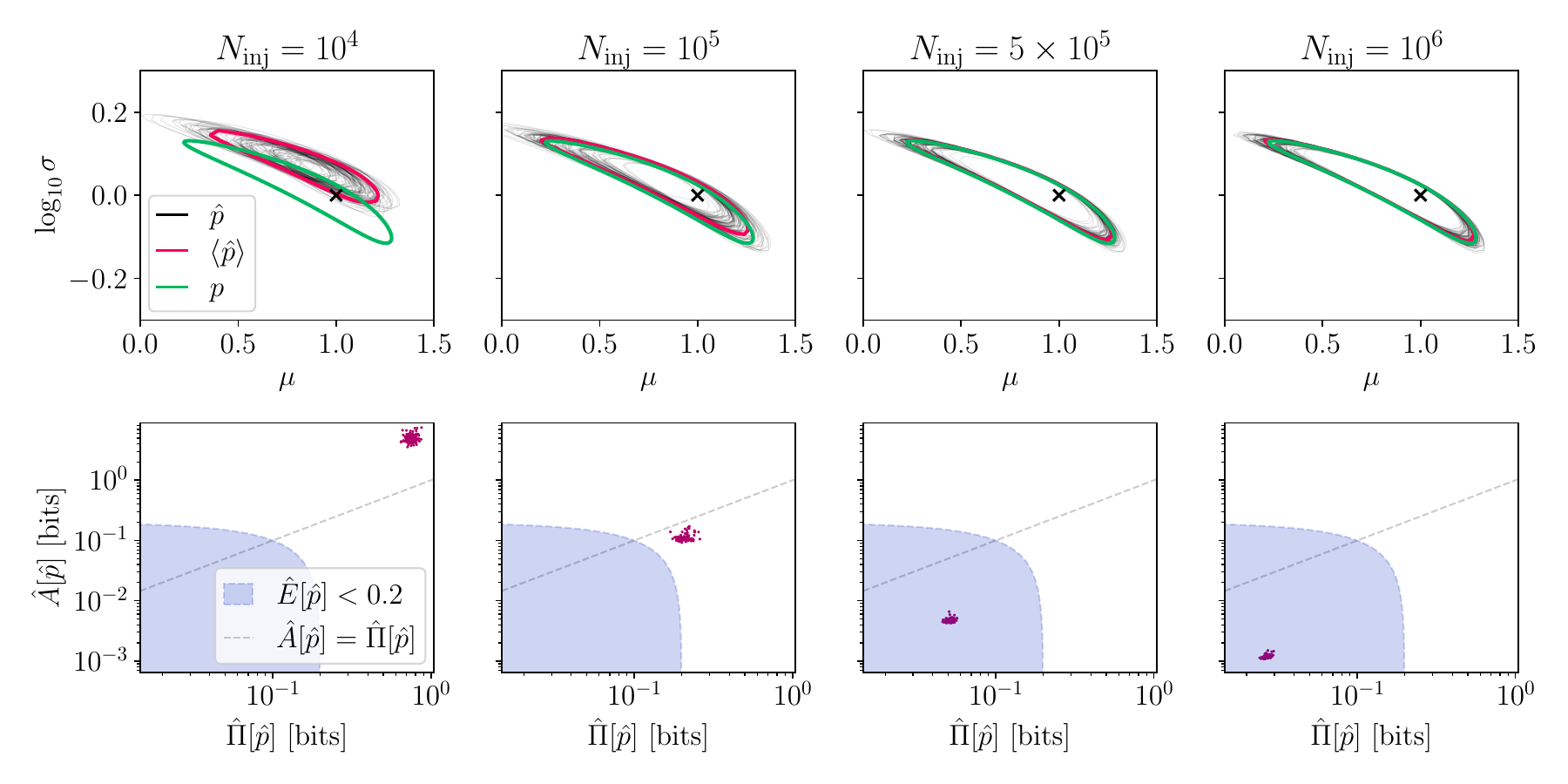}
    \caption{Same as Fig.~\ref{fig: nobs 100 corrected vt 1 gaussian example}, with $\Nobs=1000$ observations and the selection given by $d=\theta+n > 1$ and using the corrected likelihood estimator of Eq.~\ref{eq: unbiased likelihood estimator}.}
    \label{fig: nobs 1000 corrected vt 1 gaussian example}
\end{figure}

\begin{figure}[h!]
    \centering
    \includegraphics[width=\linewidth]{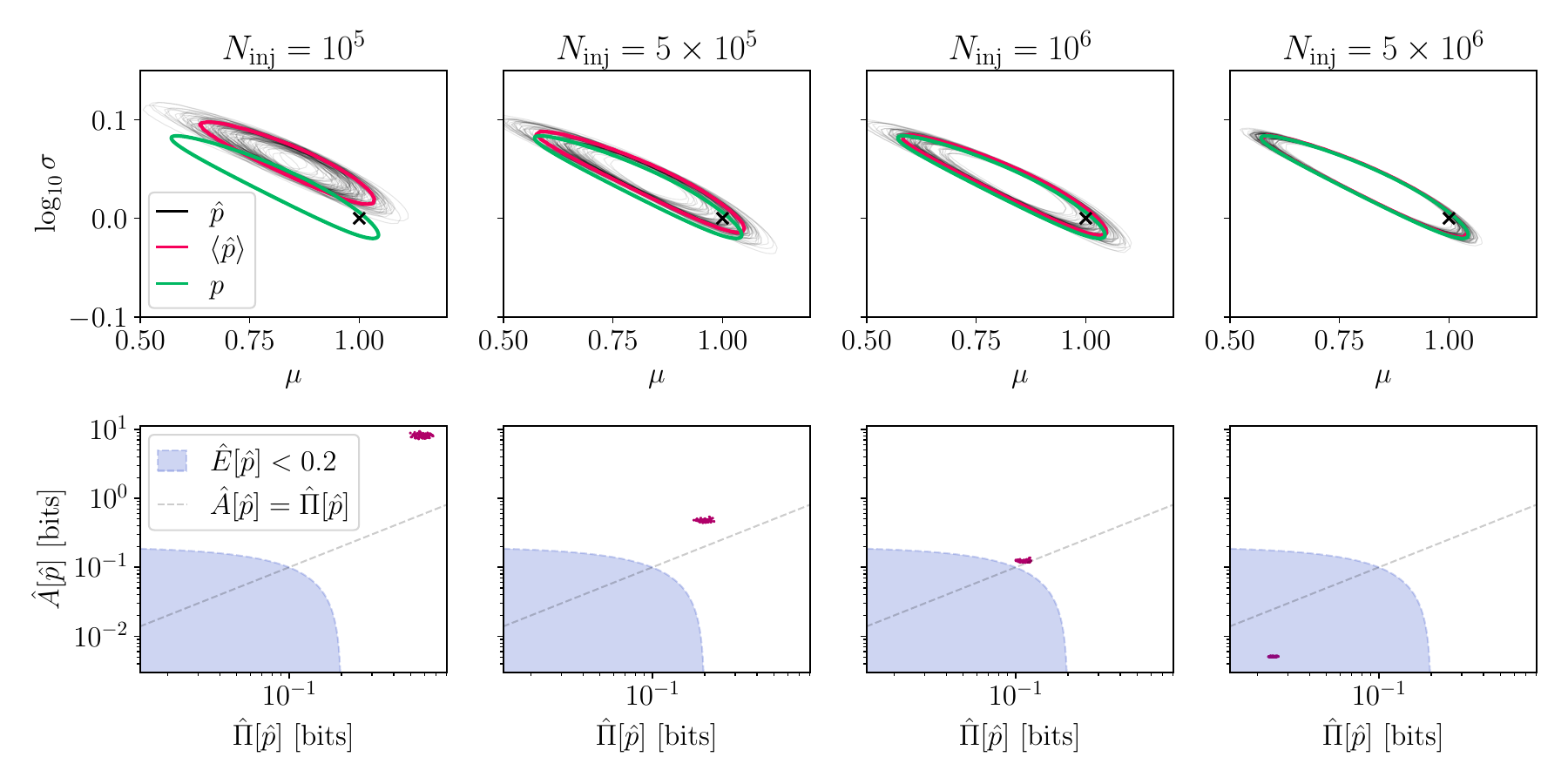}
    \caption{Same as Fig.~\ref{fig: nobs 100 corrected  vt 1 gaussian example}, with $\Nobs=5000$ observations and the selection given by $d=\theta+n > 1$ and using the corrected likelihood estimator of Eq.~\ref{eq: unbiased likelihood estimator}.}
    \label{fig: nobs 5000 corrected vt 1 gaussian example}
\end{figure}

Note that the scaling is roughly similar to the scaling arguments shown in Table~\ref{tab: scaling}. Restricting our attention to the case with selection $d=\theta+n>1$ and $\Ninj = 10^6$, in Figs.~\ref{fig: nobs 100 corrected vt 1 gaussian example},~\ref{fig: nobs 1000 corrected vt 1 gaussian example}, and~\ref{fig: nobs 5000 corrected vt 1 gaussian example} we have $\Nobs=100$, $1000$, and $5000$ and $\hat{\Pi}[\hat{p}]\sim 0.005, 0.03, 0.1$ bits: scaling roughly linearly in $\Nobs$. Note the posteriors are far from Gaussian, for $\Nobs = 100$, and so the scaling does not yet kick in.
For a more quantitative scaling analogous to our precision statistic scaling, see Fig. 3 in \citet{Essick:2022ojx}.

As for the scaling of the accuracy statistic, for $\Ninj = 10^6$ in Figs.~\ref{fig: nobs 100 corrected vt 1 gaussian example},~\ref{fig: nobs 1000 corrected vt 1 gaussian example}, and~\ref{fig: nobs 5000 corrected vt 1 gaussian example} we have $\Nobs=100$, $1000$, and $5000$ and $\hat{A}[\hat{p}]\sim 10^{-6}, 10^{-3}, 0.1$ bits, roughly consistent with the $\Nobs^3$ scaling predicted in Table~\ref{tab: scaling}.

When the selection criterion is changed, the shapes and uncertainties in the hierarchical posterior become significantly different. We consider additional inferences where selection is given by $d = \theta+ n > 3$. For the same number of observations $\Nobs$ and injections $\Ninj$, the error statistics in the inference with a stricter selection criterion are much higher.

\begin{figure}[h!]
    \centering
    \includegraphics[width=\linewidth]{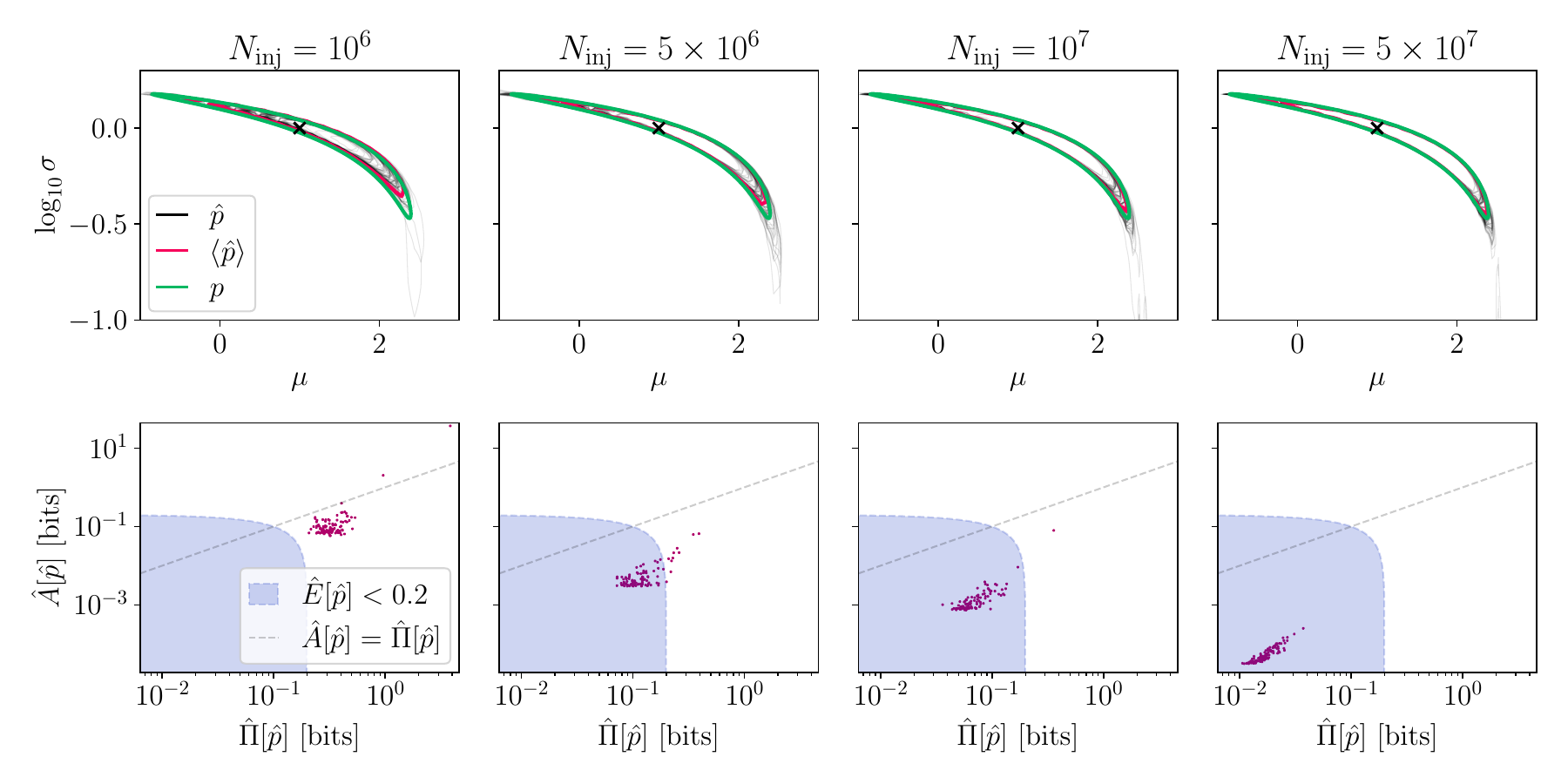}
    \caption{Same as Fig.~\ref{fig: nobs 100 corrected vt 1 gaussian example}, with $\Nobs=1000$ observations and the selection given by $d=\theta+n > 3$ and using the corrected likelihood estimator of Eq.~\ref{eq: unbiased likelihood estimator}.}
    \label{fig: nobs 1000 vt 3 gaussian example}
\end{figure}

\begin{figure}[h!]
    \centering
    \includegraphics[width=\linewidth]{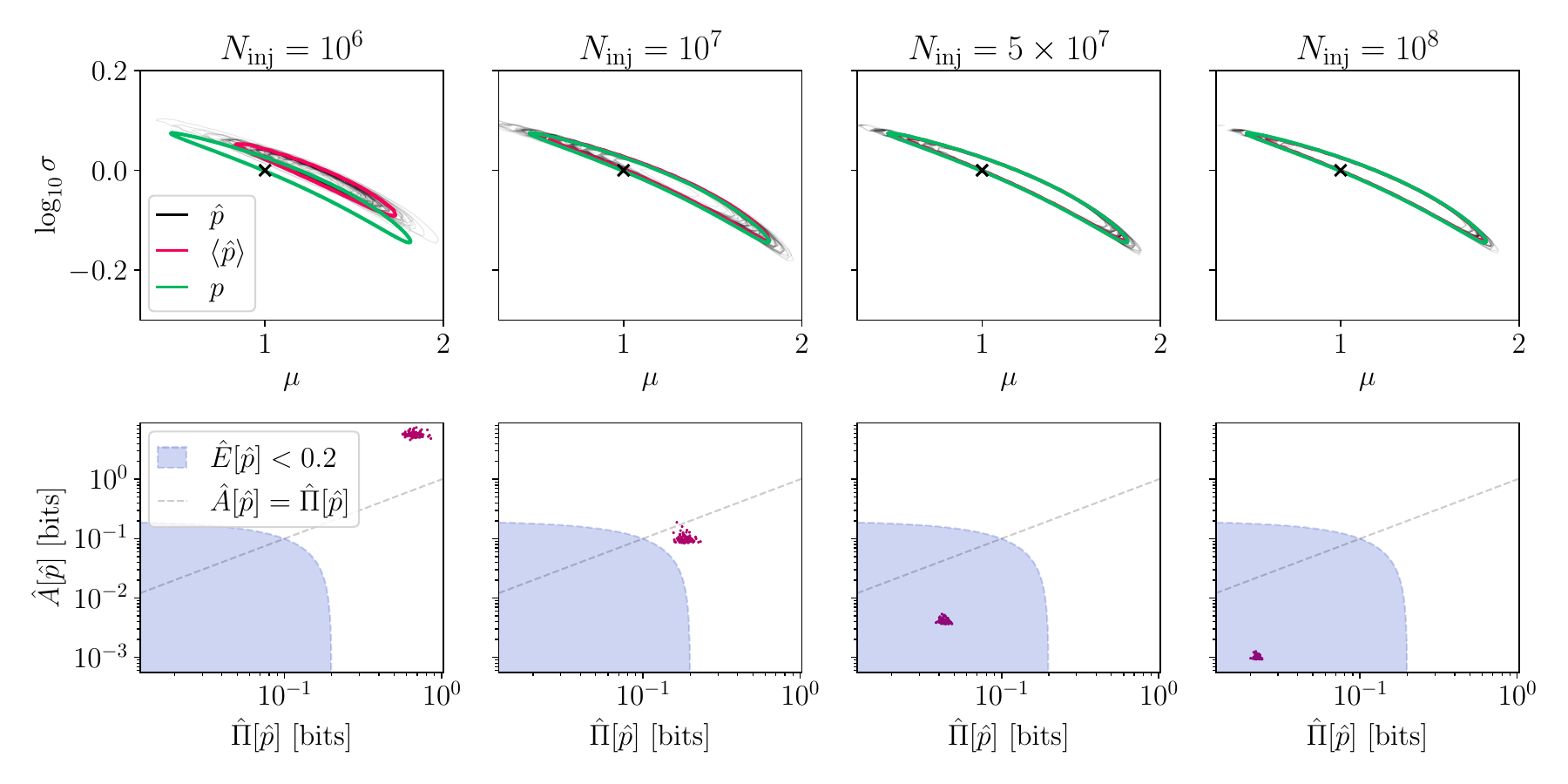}
    \caption{Same as Fig.~\ref{fig: nobs 100 corrected vt 1 gaussian example}, with $\Nobs=5000$ observations and the selection given by $d=\theta+n > 3$ and using the corrected likelihood estimator of Eq.~\ref{eq: unbiased likelihood estimator}.}
    \label{fig: nobs 5000 vt 3 gaussian example}
\end{figure}

\clearpage

\section{Examples with posterior correction}
\label{app: examples of posterior correction}

We show some examples of \textit{explicitly} correcting the posterior estimators using the general expression (Eq.~\ref{eq: modifying term}) derived in Appendix~\ref{app: correction in general}. We use the same Gaussian hierarchical inference framework described in Sec.~\ref{sec: gaussian example}. 
Instead of estimating the accuracy statistic $\hat{A}$ with Eq.~\ref{eq: accuracy statistic}, however, we calculate the bias correction with the modifying term
\beq
m(\Lambda) = 
\begin{cases}
\exp\left[-\dfrac{1}{2}\dfrac{\Nobs(\Nobs+1)\hat{\sigma}^2_{\xi}(\Lambda)}{\hat{\xi}(\Lambda)^2} + \dfrac{1}{N_{\rm samp}}\displaystyle\sum_{m=1}^{N_{\rm samp}} \hat{C}_{\ln \mathcal{L}}(\Lambda, \Lambda_m)\right] & \text{if using uncorrected likelihood in Eq.~\ref{eq: estimated hierarchical likelihood}} \\ 
\exp\left[\dfrac{1}{N_{\rm samp}}\displaystyle\sum_{m=1}^{N_{\rm samp}} \hat{C}_{\ln \mathcal{L}}(\Lambda, \Lambda_m)\right]  & \text{if using corrected likelihood in Eq.~\ref{eq: unbiased likelihood estimator}}
\end{cases}
\label{eq: bias correction term}
\eeq
where $\hat{C}_{\ln \mathcal{L}}$ is defined in Eq.~\ref{eq: population covariance with selection}, and 
\beq
\hat{p}_C(\Lambda) \propto \hat{p}(\Lambda)m(\Lambda)
\label{eq: practical posterior correction}
\eeq
is the corrected posterior estimator.

\begin{figure}
    \centering
    \includegraphics[width=\linewidth]{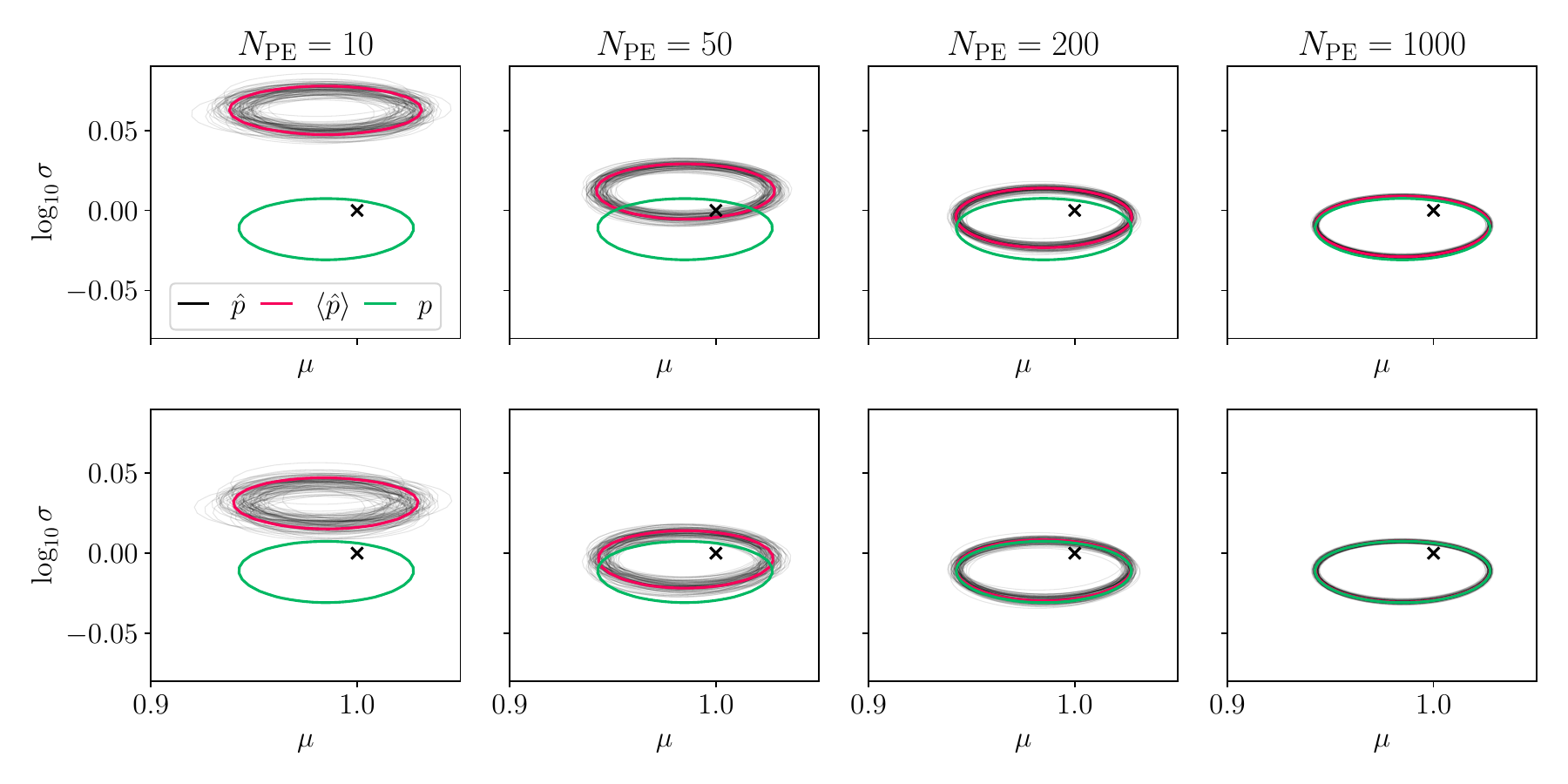}
    \caption{A comparison between the posterior estimator where $\Nobs=5000$ without correction (in the upper panels) and with correction (in the lower panels). Each black curve represents a 90\% \ac{HPD} contour for an independent posterior estimator, and the red curve represents the mean of the posterior estimators. The green curve is the true posterior computed analytically, and the black $\times$ symbol represents the true hyperparameters $\mu = 1$ and $\sigma = 1$.}
    \label{fig: posterior comparisons nobs 5000}
\end{figure}

We show an example comparing the uncorrected posterior estimators with the corrected posterior estimator when only parameter estimation uncertainty is considered, as Sec.~\ref{subsec: example without selection effects}. We show the example in Fig.~\ref{fig: posterior comparisons nobs 5000} where $\Nobs = 5000$ (where we know from Table~\ref{tab: scaling} that the bias will contribute strongly for small $\NPE$).

\begin{figure}[h!]
    \centering
    \includegraphics[width=\linewidth]{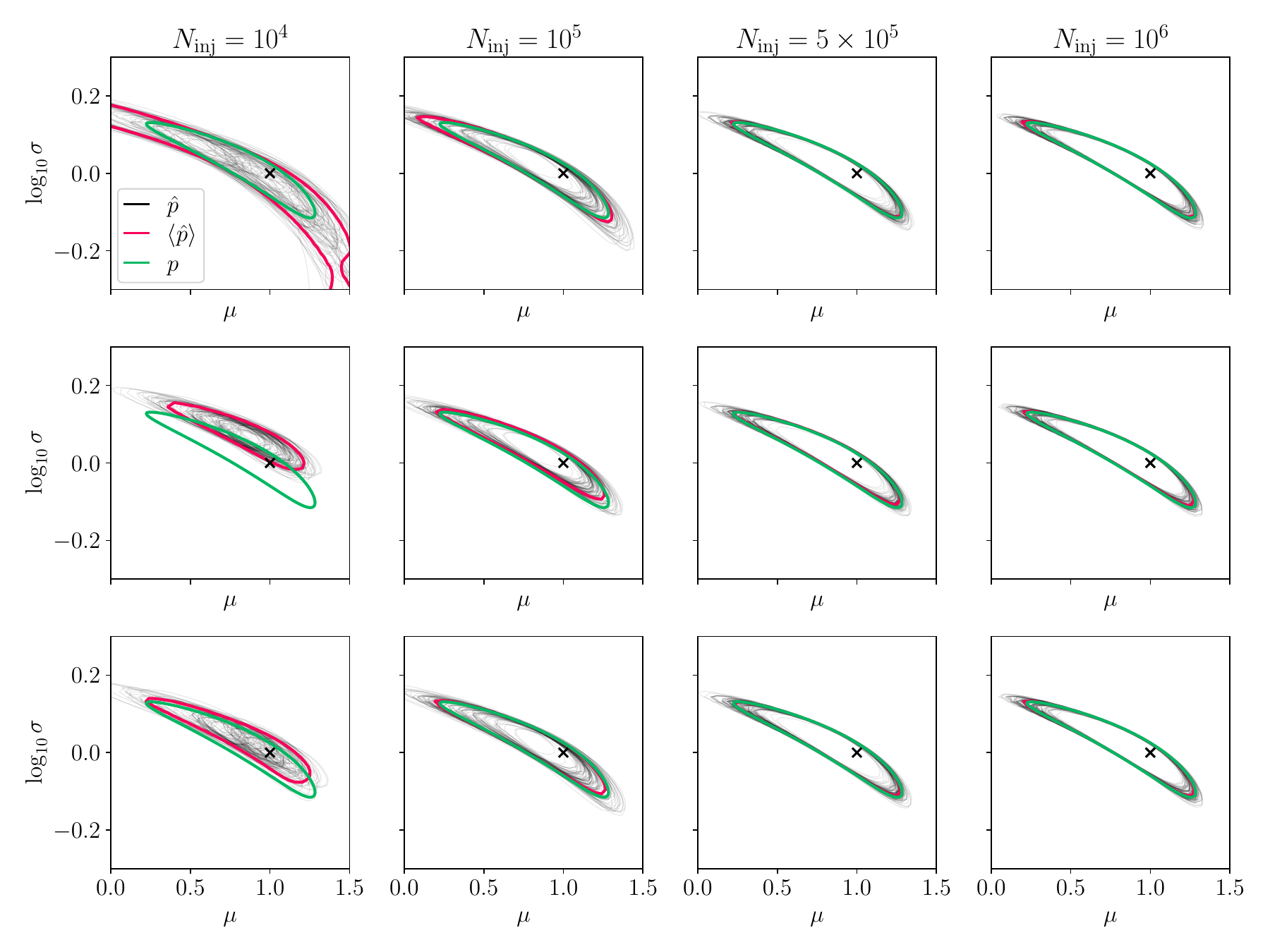}
    \caption{A comparison between the posterior estimator where $\Nobs=1000$ without any correction (in the top panels), with likelihood correction (in the central panels) and with both likelihood and posterior correction (in the bottom panels). Each black curve represents a 90\% \ac{HPD} contour for an independent posterior estimator, and the red curve represents the mean of the posterior estimators. The green curve is the true posterior computed analytically, and the black $\times$ symbol represents the true hyperparameters $\mu = 1$ and $\sigma = 1$. Here, selection is given by $d = \theta + n>1$.}
    \label{fig: selection posterior comparisons nobs 1000}
\end{figure}

\begin{figure}[h!]
    \centering
    \includegraphics[width=0.6\linewidth]{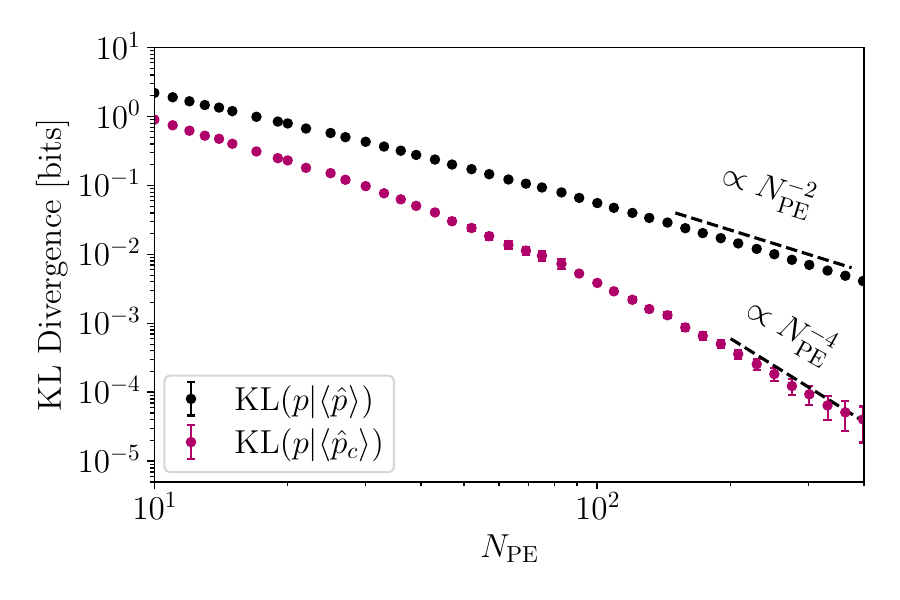}
    \caption{Empirically measured \ac{KL} divergence from true posterior $p$ and the mean of the posterior estimator $\expect{p}$. We use the same scheme as Fig.~\ref{fig: random KL fn of npe}, using $10^4$ posterior estimators for $\NPE < 90$ and $10^5$ to reduce uncertainty for $\NPE > 90$. We overlay $\NPE^{-2}$ and $\NPE^{-4}$ scaling to show that the posterior correction term removes the leading order posterior bias.}
    \label{fig: random KL fn of npe vs correction}
\end{figure}

Next, we verify that the correction indeed removes the leading order bias. In Fig.~\ref{fig: random KL fn of npe vs correction} we show an analogous figure to Fig.~\ref{fig: random KL fn of npe} where we compute the \ac{KL} divergence from the mean posterior to the true posterior $\KLD(\expect{\hat{p}}|p)$ as a function of $\NPE$ for both the uncorrected and the corrected posterior estimators. With the uncorrected posterior estimator $\hat{p}$, the bias scales as $\NPE^{-1}$ and so the \ac{KL} divergence scales with $\NPE^{-2}$. If the leading order bias in the corrected posterior estimator $\hat{p}_C$ is removed, then the residual bias in should scale with $\NPE^{-2}$ and therefore the \ac{KL} divergence should scale as $\NPE^{-4}$. Figure~\ref{fig: random KL fn of npe vs correction} shows the leading order bias is indeed removed by the correction factor.

%% file: rate_likelihood_appendix.tex
\section{Relevant quantities for using the rate-explicit likelihood}

In Sec.~\ref{sec: robustness measures} we discuss measures of the accuracy and the precision of the posterior estimator, which requires an estimator of the covariance in the log-likelihood estimator. In Appendix~\ref{app: bias in general} and~\ref{app: correction in general} we derive some general expressions for the bias and correction in the likelihood and the posterior. In this appendix, we show expressions for these quantities (which we compute directly from the general expressions derived above) when using the \textit{rate-explicit} likelihood as opposed to the rate-marginalized likelihood in Eq.~\ref{eq: rate marginalized likelihood}. The rate-explicit likelihood is
\beq
\mathcal{L}(\{d_i\}|\Lambda,N) \propto N^{\Nobs} e^{-N\xi(\Lambda)}\prod_{i=1}^{\Nobs}p(d_i|\Lambda),
\label{eq: rate explicit likelihood}
\eeq
where $N$ represents the Poisson expectation for the number of events to occur over the observing time. 
The likelihood may be estimated using Eqs.~\ref{eq: single event estimator} and \ref{eq: selection effects estimator}, but as pointed out in Sec.~\ref{subsec: monte carlo bias}, this represents a biased estimator. 

Applying Eq.~\ref{eq: general covariance term}, the variance in the log-likelihood estimator is therefore
\beq
\hat{\sigma}^2_{\ln\mathcal{L}}(\Lambda, N) = \sum_{i=1}^{\Nobs}\frac{\hat{\sigma}^2_{p_i}(\Lambda)}{\hat{p}(d_i|\Lambda)^2}
+ N^2\hat{\sigma}^2_{\xi}(\Lambda)
\eeq
where $\hat{p}(d_i|\Lambda)$ is defined in Eq.~\ref{eq: single event estimator}, $\hat{\xi}(\Lambda)$ in Eq.~\ref{eq: selection effects estimator} and the variance in a Monte Carlo integral $\hat{\sigma}^2_I$ is defined in Eq.~\ref{eq: variance estimator}. The covariance in the log-likelihood estimator is
\beq
\hat{C}_{\ln\mathcal{L}}(\Lambda, N, \Lambda', N') = \sum_{i=1}^{\Nobs}\frac{\hat{C}_{p_i}(\Lambda, \Lambda')}{\hat{p}(d_i|\Lambda)\hat{p}(d_i|\Lambda')}
+ NN'\hat{C}_{\xi}(\Lambda, \Lambda'),
\eeq
where $\hat{C}_{p_i}(\Lambda, \Lambda')$ and $\hat{C}_{\xi}(\Lambda, \Lambda')$ are defined in Eqs.~\ref{eq: single event covariance estimator} and \ref{eq: selection covariance estimator}.

Using Eq.~\ref{eq: general likelihood bias correction}, the correction for the likelihood is given as 
\beq
\hat{\mathcal{L}}(\{d_i\} | \Lambda, N) \mapsto \hat{\mathcal{L}}(\{d_i\} | \Lambda, N)\exp\left(-\frac{N^2\hat{\sigma}^2_{\xi}}{2}\right).
\label{eq: rate likelihood correction}
\eeq
Using these expressions, the error, accuracy and precision statistics may be computed with the usual Eqs.~\ref{eq: error statistic}, \ref{eq: precision statistic} and \ref{eq: accuracy statistic}, where the weights are modified if using the uncorrected likelihood estimator
\beq
w_n = 
\begin{cases}
\dfrac{N_n^2\hat{\sigma}^2_{\xi}(\Lambda)}{2} - \dfrac{1}{N_{\rm samp}}\displaystyle\sum_{m=1}^{N_{\rm samp}} \hat{C}_{\ln\mathcal{L}}(\Lambda_n, N_n, \Lambda_m, N_m) & \text{if using uncorrected likelihood in Eq.~\ref{eq: rate explicit likelihood}} \\ 
\dfrac{1}{N_{\rm samp}}\displaystyle\sum_{m=1}^{N_{\rm samp}} \hat{C}_{\ln\mathcal{L}}(\Lambda_n, N_n, \Lambda_m, N_m)  & \text{if using corrected likelihood in Eq.~\ref{eq: rate likelihood correction}},
\end{cases}
\eeq
where $N_n$ and $N_m$ are the $n^{\rm th}$ and $m^{\rm th}$ samples for the Poisson expectation hyperparameter $N$.
The posterior correction is again given by Eq.~\ref{eq: practical posterior correction}.

%% file: main.bbl
\providecommand{\noopsort}[1]{}\providecommand{\singleletter}[1]{#1}%
\begin{thebibliography}{93}%
\makeatletter
\providecommand \@ifxundefined [1]{%
 \@ifx{#1\undefined}
}%
\providecommand \@ifnum [1]{%
 \ifnum #1\expandafter \@firstoftwo
 \else \expandafter \@secondoftwo
 \fi
}%
\providecommand \@ifx [1]{%
 \ifx #1\expandafter \@firstoftwo
 \else \expandafter \@secondoftwo
 \fi
}%
\providecommand \natexlab [1]{#1}%
\providecommand \enquote  [1]{``#1''}%
\providecommand \bibnamefont  [1]{#1}%
\providecommand \bibfnamefont [1]{#1}%
\providecommand \citenamefont [1]{#1}%
\providecommand \href@noop [0]{\@secondoftwo}%
\providecommand \href [0]{\begingroup \@sanitize@url \@href}%
\providecommand \@href[1]{\@@startlink{#1}\@@href}%
\providecommand \@@href[1]{\endgroup#1\@@endlink}%
\providecommand \@sanitize@url [0]{\catcode `\\12\catcode `\$12\catcode `\&12\catcode `\#12\catcode `\^12\catcode `\_12\catcode `\%12\relax}%
\providecommand \@@startlink[1]{}%
\providecommand \@@endlink[0]{}%
\providecommand \url  [0]{\begingroup\@sanitize@url \@url }%
\providecommand \@url [1]{\endgroup\@href {#1}{\urlprefix }}%
\providecommand \urlprefix  [0]{URL }%
\providecommand \Eprint [0]{\href }%
\providecommand \doibase [0]{https://doi.org/}%
\providecommand \selectlanguage [0]{\@gobble}%
\providecommand \bibinfo  [0]{\@secondoftwo}%
\providecommand \bibfield  [0]{\@secondoftwo}%
\providecommand \translation [1]{[#1]}%
\providecommand \BibitemOpen [0]{}%
\providecommand \bibitemStop [0]{}%
\providecommand \bibitemNoStop [0]{.\EOS\space}%
\providecommand \EOS [0]{\spacefactor3000\relax}%
\providecommand \BibitemShut  [1]{\csname bibitem#1\endcsname}%
\let\auto@bib@innerbib\@empty
\bibitem [{\citenamefont {Abbott}\ \emph {et~al.}(2016{\natexlab{a}})\citenamefont {Abbott} \emph {et~al.}}]{LIGOScientific:2016aoc}%
  \BibitemOpen
  \bibfield  {author} {\bibinfo {author} {\bibfnamefont {B.~P.}\ \bibnamefont {Abbott}} \emph {et~al.} (\bibinfo {collaboration} {LIGO Scientific, Virgo}),\ }\bibfield  {title} {\bibinfo {title} {{Observation of Gravitational Waves from a Binary Black Hole Merger}},\ }\href {https://doi.org/10.1103/PhysRevLett.116.061102} {\bibfield  {journal} {\bibinfo  {journal} {Phys. Rev. Lett.}\ }\textbf {\bibinfo {volume} {116}},\ \bibinfo {pages} {061102} (\bibinfo {year} {2016}{\natexlab{a}})},\ \Eprint {https://arxiv.org/abs/1602.03837} {arXiv:1602.03837 [gr-qc]} \BibitemShut {NoStop}%
\bibitem [{\citenamefont {Abbott}\ \emph {et~al.}(2019{\natexlab{a}})\citenamefont {Abbott} \emph {et~al.}}]{LIGOScientific:2018mvr}%
  \BibitemOpen
  \bibfield  {author} {\bibinfo {author} {\bibfnamefont {B.~P.}\ \bibnamefont {Abbott}} \emph {et~al.} (\bibinfo {collaboration} {LIGO Scientific, Virgo}),\ }\bibfield  {title} {\bibinfo {title} {{GWTC-1: A Gravitational-Wave Transient Catalog of Compact Binary Mergers Observed by LIGO and Virgo during the First and Second Observing Runs}},\ }\href {https://doi.org/10.1103/PhysRevX.9.031040} {\bibfield  {journal} {\bibinfo  {journal} {Phys. Rev. X}\ }\textbf {\bibinfo {volume} {9}},\ \bibinfo {pages} {031040} (\bibinfo {year} {2019}{\natexlab{a}})},\ \Eprint {https://arxiv.org/abs/1811.12907} {arXiv:1811.12907 [astro-ph.HE]} \BibitemShut {NoStop}%
\bibitem [{\citenamefont {Abbott}\ \emph {et~al.}(2021{\natexlab{a}})\citenamefont {Abbott} \emph {et~al.}}]{LIGOScientific:2020ibl}%
  \BibitemOpen
  \bibfield  {author} {\bibinfo {author} {\bibfnamefont {R.}~\bibnamefont {Abbott}} \emph {et~al.} (\bibinfo {collaboration} {LIGO Scientific, Virgo}),\ }\bibfield  {title} {\bibinfo {title} {{GWTC-2: Compact Binary Coalescences Observed by LIGO and Virgo During the First Half of the Third Observing Run}},\ }\href {https://doi.org/10.1103/PhysRevX.11.021053} {\bibfield  {journal} {\bibinfo  {journal} {Phys. Rev. X}\ }\textbf {\bibinfo {volume} {11}},\ \bibinfo {pages} {021053} (\bibinfo {year} {2021}{\natexlab{a}})},\ \Eprint {https://arxiv.org/abs/2010.14527} {arXiv:2010.14527 [gr-qc]} \BibitemShut {NoStop}%
\bibitem [{\citenamefont {Abbott}\ \emph {et~al.}(2024)\citenamefont {Abbott} \emph {et~al.}}]{LIGOScientific:2021usb}%
  \BibitemOpen
  \bibfield  {author} {\bibinfo {author} {\bibfnamefont {R.}~\bibnamefont {Abbott}} \emph {et~al.} (\bibinfo {collaboration} {LIGO Scientific, VIRGO}),\ }\bibfield  {title} {\bibinfo {title} {{GWTC-2.1: Deep extended catalog of compact binary coalescences observed by LIGO and Virgo during the first half of the third observing run}},\ }\href {https://doi.org/10.1103/PhysRevD.109.022001} {\bibfield  {journal} {\bibinfo  {journal} {Phys. Rev. D}\ }\textbf {\bibinfo {volume} {109}},\ \bibinfo {pages} {022001} (\bibinfo {year} {2024})},\ \Eprint {https://arxiv.org/abs/2108.01045} {arXiv:2108.01045 [gr-qc]} \BibitemShut {NoStop}%
\bibitem [{\citenamefont {Abbott}\ \emph {et~al.}(2023{\natexlab{a}})\citenamefont {Abbott} \emph {et~al.}}]{KAGRA:2021vkt}%
  \BibitemOpen
  \bibfield  {author} {\bibinfo {author} {\bibfnamefont {R.}~\bibnamefont {Abbott}} \emph {et~al.} (\bibinfo {collaboration} {KAGRA, VIRGO, LIGO Scientific}),\ }\bibfield  {title} {\bibinfo {title} {{GWTC-3: Compact Binary Coalescences Observed by LIGO and Virgo during the Second Part of the Third Observing Run}},\ }\href {https://doi.org/10.1103/PhysRevX.13.041039} {\bibfield  {journal} {\bibinfo  {journal} {Phys. Rev. X}\ }\textbf {\bibinfo {volume} {13}},\ \bibinfo {pages} {041039} (\bibinfo {year} {2023}{\natexlab{a}})},\ \Eprint {https://arxiv.org/abs/2111.03606} {arXiv:2111.03606 [gr-qc]} \BibitemShut {NoStop}%
\bibitem [{\citenamefont {Abbott}\ \emph {et~al.}(2016{\natexlab{b}})\citenamefont {Abbott} \emph {et~al.}}]{KAGRA:2013rdx}%
  \BibitemOpen
  \bibfield  {author} {\bibinfo {author} {\bibfnamefont {B.~P.}\ \bibnamefont {Abbott}} \emph {et~al.} (\bibinfo {collaboration} {KAGRA, LIGO Scientific, Virgo}),\ }\bibfield  {title} {\bibinfo {title} {{Prospects for observing and localizing gravitational-wave transients with Advanced LIGO, Advanced Virgo and KAGRA}},\ }\href {https://doi.org/10.1007/s41114-020-00026-9} {\bibfield  {journal} {\bibinfo  {journal} {Living Rev. Rel.}\ }\textbf {\bibinfo {volume} {19}},\ \bibinfo {pages} {1} (\bibinfo {year} {2016}{\natexlab{b}})},\ \Eprint {https://arxiv.org/abs/1304.0670} {arXiv:1304.0670 [gr-qc]} \BibitemShut {NoStop}%
\bibitem [{\citenamefont {Vitale}\ \emph {et~al.}(2017)\citenamefont {Vitale}, \citenamefont {Lynch}, \citenamefont {Sturani},\ and\ \citenamefont {Graff}}]{Vitale:2015tea}%
  \BibitemOpen
  \bibfield  {author} {\bibinfo {author} {\bibfnamefont {S.}~\bibnamefont {Vitale}}, \bibinfo {author} {\bibfnamefont {R.}~\bibnamefont {Lynch}}, \bibinfo {author} {\bibfnamefont {R.}~\bibnamefont {Sturani}},\ and\ \bibinfo {author} {\bibfnamefont {P.}~\bibnamefont {Graff}},\ }\bibfield  {title} {\bibinfo {title} {{Use of gravitational waves to probe the formation channels of compact binaries}},\ }\href {https://doi.org/10.1088/1361-6382/aa552e} {\bibfield  {journal} {\bibinfo  {journal} {Class. Quant. Grav.}\ }\textbf {\bibinfo {volume} {34}},\ \bibinfo {pages} {03LT01} (\bibinfo {year} {2017})},\ \Eprint {https://arxiv.org/abs/1503.04307} {arXiv:1503.04307 [gr-qc]} \BibitemShut {NoStop}%
\bibitem [{\citenamefont {Abbott}\ \emph {et~al.}(2016{\natexlab{c}})\citenamefont {Abbott} \emph {et~al.}}]{LIGOScientific:2016vpg}%
  \BibitemOpen
  \bibfield  {author} {\bibinfo {author} {\bibfnamefont {B.~P.}\ \bibnamefont {Abbott}} \emph {et~al.} (\bibinfo {collaboration} {LIGO Scientific, Virgo}),\ }\bibfield  {title} {\bibinfo {title} {{Astrophysical Implications of the Binary Black-Hole Merger GW150914}},\ }\href {https://doi.org/10.3847/2041-8205/818/2/L22} {\bibfield  {journal} {\bibinfo  {journal} {Astrophys. J. Lett.}\ }\textbf {\bibinfo {volume} {818}},\ \bibinfo {pages} {L22} (\bibinfo {year} {2016}{\natexlab{c}})},\ \Eprint {https://arxiv.org/abs/1602.03846} {arXiv:1602.03846 [astro-ph.HE]} \BibitemShut {NoStop}%
\bibitem [{\citenamefont {Zevin}\ \emph {et~al.}(2017{\natexlab{a}})\citenamefont {Zevin}, \citenamefont {Pankow}, \citenamefont {Rodriguez}, \citenamefont {Sampson}, \citenamefont {Chase}, \citenamefont {Kalogera},\ and\ \citenamefont {Rasio}}]{Zevin:2017evb}%
  \BibitemOpen
  \bibfield  {author} {\bibinfo {author} {\bibfnamefont {M.}~\bibnamefont {Zevin}}, \bibinfo {author} {\bibfnamefont {C.}~\bibnamefont {Pankow}}, \bibinfo {author} {\bibfnamefont {C.~L.}\ \bibnamefont {Rodriguez}}, \bibinfo {author} {\bibfnamefont {L.}~\bibnamefont {Sampson}}, \bibinfo {author} {\bibfnamefont {E.}~\bibnamefont {Chase}}, \bibinfo {author} {\bibfnamefont {V.}~\bibnamefont {Kalogera}},\ and\ \bibinfo {author} {\bibfnamefont {F.~A.}\ \bibnamefont {Rasio}},\ }\bibfield  {title} {\bibinfo {title} {{Constraining Formation Models of Binary Black Holes with Gravitational-Wave Observations}},\ }\href {https://doi.org/10.3847/1538-4357/aa8408} {\bibfield  {journal} {\bibinfo  {journal} {Astrophys. J.}\ }\textbf {\bibinfo {volume} {846}},\ \bibinfo {pages} {82} (\bibinfo {year} {2017}{\natexlab{a}})},\ \Eprint {https://arxiv.org/abs/1704.07379} {arXiv:1704.07379 [astro-ph.HE]} \BibitemShut {NoStop}%
\bibitem [{\citenamefont {Mandel}\ and\ \citenamefont {Farmer}(2022)}]{Mandel:2018hfr}%
  \BibitemOpen
  \bibfield  {author} {\bibinfo {author} {\bibfnamefont {I.}~\bibnamefont {Mandel}}\ and\ \bibinfo {author} {\bibfnamefont {A.}~\bibnamefont {Farmer}},\ }\bibfield  {title} {\bibinfo {title} {{Merging stellar-mass binary black holes}},\ }\href {https://doi.org/10.1016/j.physrep.2022.01.003} {\bibfield  {journal} {\bibinfo  {journal} {Phys. Rept.}\ }\textbf {\bibinfo {volume} {955}},\ \bibinfo {pages} {1} (\bibinfo {year} {2022})},\ \Eprint {https://arxiv.org/abs/1806.05820} {arXiv:1806.05820 [astro-ph.HE]} \BibitemShut {NoStop}%
\bibitem [{\citenamefont {Mandel}\ and\ \citenamefont {Broekgaarden}(2022)}]{Mandel:2021smh}%
  \BibitemOpen
  \bibfield  {author} {\bibinfo {author} {\bibfnamefont {I.}~\bibnamefont {Mandel}}\ and\ \bibinfo {author} {\bibfnamefont {F.~S.}\ \bibnamefont {Broekgaarden}},\ }\bibfield  {title} {\bibinfo {title} {{Rates of compact object coalescences}},\ }\href {https://doi.org/10.1007/s41114-021-00034-3} {\bibfield  {journal} {\bibinfo  {journal} {Living Rev. Rel.}\ }\textbf {\bibinfo {volume} {25}},\ \bibinfo {pages} {1} (\bibinfo {year} {2022})},\ \Eprint {https://arxiv.org/abs/2107.14239} {arXiv:2107.14239 [astro-ph.HE]} \BibitemShut {NoStop}%
\bibitem [{\citenamefont {Abbott}\ \emph {et~al.}(2023{\natexlab{b}})\citenamefont {Abbott} \emph {et~al.}}]{KAGRA:2021duu}%
  \BibitemOpen
  \bibfield  {author} {\bibinfo {author} {\bibfnamefont {R.}~\bibnamefont {Abbott}} \emph {et~al.} (\bibinfo {collaboration} {KAGRA, VIRGO, LIGO Scientific}),\ }\bibfield  {title} {\bibinfo {title} {{Population of Merging Compact Binaries Inferred Using Gravitational Waves through GWTC-3}},\ }\href {https://doi.org/10.1103/PhysRevX.13.011048} {\bibfield  {journal} {\bibinfo  {journal} {Phys. Rev. X}\ }\textbf {\bibinfo {volume} {13}},\ \bibinfo {pages} {011048} (\bibinfo {year} {2023}{\natexlab{b}})},\ \Eprint {https://arxiv.org/abs/2111.03634} {arXiv:2111.03634 [astro-ph.HE]} \BibitemShut {NoStop}%
\bibitem [{\citenamefont {Schutz}(1986)}]{Schutz:1986gp}%
  \BibitemOpen
  \bibfield  {author} {\bibinfo {author} {\bibfnamefont {B.~F.}\ \bibnamefont {Schutz}},\ }\bibfield  {title} {\bibinfo {title} {{Determining the Hubble Constant from Gravitational Wave Observations}},\ }\href {https://doi.org/10.1038/323310a0} {\bibfield  {journal} {\bibinfo  {journal} {Nature}\ }\textbf {\bibinfo {volume} {323}},\ \bibinfo {pages} {310} (\bibinfo {year} {1986})}\BibitemShut {NoStop}%
\bibitem [{\citenamefont {Holz}\ and\ \citenamefont {Hughes}(2005)}]{Holz:2005df}%
  \BibitemOpen
  \bibfield  {author} {\bibinfo {author} {\bibfnamefont {D.~E.}\ \bibnamefont {Holz}}\ and\ \bibinfo {author} {\bibfnamefont {S.~A.}\ \bibnamefont {Hughes}},\ }\bibfield  {title} {\bibinfo {title} {{Using gravitational-wave standard sirens}},\ }\href {https://doi.org/10.1086/431341} {\bibfield  {journal} {\bibinfo  {journal} {Astrophys. J.}\ }\textbf {\bibinfo {volume} {629}},\ \bibinfo {pages} {15} (\bibinfo {year} {2005})},\ \Eprint {https://arxiv.org/abs/astro-ph/0504616} {arXiv:astro-ph/0504616} \BibitemShut {NoStop}%
\bibitem [{\citenamefont {Messenger}\ and\ \citenamefont {Read}(2012)}]{Messenger:2011gi}%
  \BibitemOpen
  \bibfield  {author} {\bibinfo {author} {\bibfnamefont {C.}~\bibnamefont {Messenger}}\ and\ \bibinfo {author} {\bibfnamefont {J.}~\bibnamefont {Read}},\ }\bibfield  {title} {\bibinfo {title} {{Measuring a cosmological distance-redshift relationship using only gravitational wave observations of binary neutron star coalescences}},\ }\href {https://doi.org/10.1103/PhysRevLett.108.091101} {\bibfield  {journal} {\bibinfo  {journal} {Phys. Rev. Lett.}\ }\textbf {\bibinfo {volume} {108}},\ \bibinfo {pages} {091101} (\bibinfo {year} {2012})},\ \Eprint {https://arxiv.org/abs/1107.5725} {arXiv:1107.5725 [gr-qc]} \BibitemShut {NoStop}%
\bibitem [{\citenamefont {Abbott}\ \emph {et~al.}(2017)\citenamefont {Abbott} \emph {et~al.}}]{LIGOScientific:2017adf}%
  \BibitemOpen
  \bibfield  {author} {\bibinfo {author} {\bibfnamefont {B.~P.}\ \bibnamefont {Abbott}} \emph {et~al.} (\bibinfo {collaboration} {LIGO Scientific, Virgo, 1M2H, Dark Energy Camera GW-E, DES, DLT40, Las Cumbres Observatory, VINROUGE, MASTER}),\ }\bibfield  {title} {\bibinfo {title} {{A gravitational-wave standard siren measurement of the Hubble constant}},\ }\href {https://doi.org/10.1038/nature24471} {\bibfield  {journal} {\bibinfo  {journal} {Nature}\ }\textbf {\bibinfo {volume} {551}},\ \bibinfo {pages} {85} (\bibinfo {year} {2017})},\ \Eprint {https://arxiv.org/abs/1710.05835} {arXiv:1710.05835 [astro-ph.CO]} \BibitemShut {NoStop}%
\bibitem [{\citenamefont {Christensen}(2019)}]{Christensen:2018iqi}%
  \BibitemOpen
  \bibfield  {author} {\bibinfo {author} {\bibfnamefont {N.}~\bibnamefont {Christensen}},\ }\bibfield  {title} {\bibinfo {title} {{Stochastic Gravitational Wave Backgrounds}},\ }\href {https://doi.org/10.1088/1361-6633/aae6b5} {\bibfield  {journal} {\bibinfo  {journal} {Rept. Prog. Phys.}\ }\textbf {\bibinfo {volume} {82}},\ \bibinfo {pages} {016903} (\bibinfo {year} {2019})},\ \Eprint {https://arxiv.org/abs/1811.08797} {arXiv:1811.08797 [gr-qc]} \BibitemShut {NoStop}%
\bibitem [{\citenamefont {Ezquiaga}\ and\ \citenamefont {Holz}(2022)}]{Ezquiaga:2022zkx}%
  \BibitemOpen
  \bibfield  {author} {\bibinfo {author} {\bibfnamefont {J.~M.}\ \bibnamefont {Ezquiaga}}\ and\ \bibinfo {author} {\bibfnamefont {D.~E.}\ \bibnamefont {Holz}},\ }\bibfield  {title} {\bibinfo {title} {{Spectral Sirens: Cosmology from the Full Mass Distribution of Compact Binaries}},\ }\href {https://doi.org/10.1103/PhysRevLett.129.061102} {\bibfield  {journal} {\bibinfo  {journal} {Phys. Rev. Lett.}\ }\textbf {\bibinfo {volume} {129}},\ \bibinfo {pages} {061102} (\bibinfo {year} {2022})},\ \Eprint {https://arxiv.org/abs/2202.08240} {arXiv:2202.08240 [astro-ph.CO]} \BibitemShut {NoStop}%
\bibitem [{\citenamefont {Gair}\ \emph {et~al.}(2023)\citenamefont {Gair} \emph {et~al.}}]{Gair:2022zsa}%
  \BibitemOpen
  \bibfield  {author} {\bibinfo {author} {\bibfnamefont {J.~R.}\ \bibnamefont {Gair}} \emph {et~al.},\ }\bibfield  {title} {\bibinfo {title} {{The Hitchhiker\textquoteright{}s Guide to the Galaxy Catalog Approach for Dark Siren Gravitational-wave Cosmology}},\ }\href {https://doi.org/10.3847/1538-3881/acca78} {\bibfield  {journal} {\bibinfo  {journal} {Astron. J.}\ }\textbf {\bibinfo {volume} {166}},\ \bibinfo {pages} {22} (\bibinfo {year} {2023})},\ \Eprint {https://arxiv.org/abs/2212.08694} {arXiv:2212.08694 [gr-qc]} \BibitemShut {NoStop}%
\bibitem [{\citenamefont {Baym}\ \emph {et~al.}(2018)\citenamefont {Baym}, \citenamefont {Hatsuda}, \citenamefont {Kojo}, \citenamefont {Powell}, \citenamefont {Song},\ and\ \citenamefont {Takatsuka}}]{Baym:2017whm}%
  \BibitemOpen
  \bibfield  {author} {\bibinfo {author} {\bibfnamefont {G.}~\bibnamefont {Baym}}, \bibinfo {author} {\bibfnamefont {T.}~\bibnamefont {Hatsuda}}, \bibinfo {author} {\bibfnamefont {T.}~\bibnamefont {Kojo}}, \bibinfo {author} {\bibfnamefont {P.~D.}\ \bibnamefont {Powell}}, \bibinfo {author} {\bibfnamefont {Y.}~\bibnamefont {Song}},\ and\ \bibinfo {author} {\bibfnamefont {T.}~\bibnamefont {Takatsuka}},\ }\bibfield  {title} {\bibinfo {title} {{From hadrons to quarks in neutron stars: a review}},\ }\href {https://doi.org/10.1088/1361-6633/aaae14} {\bibfield  {journal} {\bibinfo  {journal} {Rept. Prog. Phys.}\ }\textbf {\bibinfo {volume} {81}},\ \bibinfo {pages} {056902} (\bibinfo {year} {2018})},\ \Eprint {https://arxiv.org/abs/1707.04966} {arXiv:1707.04966 [astro-ph.HE]} \BibitemShut {NoStop}%
\bibitem [{\citenamefont {Capano}\ \emph {et~al.}(2020)\citenamefont {Capano}, \citenamefont {Tews}, \citenamefont {Brown}, \citenamefont {Margalit}, \citenamefont {De}, \citenamefont {Kumar}, \citenamefont {Brown}, \citenamefont {Krishnan},\ and\ \citenamefont {Reddy}}]{Capano:2019eae}%
  \BibitemOpen
  \bibfield  {author} {\bibinfo {author} {\bibfnamefont {C.~D.}\ \bibnamefont {Capano}}, \bibinfo {author} {\bibfnamefont {I.}~\bibnamefont {Tews}}, \bibinfo {author} {\bibfnamefont {S.~M.}\ \bibnamefont {Brown}}, \bibinfo {author} {\bibfnamefont {B.}~\bibnamefont {Margalit}}, \bibinfo {author} {\bibfnamefont {S.}~\bibnamefont {De}}, \bibinfo {author} {\bibfnamefont {S.}~\bibnamefont {Kumar}}, \bibinfo {author} {\bibfnamefont {D.~A.}\ \bibnamefont {Brown}}, \bibinfo {author} {\bibfnamefont {B.}~\bibnamefont {Krishnan}},\ and\ \bibinfo {author} {\bibfnamefont {S.}~\bibnamefont {Reddy}},\ }\bibfield  {title} {\bibinfo {title} {{Stringent constraints on neutron-star radii from multimessenger observations and nuclear theory}},\ }\href {https://doi.org/10.1038/s41550-020-1014-6} {\bibfield  {journal} {\bibinfo  {journal} {Nature Astron.}\ }\textbf {\bibinfo {volume} {4}},\ \bibinfo {pages} {625} (\bibinfo {year} {2020})},\ \Eprint {https://arxiv.org/abs/1908.10352} {arXiv:1908.10352 [astro-ph.HE]} \BibitemShut
  {NoStop}%
\bibitem [{\citenamefont {Dietrich}\ \emph {et~al.}(2020)\citenamefont {Dietrich}, \citenamefont {Coughlin}, \citenamefont {Pang}, \citenamefont {Bulla}, \citenamefont {Heinzel}, \citenamefont {Issa}, \citenamefont {Tews},\ and\ \citenamefont {Antier}}]{Dietrich:2020efo}%
  \BibitemOpen
  \bibfield  {author} {\bibinfo {author} {\bibfnamefont {T.}~\bibnamefont {Dietrich}}, \bibinfo {author} {\bibfnamefont {M.~W.}\ \bibnamefont {Coughlin}}, \bibinfo {author} {\bibfnamefont {P.~T.~H.}\ \bibnamefont {Pang}}, \bibinfo {author} {\bibfnamefont {M.}~\bibnamefont {Bulla}}, \bibinfo {author} {\bibfnamefont {J.}~\bibnamefont {Heinzel}}, \bibinfo {author} {\bibfnamefont {L.}~\bibnamefont {Issa}}, \bibinfo {author} {\bibfnamefont {I.}~\bibnamefont {Tews}},\ and\ \bibinfo {author} {\bibfnamefont {S.}~\bibnamefont {Antier}},\ }\bibfield  {title} {\bibinfo {title} {{Multimessenger constraints on the neutron-star equation of state and the Hubble constant}},\ }\href {https://doi.org/10.1126/science.abb4317} {\bibfield  {journal} {\bibinfo  {journal} {Science}\ }\textbf {\bibinfo {volume} {370}},\ \bibinfo {pages} {1450} (\bibinfo {year} {2020})},\ \Eprint {https://arxiv.org/abs/2002.11355} {arXiv:2002.11355 [astro-ph.HE]} \BibitemShut {NoStop}%
\bibitem [{\citenamefont {Chatziioannou}(2020)}]{Chatziioannou:2020pqz}%
  \BibitemOpen
  \bibfield  {author} {\bibinfo {author} {\bibfnamefont {K.}~\bibnamefont {Chatziioannou}},\ }\bibfield  {title} {\bibinfo {title} {{Neutron star tidal deformability and equation of state constraints}},\ }\href {https://doi.org/10.1007/s10714-020-02754-3} {\bibfield  {journal} {\bibinfo  {journal} {Gen. Rel. Grav.}\ }\textbf {\bibinfo {volume} {52}},\ \bibinfo {pages} {109} (\bibinfo {year} {2020})},\ \Eprint {https://arxiv.org/abs/2006.03168} {arXiv:2006.03168 [gr-qc]} \BibitemShut {NoStop}%
\bibitem [{\citenamefont {Landry}\ \emph {et~al.}(2020)\citenamefont {Landry}, \citenamefont {Essick},\ and\ \citenamefont {Chatziioannou}}]{Landry:2020vaw}%
  \BibitemOpen
  \bibfield  {author} {\bibinfo {author} {\bibfnamefont {P.}~\bibnamefont {Landry}}, \bibinfo {author} {\bibfnamefont {R.}~\bibnamefont {Essick}},\ and\ \bibinfo {author} {\bibfnamefont {K.}~\bibnamefont {Chatziioannou}},\ }\bibfield  {title} {\bibinfo {title} {{Nonparametric constraints on neutron star matter with existing and upcoming gravitational wave and pulsar observations}},\ }\href {https://doi.org/10.1103/PhysRevD.101.123007} {\bibfield  {journal} {\bibinfo  {journal} {Phys. Rev. D}\ }\textbf {\bibinfo {volume} {101}},\ \bibinfo {pages} {123007} (\bibinfo {year} {2020})},\ \Eprint {https://arxiv.org/abs/2003.04880} {arXiv:2003.04880 [astro-ph.HE]} \BibitemShut {NoStop}%
\bibitem [{\citenamefont {Mirshekari}\ \emph {et~al.}(2012)\citenamefont {Mirshekari}, \citenamefont {Yunes},\ and\ \citenamefont {Will}}]{Mirshekari:2011yq}%
  \BibitemOpen
  \bibfield  {author} {\bibinfo {author} {\bibfnamefont {S.}~\bibnamefont {Mirshekari}}, \bibinfo {author} {\bibfnamefont {N.}~\bibnamefont {Yunes}},\ and\ \bibinfo {author} {\bibfnamefont {C.~M.}\ \bibnamefont {Will}},\ }\bibfield  {title} {\bibinfo {title} {{Constraining Generic Lorentz Violation and the Speed of the Graviton with Gravitational Waves}},\ }\href {https://doi.org/10.1103/PhysRevD.85.024041} {\bibfield  {journal} {\bibinfo  {journal} {Phys. Rev. D}\ }\textbf {\bibinfo {volume} {85}},\ \bibinfo {pages} {024041} (\bibinfo {year} {2012})},\ \Eprint {https://arxiv.org/abs/1110.2720} {arXiv:1110.2720 [gr-qc]} \BibitemShut {NoStop}%
\bibitem [{\citenamefont {Will}(2014)}]{Will:2014kxa}%
  \BibitemOpen
  \bibfield  {author} {\bibinfo {author} {\bibfnamefont {C.~M.}\ \bibnamefont {Will}},\ }\bibfield  {title} {\bibinfo {title} {{The Confrontation between General Relativity and Experiment}},\ }\href {https://doi.org/10.12942/lrr-2014-4} {\bibfield  {journal} {\bibinfo  {journal} {Living Rev. Rel.}\ }\textbf {\bibinfo {volume} {17}},\ \bibinfo {pages} {4} (\bibinfo {year} {2014})},\ \Eprint {https://arxiv.org/abs/1403.7377} {arXiv:1403.7377 [gr-qc]} \BibitemShut {NoStop}%
\bibitem [{\citenamefont {Abbott}\ \emph {et~al.}(2016{\natexlab{d}})\citenamefont {Abbott} \emph {et~al.}}]{LIGOScientific:2016lio}%
  \BibitemOpen
  \bibfield  {author} {\bibinfo {author} {\bibfnamefont {B.~P.}\ \bibnamefont {Abbott}} \emph {et~al.} (\bibinfo {collaboration} {LIGO Scientific, Virgo}),\ }\bibfield  {title} {\bibinfo {title} {{Tests of general relativity with GW150914}},\ }\href {https://doi.org/10.1103/PhysRevLett.116.221101} {\bibfield  {journal} {\bibinfo  {journal} {Phys. Rev. Lett.}\ }\textbf {\bibinfo {volume} {116}},\ \bibinfo {pages} {221101} (\bibinfo {year} {2016}{\natexlab{d}})},\ \bibinfo {note} {[Erratum: Phys.Rev.Lett. 121, 129902 (2018)]},\ \Eprint {https://arxiv.org/abs/1602.03841} {arXiv:1602.03841 [gr-qc]} \BibitemShut {NoStop}%
\bibitem [{\citenamefont {Abbott}\ \emph {et~al.}(2019{\natexlab{b}})\citenamefont {Abbott} \emph {et~al.}}]{LIGOScientific:2018dkp}%
  \BibitemOpen
  \bibfield  {author} {\bibinfo {author} {\bibfnamefont {B.~P.}\ \bibnamefont {Abbott}} \emph {et~al.} (\bibinfo {collaboration} {LIGO Scientific, Virgo}),\ }\bibfield  {title} {\bibinfo {title} {{Tests of General Relativity with GW170817}},\ }\href {https://doi.org/10.1103/PhysRevLett.123.011102} {\bibfield  {journal} {\bibinfo  {journal} {Phys. Rev. Lett.}\ }\textbf {\bibinfo {volume} {123}},\ \bibinfo {pages} {011102} (\bibinfo {year} {2019}{\natexlab{b}})},\ \Eprint {https://arxiv.org/abs/1811.00364} {arXiv:1811.00364 [gr-qc]} \BibitemShut {NoStop}%
\bibitem [{\citenamefont {Abbott}\ \emph {et~al.}(2021{\natexlab{b}})\citenamefont {Abbott} \emph {et~al.}}]{LIGOScientific:2021sio}%
  \BibitemOpen
  \bibfield  {author} {\bibinfo {author} {\bibfnamefont {R.}~\bibnamefont {Abbott}} \emph {et~al.} (\bibinfo {collaboration} {LIGO Scientific, VIRGO, KAGRA}),\ }\bibfield  {title} {\bibinfo {title} {{Tests of General Relativity with GWTC-3}},\ }\href@noop {} {\  (\bibinfo {year} {2021}{\natexlab{b}})},\ \Eprint {https://arxiv.org/abs/2112.06861} {arXiv:2112.06861 [gr-qc]} \BibitemShut {NoStop}%
\bibitem [{\citenamefont {Del~Pozzo}(2012)}]{DelPozzo:2011vcw}%
  \BibitemOpen
  \bibfield  {author} {\bibinfo {author} {\bibfnamefont {W.}~\bibnamefont {Del~Pozzo}},\ }\bibfield  {title} {\bibinfo {title} {{Inference of the cosmological parameters from gravitational waves: application to second generation interferometers}},\ }\href {https://doi.org/10.1103/PhysRevD.86.043011} {\bibfield  {journal} {\bibinfo  {journal} {Phys. Rev. D}\ }\textbf {\bibinfo {volume} {86}},\ \bibinfo {pages} {043011} (\bibinfo {year} {2012})},\ \Eprint {https://arxiv.org/abs/1108.1317} {arXiv:1108.1317 [astro-ph.CO]} \BibitemShut {NoStop}%
\bibitem [{\citenamefont {Del~Pozzo}\ \emph {et~al.}(2013)\citenamefont {Del~Pozzo}, \citenamefont {Li}, \citenamefont {Agathos}, \citenamefont {Van Den~Broeck},\ and\ \citenamefont {Vitale}}]{DelPozzo:2013ala}%
  \BibitemOpen
  \bibfield  {author} {\bibinfo {author} {\bibfnamefont {W.}~\bibnamefont {Del~Pozzo}}, \bibinfo {author} {\bibfnamefont {T.~G.~F.}\ \bibnamefont {Li}}, \bibinfo {author} {\bibfnamefont {M.}~\bibnamefont {Agathos}}, \bibinfo {author} {\bibfnamefont {C.}~\bibnamefont {Van Den~Broeck}},\ and\ \bibinfo {author} {\bibfnamefont {S.}~\bibnamefont {Vitale}},\ }\bibfield  {title} {\bibinfo {title} {{Demonstrating the feasibility of probing the neutron star equation of state with second-generation gravitational wave detectors}},\ }\href {https://doi.org/10.1103/PhysRevLett.111.071101} {\bibfield  {journal} {\bibinfo  {journal} {Phys. Rev. Lett.}\ }\textbf {\bibinfo {volume} {111}},\ \bibinfo {pages} {071101} (\bibinfo {year} {2013})},\ \Eprint {https://arxiv.org/abs/1307.8338} {arXiv:1307.8338 [gr-qc]} \BibitemShut {NoStop}%
\bibitem [{\citenamefont {Isi}\ \emph {et~al.}(2019)\citenamefont {Isi}, \citenamefont {Chatziioannou},\ and\ \citenamefont {Farr}}]{Isi:2019asy}%
  \BibitemOpen
  \bibfield  {author} {\bibinfo {author} {\bibfnamefont {M.}~\bibnamefont {Isi}}, \bibinfo {author} {\bibfnamefont {K.}~\bibnamefont {Chatziioannou}},\ and\ \bibinfo {author} {\bibfnamefont {W.~M.}\ \bibnamefont {Farr}},\ }\bibfield  {title} {\bibinfo {title} {{Hierarchical test of general relativity with gravitational waves}},\ }\href {https://doi.org/10.1103/PhysRevLett.123.121101} {\bibfield  {journal} {\bibinfo  {journal} {Phys. Rev. Lett.}\ }\textbf {\bibinfo {volume} {123}},\ \bibinfo {pages} {121101} (\bibinfo {year} {2019})},\ \Eprint {https://arxiv.org/abs/1904.08011} {arXiv:1904.08011 [gr-qc]} \BibitemShut {NoStop}%
\bibitem [{\citenamefont {Essick}\ and\ \citenamefont {Fishbach}(2024)}]{Essick:2023upv}%
  \BibitemOpen
  \bibfield  {author} {\bibinfo {author} {\bibfnamefont {R.}~\bibnamefont {Essick}}\ and\ \bibinfo {author} {\bibfnamefont {M.}~\bibnamefont {Fishbach}},\ }\bibfield  {title} {\bibinfo {title} {{Ensuring Consistency between Noise and Detection in Hierarchical Bayesian Inference}},\ }\href {https://doi.org/10.3847/1538-4357/ad1604} {\bibfield  {journal} {\bibinfo  {journal} {Astrophys. J.}\ }\textbf {\bibinfo {volume} {962}},\ \bibinfo {pages} {169} (\bibinfo {year} {2024})},\ \Eprint {https://arxiv.org/abs/2310.02017} {arXiv:2310.02017 [gr-qc]} \BibitemShut {NoStop}%
\bibitem [{\citenamefont {Moore}\ \emph {et~al.}(2015)\citenamefont {Moore}, \citenamefont {Cole},\ and\ \citenamefont {Berry}}]{Moore:2014lga}%
  \BibitemOpen
  \bibfield  {author} {\bibinfo {author} {\bibfnamefont {C.~J.}\ \bibnamefont {Moore}}, \bibinfo {author} {\bibfnamefont {R.~H.}\ \bibnamefont {Cole}},\ and\ \bibinfo {author} {\bibfnamefont {C.~P.~L.}\ \bibnamefont {Berry}},\ }\bibfield  {title} {\bibinfo {title} {{Gravitational-wave sensitivity curves}},\ }\href {https://doi.org/10.1088/0264-9381/32/1/015014} {\bibfield  {journal} {\bibinfo  {journal} {Class. Quant. Grav.}\ }\textbf {\bibinfo {volume} {32}},\ \bibinfo {pages} {015014} (\bibinfo {year} {2015})},\ \Eprint {https://arxiv.org/abs/1408.0740} {arXiv:1408.0740 [gr-qc]} \BibitemShut {NoStop}%
\bibitem [{\citenamefont {Abbott}\ \emph {et~al.}(2020{\natexlab{a}})\citenamefont {Abbott} \emph {et~al.}}]{LIGOScientific:2019hgc}%
  \BibitemOpen
  \bibfield  {author} {\bibinfo {author} {\bibfnamefont {B.~P.}\ \bibnamefont {Abbott}} \emph {et~al.} (\bibinfo {collaboration} {LIGO Scientific, Virgo}),\ }\bibfield  {title} {\bibinfo {title} {{A guide to LIGO\textendash{}Virgo detector noise and extraction of transient gravitational-wave signals}},\ }\href {https://doi.org/10.1088/1361-6382/ab685e} {\bibfield  {journal} {\bibinfo  {journal} {Class. Quant. Grav.}\ }\textbf {\bibinfo {volume} {37}},\ \bibinfo {pages} {055002} (\bibinfo {year} {2020}{\natexlab{a}})},\ \Eprint {https://arxiv.org/abs/1908.11170} {arXiv:1908.11170 [gr-qc]} \BibitemShut {NoStop}%
\bibitem [{\citenamefont {Pankow}\ \emph {et~al.}(2018)\citenamefont {Pankow} \emph {et~al.}}]{Pankow:2018qpo}%
  \BibitemOpen
  \bibfield  {author} {\bibinfo {author} {\bibfnamefont {C.}~\bibnamefont {Pankow}} \emph {et~al.},\ }\bibfield  {title} {\bibinfo {title} {{Mitigation of the instrumental noise transient in gravitational-wave data surrounding GW170817}},\ }\href {https://doi.org/10.1103/PhysRevD.98.084016} {\bibfield  {journal} {\bibinfo  {journal} {Phys. Rev. D}\ }\textbf {\bibinfo {volume} {98}},\ \bibinfo {pages} {084016} (\bibinfo {year} {2018})},\ \Eprint {https://arxiv.org/abs/1808.03619} {arXiv:1808.03619 [gr-qc]} \BibitemShut {NoStop}%
\bibitem [{\citenamefont {Chatziioannou}\ \emph {et~al.}(2021)\citenamefont {Chatziioannou}, \citenamefont {Cornish}, \citenamefont {Wijngaarden},\ and\ \citenamefont {Littenberg}}]{Chatziioannou:2021ezd}%
  \BibitemOpen
  \bibfield  {author} {\bibinfo {author} {\bibfnamefont {K.}~\bibnamefont {Chatziioannou}}, \bibinfo {author} {\bibfnamefont {N.}~\bibnamefont {Cornish}}, \bibinfo {author} {\bibfnamefont {M.}~\bibnamefont {Wijngaarden}},\ and\ \bibinfo {author} {\bibfnamefont {T.~B.}\ \bibnamefont {Littenberg}},\ }\bibfield  {title} {\bibinfo {title} {{Modeling compact binary signals and instrumental glitches in gravitational wave data}},\ }\href {https://doi.org/10.1103/PhysRevD.103.044013} {\bibfield  {journal} {\bibinfo  {journal} {Phys. Rev. D}\ }\textbf {\bibinfo {volume} {103}},\ \bibinfo {pages} {044013} (\bibinfo {year} {2021})},\ \Eprint {https://arxiv.org/abs/2101.01200} {arXiv:2101.01200 [gr-qc]} \BibitemShut {NoStop}%
\bibitem [{\citenamefont {Hourihane}\ \emph {et~al.}(2022)\citenamefont {Hourihane}, \citenamefont {Chatziioannou}, \citenamefont {Wijngaarden}, \citenamefont {Davis}, \citenamefont {Littenberg},\ and\ \citenamefont {Cornish}}]{Hourihane:2022doe}%
  \BibitemOpen
  \bibfield  {author} {\bibinfo {author} {\bibfnamefont {S.}~\bibnamefont {Hourihane}}, \bibinfo {author} {\bibfnamefont {K.}~\bibnamefont {Chatziioannou}}, \bibinfo {author} {\bibfnamefont {M.}~\bibnamefont {Wijngaarden}}, \bibinfo {author} {\bibfnamefont {D.}~\bibnamefont {Davis}}, \bibinfo {author} {\bibfnamefont {T.}~\bibnamefont {Littenberg}},\ and\ \bibinfo {author} {\bibfnamefont {N.}~\bibnamefont {Cornish}},\ }\bibfield  {title} {\bibinfo {title} {{Accurate modeling and mitigation of overlapping signals and glitches in gravitational-wave data}},\ }\href {https://doi.org/10.1103/PhysRevD.106.042006} {\bibfield  {journal} {\bibinfo  {journal} {Phys. Rev. D}\ }\textbf {\bibinfo {volume} {106}},\ \bibinfo {pages} {042006} (\bibinfo {year} {2022})},\ \Eprint {https://arxiv.org/abs/2205.13580} {arXiv:2205.13580 [gr-qc]} \BibitemShut {NoStop}%
\bibitem [{\citenamefont {Davis}\ \emph {et~al.}(2022)\citenamefont {Davis}, \citenamefont {Littenberg}, \citenamefont {Romero-Shaw}, \citenamefont {Millhouse}, \citenamefont {McIver}, \citenamefont {Di~Renzo},\ and\ \citenamefont {Ashton}}]{Davis:2022ird}%
  \BibitemOpen
  \bibfield  {author} {\bibinfo {author} {\bibfnamefont {D.}~\bibnamefont {Davis}}, \bibinfo {author} {\bibfnamefont {T.~B.}\ \bibnamefont {Littenberg}}, \bibinfo {author} {\bibfnamefont {I.~M.}\ \bibnamefont {Romero-Shaw}}, \bibinfo {author} {\bibfnamefont {M.}~\bibnamefont {Millhouse}}, \bibinfo {author} {\bibfnamefont {J.}~\bibnamefont {McIver}}, \bibinfo {author} {\bibfnamefont {F.}~\bibnamefont {Di~Renzo}},\ and\ \bibinfo {author} {\bibfnamefont {G.}~\bibnamefont {Ashton}},\ }\bibfield  {title} {\bibinfo {title} {{Subtracting glitches from gravitational-wave detector data during the third LIGO-Virgo observing run}},\ }\href {https://doi.org/10.1088/1361-6382/aca238} {\bibfield  {journal} {\bibinfo  {journal} {Class. Quant. Grav.}\ }\textbf {\bibinfo {volume} {39}},\ \bibinfo {pages} {245013} (\bibinfo {year} {2022})},\ \Eprint {https://arxiv.org/abs/2207.03429} {arXiv:2207.03429 [astro-ph.IM]} \BibitemShut {NoStop}%
\bibitem [{\citenamefont {Payne}\ \emph {et~al.}(2022)\citenamefont {Payne}, \citenamefont {Hourihane}, \citenamefont {Golomb}, \citenamefont {Udall}, \citenamefont {Udall}, \citenamefont {Davis},\ and\ \citenamefont {Chatziioannou}}]{Payne:2022spz}%
  \BibitemOpen
  \bibfield  {author} {\bibinfo {author} {\bibfnamefont {E.}~\bibnamefont {Payne}}, \bibinfo {author} {\bibfnamefont {S.}~\bibnamefont {Hourihane}}, \bibinfo {author} {\bibfnamefont {J.}~\bibnamefont {Golomb}}, \bibinfo {author} {\bibfnamefont {R.}~\bibnamefont {Udall}}, \bibinfo {author} {\bibfnamefont {R.}~\bibnamefont {Udall}}, \bibinfo {author} {\bibfnamefont {D.}~\bibnamefont {Davis}},\ and\ \bibinfo {author} {\bibfnamefont {K.}~\bibnamefont {Chatziioannou}},\ }\bibfield  {title} {\bibinfo {title} {{Curious case of GW200129: Interplay between spin-precession inference and data-quality issues}},\ }\href {https://doi.org/10.1103/PhysRevD.106.104017} {\bibfield  {journal} {\bibinfo  {journal} {Phys. Rev. D}\ }\textbf {\bibinfo {volume} {106}},\ \bibinfo {pages} {104017} (\bibinfo {year} {2022})},\ \Eprint {https://arxiv.org/abs/2206.11932} {arXiv:2206.11932 [gr-qc]} \BibitemShut {NoStop}%
\bibitem [{\citenamefont {Ghonge}\ \emph {et~al.}(2024)\citenamefont {Ghonge}, \citenamefont {Brandt}, \citenamefont {Sullivan}, \citenamefont {Millhouse}, \citenamefont {Chatziioannou}, \citenamefont {Clark}, \citenamefont {Littenberg}, \citenamefont {Cornish}, \citenamefont {Hourihane},\ and\ \citenamefont {Cadonati}}]{Ghonge:2023ksb}%
  \BibitemOpen
  \bibfield  {author} {\bibinfo {author} {\bibfnamefont {S.}~\bibnamefont {Ghonge}}, \bibinfo {author} {\bibfnamefont {J.}~\bibnamefont {Brandt}}, \bibinfo {author} {\bibfnamefont {J.~M.}\ \bibnamefont {Sullivan}}, \bibinfo {author} {\bibfnamefont {M.}~\bibnamefont {Millhouse}}, \bibinfo {author} {\bibfnamefont {K.}~\bibnamefont {Chatziioannou}}, \bibinfo {author} {\bibfnamefont {J.~A.}\ \bibnamefont {Clark}}, \bibinfo {author} {\bibfnamefont {T.}~\bibnamefont {Littenberg}}, \bibinfo {author} {\bibfnamefont {N.}~\bibnamefont {Cornish}}, \bibinfo {author} {\bibfnamefont {S.}~\bibnamefont {Hourihane}},\ and\ \bibinfo {author} {\bibfnamefont {L.}~\bibnamefont {Cadonati}},\ }\bibfield  {title} {\bibinfo {title} {{Assessing and mitigating the impact of glitches on gravitational-wave parameter estimation: A model agnostic approach}},\ }\href {https://doi.org/10.1103/PhysRevD.110.122002} {\bibfield  {journal} {\bibinfo  {journal} {Phys. Rev. D}\ }\textbf {\bibinfo {volume} {110}},\ \bibinfo {pages} {122002}
  (\bibinfo {year} {2024})},\ \Eprint {https://arxiv.org/abs/2311.09159} {arXiv:2311.09159 [gr-qc]} \BibitemShut {NoStop}%
\bibitem [{\citenamefont {Macas}\ \emph {et~al.}(2024)\citenamefont {Macas}, \citenamefont {Lundgren},\ and\ \citenamefont {Ashton}}]{Macas:2023wiw}%
  \BibitemOpen
  \bibfield  {author} {\bibinfo {author} {\bibfnamefont {R.}~\bibnamefont {Macas}}, \bibinfo {author} {\bibfnamefont {A.}~\bibnamefont {Lundgren}},\ and\ \bibinfo {author} {\bibfnamefont {G.}~\bibnamefont {Ashton}},\ }\bibfield  {title} {\bibinfo {title} {{Revisiting the evidence for precession in GW200129 with machine learning noise mitigation}},\ }\href {https://doi.org/10.1103/PhysRevD.109.062006} {\bibfield  {journal} {\bibinfo  {journal} {Phys. Rev. D}\ }\textbf {\bibinfo {volume} {109}},\ \bibinfo {pages} {062006} (\bibinfo {year} {2024})},\ \Eprint {https://arxiv.org/abs/2311.09921} {arXiv:2311.09921 [gr-qc]} \BibitemShut {NoStop}%
\bibitem [{\citenamefont {Udall}\ \emph {et~al.}(2025)\citenamefont {Udall}, \citenamefont {Hourihane}, \citenamefont {Miller}, \citenamefont {Davis}, \citenamefont {Chatziioannou}, \citenamefont {Isi},\ and\ \citenamefont {Deshong}}]{Udall:2024ovp}%
  \BibitemOpen
  \bibfield  {author} {\bibinfo {author} {\bibfnamefont {R.}~\bibnamefont {Udall}}, \bibinfo {author} {\bibfnamefont {S.}~\bibnamefont {Hourihane}}, \bibinfo {author} {\bibfnamefont {S.}~\bibnamefont {Miller}}, \bibinfo {author} {\bibfnamefont {D.}~\bibnamefont {Davis}}, \bibinfo {author} {\bibfnamefont {K.}~\bibnamefont {Chatziioannou}}, \bibinfo {author} {\bibfnamefont {M.}~\bibnamefont {Isi}},\ and\ \bibinfo {author} {\bibfnamefont {H.}~\bibnamefont {Deshong}},\ }\bibfield  {title} {\bibinfo {title} {{Antialigned spin of GW191109: Glitch mitigation and its implications}},\ }\href {https://doi.org/10.1103/PhysRevD.111.024046} {\bibfield  {journal} {\bibinfo  {journal} {Phys. Rev. D}\ }\textbf {\bibinfo {volume} {111}},\ \bibinfo {pages} {024046} (\bibinfo {year} {2025})},\ \Eprint {https://arxiv.org/abs/2409.03912} {arXiv:2409.03912 [gr-qc]} \BibitemShut {NoStop}%
\bibitem [{\citenamefont {Cabero}\ \emph {et~al.}(2019)\citenamefont {Cabero} \emph {et~al.}}]{Cabero:2019orq}%
  \BibitemOpen
  \bibfield  {author} {\bibinfo {author} {\bibfnamefont {M.}~\bibnamefont {Cabero}} \emph {et~al.},\ }\bibfield  {title} {\bibinfo {title} {{Blip glitches in Advanced LIGO data}},\ }\href {https://doi.org/10.1088/1361-6382/ab2e14} {\bibfield  {journal} {\bibinfo  {journal} {Class. Quant. Grav.}\ }\textbf {\bibinfo {volume} {36}},\ \bibinfo {pages} {15} (\bibinfo {year} {2019})},\ \Eprint {https://arxiv.org/abs/1901.05093} {arXiv:1901.05093 [physics.ins-det]} \BibitemShut {NoStop}%
\bibitem [{\citenamefont {Zevin}\ \emph {et~al.}(2017{\natexlab{b}})\citenamefont {Zevin} \emph {et~al.}}]{Zevin:2016qwy}%
  \BibitemOpen
  \bibfield  {author} {\bibinfo {author} {\bibfnamefont {M.}~\bibnamefont {Zevin}} \emph {et~al.},\ }\bibfield  {title} {\bibinfo {title} {{Gravity Spy: Integrating Advanced LIGO Detector Characterization, Machine Learning, and Citizen Science}},\ }\href {https://doi.org/10.1088/1361-6382/aa5cea} {\bibfield  {journal} {\bibinfo  {journal} {Class. Quant. Grav.}\ }\textbf {\bibinfo {volume} {34}},\ \bibinfo {pages} {064003} (\bibinfo {year} {2017}{\natexlab{b}})},\ \Eprint {https://arxiv.org/abs/1611.04596} {arXiv:1611.04596 [gr-qc]} \BibitemShut {NoStop}%
\bibitem [{\citenamefont {Davis}\ \emph {et~al.}(2021)\citenamefont {Davis} \emph {et~al.}}]{LIGO:2021ppb}%
  \BibitemOpen
  \bibfield  {author} {\bibinfo {author} {\bibfnamefont {D.}~\bibnamefont {Davis}} \emph {et~al.} (\bibinfo {collaboration} {LIGO}),\ }\bibfield  {title} {\bibinfo {title} {{LIGO detector characterization in the second and third observing runs}},\ }\href {https://doi.org/10.1088/1361-6382/abfd85} {\bibfield  {journal} {\bibinfo  {journal} {Class. Quant. Grav.}\ }\textbf {\bibinfo {volume} {38}},\ \bibinfo {pages} {135014} (\bibinfo {year} {2021})},\ \Eprint {https://arxiv.org/abs/2101.11673} {arXiv:2101.11673 [astro-ph.IM]} \BibitemShut {NoStop}%
\bibitem [{\citenamefont {Soni}\ \emph {et~al.}(2021)\citenamefont {Soni} \emph {et~al.}}]{Soni:2021cjy}%
  \BibitemOpen
  \bibfield  {author} {\bibinfo {author} {\bibfnamefont {S.}~\bibnamefont {Soni}} \emph {et~al.},\ }\bibfield  {title} {\bibinfo {title} {{Discovering features in gravitational-wave data through detector characterization, citizen science and machine learning}},\ }\href {https://doi.org/10.1088/1361-6382/ac1ccb} {\bibfield  {journal} {\bibinfo  {journal} {Class. Quant. Grav.}\ }\textbf {\bibinfo {volume} {38}},\ \bibinfo {pages} {195016} (\bibinfo {year} {2021})},\ \Eprint {https://arxiv.org/abs/2103.12104} {arXiv:2103.12104 [gr-qc]} \BibitemShut {NoStop}%
\bibitem [{\citenamefont {Acernese}\ \emph {et~al.}(2023)\citenamefont {Acernese} \emph {et~al.}}]{Virgo:2022ysc}%
  \BibitemOpen
  \bibfield  {author} {\bibinfo {author} {\bibfnamefont {F.}~\bibnamefont {Acernese}} \emph {et~al.} (\bibinfo {collaboration} {Virgo}),\ }\bibfield  {title} {\bibinfo {title} {{Virgo detector characterization and data quality: results from the O3 run}},\ }\href {https://doi.org/10.1088/1361-6382/acd92d} {\bibfield  {journal} {\bibinfo  {journal} {Class. Quant. Grav.}\ }\textbf {\bibinfo {volume} {40}},\ \bibinfo {pages} {185006} (\bibinfo {year} {2023})},\ \Eprint {https://arxiv.org/abs/2210.15633} {arXiv:2210.15633 [gr-qc]} \BibitemShut {NoStop}%
\bibitem [{\citenamefont {Akutsu}\ \emph {et~al.}(2021)\citenamefont {Akutsu} \emph {et~al.}}]{KAGRA:2020agh}%
  \BibitemOpen
  \bibfield  {author} {\bibinfo {author} {\bibfnamefont {T.}~\bibnamefont {Akutsu}} \emph {et~al.} (\bibinfo {collaboration} {KAGRA}),\ }\bibfield  {title} {\bibinfo {title} {{Overview of KAGRA: Calibration, detector characterization, physical environmental monitors, and the geophysics interferometer}},\ }\href {https://doi.org/10.1093/ptep/ptab018} {\bibfield  {journal} {\bibinfo  {journal} {PTEP}\ }\textbf {\bibinfo {volume} {2021}},\ \bibinfo {pages} {05A102} (\bibinfo {year} {2021})},\ \Eprint {https://arxiv.org/abs/2009.09305} {arXiv:2009.09305 [gr-qc]} \BibitemShut {NoStop}%
\bibitem [{\citenamefont {Ashton}\ \emph {et~al.}(2022)\citenamefont {Ashton}, \citenamefont {Thiele}, \citenamefont {Lecoeuche}, \citenamefont {McIver},\ and\ \citenamefont {Nuttall}}]{Ashton:2021tvz}%
  \BibitemOpen
  \bibfield  {author} {\bibinfo {author} {\bibfnamefont {G.}~\bibnamefont {Ashton}}, \bibinfo {author} {\bibfnamefont {S.}~\bibnamefont {Thiele}}, \bibinfo {author} {\bibfnamefont {Y.}~\bibnamefont {Lecoeuche}}, \bibinfo {author} {\bibfnamefont {J.}~\bibnamefont {McIver}},\ and\ \bibinfo {author} {\bibfnamefont {L.~K.}\ \bibnamefont {Nuttall}},\ }\bibfield  {title} {\bibinfo {title} {{Parameterised population models of transient non-Gaussian noise in the LIGO gravitational-wave detectors}},\ }\href {https://doi.org/10.1088/1361-6382/ac8094} {\bibfield  {journal} {\bibinfo  {journal} {Class. Quant. Grav.}\ }\textbf {\bibinfo {volume} {39}},\ \bibinfo {pages} {175004} (\bibinfo {year} {2022})},\ \Eprint {https://arxiv.org/abs/2110.02689} {arXiv:2110.02689 [gr-qc]} \BibitemShut {NoStop}%
\bibitem [{\citenamefont {Magee}\ \emph {et~al.}(2024)\citenamefont {Magee}, \citenamefont {Sharma}, \citenamefont {Agrawal},\ and\ \citenamefont {Udall}}]{Magee:2024xiw}%
  \BibitemOpen
  \bibfield  {author} {\bibinfo {author} {\bibfnamefont {R.}~\bibnamefont {Magee}}, \bibinfo {author} {\bibfnamefont {R.}~\bibnamefont {Sharma}}, \bibinfo {author} {\bibfnamefont {A.}~\bibnamefont {Agrawal}},\ and\ \bibinfo {author} {\bibfnamefont {R.}~\bibnamefont {Udall}},\ }\bibfield  {title} {\bibinfo {title} {{Mitigating the impact of noise transients in gravitational-wave searches using reduced basis timeseries and convolutional neural networks}},\ }\href@noop {} {\  (\bibinfo {year} {2024})},\ \Eprint {https://arxiv.org/abs/2410.15513} {arXiv:2410.15513 [astro-ph.IM]} \BibitemShut {NoStop}%
\bibitem [{\citenamefont {Cheng}\ \emph {et~al.}(2023)\citenamefont {Cheng}, \citenamefont {Zevin},\ and\ \citenamefont {Vitale}}]{Cheng:2023ddt}%
  \BibitemOpen
  \bibfield  {author} {\bibinfo {author} {\bibfnamefont {A.~Q.}\ \bibnamefont {Cheng}}, \bibinfo {author} {\bibfnamefont {M.}~\bibnamefont {Zevin}},\ and\ \bibinfo {author} {\bibfnamefont {S.}~\bibnamefont {Vitale}},\ }\bibfield  {title} {\bibinfo {title} {{What You Don\textquoteright{}t Know Can Hurt You: Use and Abuse of Astrophysical Models in Gravitational-wave Population Analyses}},\ }\href {https://doi.org/10.3847/1538-4357/aced98} {\bibfield  {journal} {\bibinfo  {journal} {Astrophys. J.}\ }\textbf {\bibinfo {volume} {955}},\ \bibinfo {pages} {127} (\bibinfo {year} {2023})},\ \Eprint {https://arxiv.org/abs/2307.03129} {arXiv:2307.03129 [astro-ph.HE]} \BibitemShut {NoStop}%
\bibitem [{\citenamefont {Alvarez-Lopez}\ \emph {et~al.}(2025)\citenamefont {Alvarez-Lopez}, \citenamefont {Heinzel}, \citenamefont {Mould},\ and\ \citenamefont {Vitale}}]{Alvarez-Lopez:2025ltt}%
  \BibitemOpen
  \bibfield  {author} {\bibinfo {author} {\bibfnamefont {S.}~\bibnamefont {Alvarez-Lopez}}, \bibinfo {author} {\bibfnamefont {J.}~\bibnamefont {Heinzel}}, \bibinfo {author} {\bibfnamefont {M.}~\bibnamefont {Mould}},\ and\ \bibinfo {author} {\bibfnamefont {S.}~\bibnamefont {Vitale}},\ }\bibfield  {title} {\bibinfo {title} {{Nowhere left to hide: revealing realistic gravitational-wave populations in high dimensions and high resolution with PixelPop}},\ }\href@noop {} {\  (\bibinfo {year} {2025})},\ \Eprint {https://arxiv.org/abs/2506.20731} {arXiv:2506.20731 [astro-ph.HE]} \BibitemShut {NoStop}%
\bibitem [{\citenamefont {Golomb}\ and\ \citenamefont {Talbot}(2023)}]{Golomb:2022bon}%
  \BibitemOpen
  \bibfield  {author} {\bibinfo {author} {\bibfnamefont {J.}~\bibnamefont {Golomb}}\ and\ \bibinfo {author} {\bibfnamefont {C.}~\bibnamefont {Talbot}},\ }\bibfield  {title} {\bibinfo {title} {{Searching for structure in the binary black hole spin distribution}},\ }\href {https://doi.org/10.1103/PhysRevD.108.103009} {\bibfield  {journal} {\bibinfo  {journal} {Phys. Rev. D}\ }\textbf {\bibinfo {volume} {108}},\ \bibinfo {pages} {103009} (\bibinfo {year} {2023})},\ \Eprint {https://arxiv.org/abs/2210.12287} {arXiv:2210.12287 [astro-ph.HE]} \BibitemShut {NoStop}%
\bibitem [{\citenamefont {Talbot}\ and\ \citenamefont {Golomb}(2023)}]{Talbot:2023pex}%
  \BibitemOpen
  \bibfield  {author} {\bibinfo {author} {\bibfnamefont {C.}~\bibnamefont {Talbot}}\ and\ \bibinfo {author} {\bibfnamefont {J.}~\bibnamefont {Golomb}},\ }\bibfield  {title} {\bibinfo {title} {{Growing pains: understanding the impact of likelihood uncertainty on hierarchical Bayesian inference for gravitational-wave astronomy}},\ }\href {https://doi.org/10.1093/mnras/stad2968} {\bibfield  {journal} {\bibinfo  {journal} {Mon. Not. Roy. Astron. Soc.}\ }\textbf {\bibinfo {volume} {526}},\ \bibinfo {pages} {3495} (\bibinfo {year} {2023})},\ \Eprint {https://arxiv.org/abs/2304.06138} {arXiv:2304.06138 [astro-ph.IM]} \BibitemShut {NoStop}%
\bibitem [{\citenamefont {Farr}(2019)}]{Farr:2019rap}%
  \BibitemOpen
  \bibfield  {author} {\bibinfo {author} {\bibfnamefont {W.~M.}\ \bibnamefont {Farr}},\ }\bibfield  {title} {\bibinfo {title} {{Accuracy Requirements for Empirically-Measured Selection Functions}},\ }\href {https://doi.org/10.3847/2515-5172/ab1d5f} {\bibfield  {journal} {\bibinfo  {journal} {Research Notes of the AAS}\ }\textbf {\bibinfo {volume} {3}},\ \bibinfo {pages} {66} (\bibinfo {year} {2019})},\ \Eprint {https://arxiv.org/abs/1904.10879} {arXiv:1904.10879 [astro-ph.IM]} \BibitemShut {NoStop}%
\bibitem [{\citenamefont {Essick}\ and\ \citenamefont {Farr}(2022)}]{Essick:2022ojx}%
  \BibitemOpen
  \bibfield  {author} {\bibinfo {author} {\bibfnamefont {R.}~\bibnamefont {Essick}}\ and\ \bibinfo {author} {\bibfnamefont {W.}~\bibnamefont {Farr}},\ }\bibfield  {title} {\bibinfo {title} {{Precision Requirements for Monte Carlo Sums within Hierarchical Bayesian Inference}},\ }\href@noop {} {\  (\bibinfo {year} {2022})},\ \Eprint {https://arxiv.org/abs/2204.00461} {arXiv:2204.00461 [astro-ph.IM]} \BibitemShut {NoStop}%
\bibitem [{\citenamefont {Farr}\ \emph {et~al.}(2015)\citenamefont {Farr}, \citenamefont {Gair}, \citenamefont {Mandel},\ and\ \citenamefont {Cutler}}]{Farr:2013yna}%
  \BibitemOpen
  \bibfield  {author} {\bibinfo {author} {\bibfnamefont {W.~M.}\ \bibnamefont {Farr}}, \bibinfo {author} {\bibfnamefont {J.~R.}\ \bibnamefont {Gair}}, \bibinfo {author} {\bibfnamefont {I.}~\bibnamefont {Mandel}},\ and\ \bibinfo {author} {\bibfnamefont {C.}~\bibnamefont {Cutler}},\ }\bibfield  {title} {\bibinfo {title} {{Counting And Confusion: Bayesian Rate Estimation With Multiple Populations}},\ }\href {https://doi.org/10.1103/PhysRevD.91.023005} {\bibfield  {journal} {\bibinfo  {journal} {Phys. Rev. D}\ }\textbf {\bibinfo {volume} {91}},\ \bibinfo {pages} {023005} (\bibinfo {year} {2015})},\ \Eprint {https://arxiv.org/abs/1302.5341} {arXiv:1302.5341 [astro-ph.IM]} \BibitemShut {NoStop}%
\bibitem [{\citenamefont {Mandel}\ \emph {et~al.}(2019)\citenamefont {Mandel}, \citenamefont {Farr},\ and\ \citenamefont {Gair}}]{Mandel:2018mve}%
  \BibitemOpen
  \bibfield  {author} {\bibinfo {author} {\bibfnamefont {I.}~\bibnamefont {Mandel}}, \bibinfo {author} {\bibfnamefont {W.~M.}\ \bibnamefont {Farr}},\ and\ \bibinfo {author} {\bibfnamefont {J.~R.}\ \bibnamefont {Gair}},\ }\bibfield  {title} {\bibinfo {title} {{Extracting distribution parameters from multiple uncertain observations with selection biases}},\ }\href {https://doi.org/10.1093/mnras/stz896} {\bibfield  {journal} {\bibinfo  {journal} {Mon. Not. Roy. Astron. Soc.}\ }\textbf {\bibinfo {volume} {486}},\ \bibinfo {pages} {1086} (\bibinfo {year} {2019})},\ \Eprint {https://arxiv.org/abs/1809.02063} {arXiv:1809.02063 [physics.data-an]} \BibitemShut {NoStop}%
\bibitem [{\citenamefont {Vitale}\ \emph {et~al.}(2020)\citenamefont {Vitale}, \citenamefont {Gerosa}, \citenamefont {Farr},\ and\ \citenamefont {Taylor}}]{Vitale2020}%
  \BibitemOpen
  \bibfield  {author} {\bibinfo {author} {\bibfnamefont {S.}~\bibnamefont {Vitale}}, \bibinfo {author} {\bibfnamefont {D.}~\bibnamefont {Gerosa}}, \bibinfo {author} {\bibfnamefont {W.~M.}\ \bibnamefont {Farr}},\ and\ \bibinfo {author} {\bibfnamefont {S.~R.}\ \bibnamefont {Taylor}},\ }\bibinfo {title} {Inferring the properties of a population of compact binaries in presence of selection effects},\ in\ \href {https://doi.org/10.1007/978-981-15-4702-7_45-1} {\emph {\bibinfo {booktitle} {Handbook of Gravitational Wave Astronomy}}},\ \bibinfo {editor} {edited by\ \bibinfo {editor} {\bibfnamefont {C.}~\bibnamefont {Bambi}}, \bibinfo {editor} {\bibfnamefont {S.}~\bibnamefont {Katsanevas}},\ and\ \bibinfo {editor} {\bibfnamefont {K.~D.}\ \bibnamefont {Kokkotas}}}\ (\bibinfo  {publisher} {Springer Singapore},\ \bibinfo {address} {Singapore},\ \bibinfo {year} {2020})\ pp.\ \bibinfo {pages} {1--60}\BibitemShut {NoStop}%
\bibitem [{\citenamefont {Allen}\ \emph {et~al.}(2012)\citenamefont {Allen}, \citenamefont {Anderson}, \citenamefont {Brady}, \citenamefont {Brown},\ and\ \citenamefont {Creighton}}]{Allen:2005fk}%
  \BibitemOpen
  \bibfield  {author} {\bibinfo {author} {\bibfnamefont {B.}~\bibnamefont {Allen}}, \bibinfo {author} {\bibfnamefont {W.~G.}\ \bibnamefont {Anderson}}, \bibinfo {author} {\bibfnamefont {P.~R.}\ \bibnamefont {Brady}}, \bibinfo {author} {\bibfnamefont {D.~A.}\ \bibnamefont {Brown}},\ and\ \bibinfo {author} {\bibfnamefont {J.~D.~E.}\ \bibnamefont {Creighton}},\ }\bibfield  {title} {\bibinfo {title} {{FINDCHIRP: An Algorithm for detection of gravitational waves from inspiraling compact binaries}},\ }\href {https://doi.org/10.1103/PhysRevD.85.122006} {\bibfield  {journal} {\bibinfo  {journal} {Phys. Rev. D}\ }\textbf {\bibinfo {volume} {85}},\ \bibinfo {pages} {122006} (\bibinfo {year} {2012})},\ \Eprint {https://arxiv.org/abs/gr-qc/0509116} {arXiv:gr-qc/0509116} \BibitemShut {NoStop}%
\bibitem [{\citenamefont {Usman}\ \emph {et~al.}(2016)\citenamefont {Usman} \emph {et~al.}}]{Usman:2015kfa}%
  \BibitemOpen
  \bibfield  {author} {\bibinfo {author} {\bibfnamefont {S.~A.}\ \bibnamefont {Usman}} \emph {et~al.},\ }\bibfield  {title} {\bibinfo {title} {{The PyCBC search for gravitational waves from compact binary coalescence}},\ }\href {https://doi.org/10.1088/0264-9381/33/21/215004} {\bibfield  {journal} {\bibinfo  {journal} {Class. Quant. Grav.}\ }\textbf {\bibinfo {volume} {33}},\ \bibinfo {pages} {215004} (\bibinfo {year} {2016})},\ \Eprint {https://arxiv.org/abs/1508.02357} {arXiv:1508.02357 [gr-qc]} \BibitemShut {NoStop}%
\bibitem [{\citenamefont {Messick}\ \emph {et~al.}(2017)\citenamefont {Messick} \emph {et~al.}}]{Messick:2016aqy}%
  \BibitemOpen
  \bibfield  {author} {\bibinfo {author} {\bibfnamefont {C.}~\bibnamefont {Messick}} \emph {et~al.},\ }\bibfield  {title} {\bibinfo {title} {{Analysis Framework for the Prompt Discovery of Compact Binary Mergers in Gravitational-wave Data}},\ }\href {https://doi.org/10.1103/PhysRevD.95.042001} {\bibfield  {journal} {\bibinfo  {journal} {Phys. Rev. D}\ }\textbf {\bibinfo {volume} {95}},\ \bibinfo {pages} {042001} (\bibinfo {year} {2017})},\ \Eprint {https://arxiv.org/abs/1604.04324} {arXiv:1604.04324 [astro-ph.IM]} \BibitemShut {NoStop}%
\bibitem [{\citenamefont {Tiwari}(2018)}]{Tiwari:2017ndi}%
  \BibitemOpen
  \bibfield  {author} {\bibinfo {author} {\bibfnamefont {V.}~\bibnamefont {Tiwari}},\ }\bibfield  {title} {\bibinfo {title} {{Estimation of the Sensitive Volume for Gravitational-wave Source Populations Using Weighted Monte Carlo Integration}},\ }\href {https://doi.org/10.1088/1361-6382/aac89d} {\bibfield  {journal} {\bibinfo  {journal} {Class. Quant. Grav.}\ }\textbf {\bibinfo {volume} {35}},\ \bibinfo {pages} {145009} (\bibinfo {year} {2018})},\ \Eprint {https://arxiv.org/abs/1712.00482} {arXiv:1712.00482 [astro-ph.HE]} \BibitemShut {NoStop}%
\bibitem [{\citenamefont {MacKay}(2003)}]{MacKay2003}%
  \BibitemOpen
  \bibfield  {author} {\bibinfo {author} {\bibfnamefont {D.~J.~C.}\ \bibnamefont {MacKay}},\ }\href@noop {} {\emph {\bibinfo {title} {Information Theory, Inference, and Learning Algorithms}}}\ (\bibinfo  {publisher} {Copyright Cambridge University Press},\ \bibinfo {year} {2003})\BibitemShut {NoStop}%
\bibitem [{\citenamefont {Kendall}\ and\ \citenamefont {Stuart}(1977)}]{kendall1977advanced}%
  \BibitemOpen
  \bibfield  {author} {\bibinfo {author} {\bibfnamefont {M.}~\bibnamefont {Kendall}}\ and\ \bibinfo {author} {\bibfnamefont {A.}~\bibnamefont {Stuart}},\ }\href {https://books.google.com/books?id=BUoAwjZ4rfkC} {\emph {\bibinfo {title} {The Advanced Theory of Statistics: Distribution theory}}},\ Distribution Theory\ (\bibinfo  {publisher} {Macmillan},\ \bibinfo {year} {1977})\BibitemShut {NoStop}%
\bibitem [{\citenamefont {Kish}(1965)}]{kish1965survey}%
  \BibitemOpen
  \bibfield  {author} {\bibinfo {author} {\bibfnamefont {L.}~\bibnamefont {Kish}},\ }\href {https://books.google.com/books?id=yoppAAAAMAAJ} {\emph {\bibinfo {title} {Survey Sampling}}}\ (\bibinfo  {publisher} {Wiley},\ \bibinfo {year} {1965})\BibitemShut {NoStop}%
\bibitem [{\citenamefont {Gair}\ \emph {et~al.}(2022)\citenamefont {Gair}, \citenamefont {Antonelli},\ and\ \citenamefont {Barbieri}}]{Gair:2022fsj}%
  \BibitemOpen
  \bibfield  {author} {\bibinfo {author} {\bibfnamefont {J.~R.}\ \bibnamefont {Gair}}, \bibinfo {author} {\bibfnamefont {A.}~\bibnamefont {Antonelli}},\ and\ \bibinfo {author} {\bibfnamefont {R.}~\bibnamefont {Barbieri}},\ }\bibfield  {title} {\bibinfo {title} {{A Fisher matrix for gravitational-wave population inference}},\ }\href {https://doi.org/10.1093/mnras/stac3560} {\bibfield  {journal} {\bibinfo  {journal} {Mon. Not. Roy. Astron. Soc.}\ }\textbf {\bibinfo {volume} {519}},\ \bibinfo {pages} {2736} (\bibinfo {year} {2022})},\ \Eprint {https://arxiv.org/abs/2205.07893} {arXiv:2205.07893 [gr-qc]} \BibitemShut {NoStop}%
\bibitem [{\citenamefont {Kleijn}\ and\ \citenamefont {van~der Vaart}(2012)}]{Kleijn:2012bvm}%
  \BibitemOpen
  \bibfield  {author} {\bibinfo {author} {\bibfnamefont {B.}~\bibnamefont {Kleijn}}\ and\ \bibinfo {author} {\bibfnamefont {A.}~\bibnamefont {van~der Vaart}},\ }\bibfield  {title} {\bibinfo {title} {{The Bernstein-Von-Mises theorem under misspecification}},\ }\href {https://doi.org/10.1214/12-EJS675} {\bibfield  {journal} {\bibinfo  {journal} {Electronic Journal of Statistics}\ }\textbf {\bibinfo {volume} {6}},\ \bibinfo {pages} {354 } (\bibinfo {year} {2012})}\BibitemShut {NoStop}%
\bibitem [{\citenamefont {Freedman}(1999)}]{Freedman:1999bvm}%
  \BibitemOpen
  \bibfield  {author} {\bibinfo {author} {\bibfnamefont {D.}~\bibnamefont {Freedman}},\ }\bibfield  {title} {\bibinfo {title} {On the bernstein-von mises theorem with infinite-dimensional parameters},\ }\href {http://www.jstor.org/stable/120155} {\bibfield  {journal} {\bibinfo  {journal} {The Annals of Statistics}\ }\textbf {\bibinfo {volume} {27}},\ \bibinfo {pages} {1119} (\bibinfo {year} {1999})}\BibitemShut {NoStop}%
\bibitem [{\citenamefont {Miller}\ and\ \citenamefont {Harrison}(2013)}]{Miller:2013ipy}%
  \BibitemOpen
  \bibfield  {author} {\bibinfo {author} {\bibfnamefont {J.~W.}\ \bibnamefont {Miller}}\ and\ \bibinfo {author} {\bibfnamefont {M.~T.}\ \bibnamefont {Harrison}},\ }\bibfield  {title} {\bibinfo {title} {Inconsistency of pitman-yor process mixtures for the number of components},\ }\href {https://arxiv.org/abs/1309.0024} {\  (\bibinfo {year} {2013})},\ \Eprint {https://arxiv.org/abs/1309.0024} {arXiv:1309.0024 [math.ST]} \BibitemShut {NoStop}%
\bibitem [{\citenamefont {Ghosal}\ and\ \citenamefont {van~der Vaart}(2017)}]{ghosal2017fundamentals}%
  \BibitemOpen
  \bibfield  {author} {\bibinfo {author} {\bibfnamefont {S.}~\bibnamefont {Ghosal}}\ and\ \bibinfo {author} {\bibfnamefont {A.}~\bibnamefont {van~der Vaart}},\ }\href {https://books.google.com/books?id=cs8oDwAAQBAJ} {\emph {\bibinfo {title} {Fundamentals of Nonparametric Bayesian Inference}}},\ Cambridge Series in Statistical and Probabilistic Mathematics\ (\bibinfo  {publisher} {Cambridge University Press},\ \bibinfo {year} {2017})\BibitemShut {NoStop}%
\bibitem [{\citenamefont {Vitale}\ and\ \citenamefont {Mould}(2025)}]{Vitale:2025lms}%
  \BibitemOpen
  \bibfield  {author} {\bibinfo {author} {\bibfnamefont {S.}~\bibnamefont {Vitale}}\ and\ \bibinfo {author} {\bibfnamefont {M.}~\bibnamefont {Mould}},\ }\bibfield  {title} {\bibinfo {title} {{The Long Road to Alignment: Measuring Black Hole Spin Orientation with Expanding Gravitational-Wave Datasets}},\ }\href@noop {} {\  (\bibinfo {year} {2025})},\ \Eprint {https://arxiv.org/abs/2505.14875} {arXiv:2505.14875 [astro-ph.HE]} \BibitemShut {NoStop}%
\bibitem [{\citenamefont {Talbot}\ \emph {et~al.}(2019)\citenamefont {Talbot}, \citenamefont {Smith}, \citenamefont {Thrane},\ and\ \citenamefont {Poole}}]{Talbot:2019okv}%
  \BibitemOpen
  \bibfield  {author} {\bibinfo {author} {\bibfnamefont {C.}~\bibnamefont {Talbot}}, \bibinfo {author} {\bibfnamefont {R.}~\bibnamefont {Smith}}, \bibinfo {author} {\bibfnamefont {E.}~\bibnamefont {Thrane}},\ and\ \bibinfo {author} {\bibfnamefont {G.~B.}\ \bibnamefont {Poole}},\ }\bibfield  {title} {\bibinfo {title} {{Parallelized Inference for Gravitational-Wave Astronomy}},\ }\href {https://doi.org/10.1103/PhysRevD.100.043030} {\bibfield  {journal} {\bibinfo  {journal} {Phys. Rev. D}\ }\textbf {\bibinfo {volume} {100}},\ \bibinfo {pages} {043030} (\bibinfo {year} {2019})},\ \Eprint {https://arxiv.org/abs/1904.02863} {arXiv:1904.02863 [astro-ph.IM]} \BibitemShut {NoStop}%
\bibitem [{\citenamefont {Talbot}\ \emph {et~al.}(2025)\citenamefont {Talbot}, \citenamefont {Farah}, \citenamefont {Galaudage}, \citenamefont {Golomb},\ and\ \citenamefont {Tong}}]{Talbot:2024yqw}%
  \BibitemOpen
  \bibfield  {author} {\bibinfo {author} {\bibfnamefont {C.}~\bibnamefont {Talbot}}, \bibinfo {author} {\bibfnamefont {A.}~\bibnamefont {Farah}}, \bibinfo {author} {\bibfnamefont {S.}~\bibnamefont {Galaudage}}, \bibinfo {author} {\bibfnamefont {J.}~\bibnamefont {Golomb}},\ and\ \bibinfo {author} {\bibfnamefont {H.}~\bibnamefont {Tong}},\ }\bibfield  {title} {\bibinfo {title} {{GWPopulation: Hardware agnostic population inference for compact binaries and beyond}},\ }\href {https://doi.org/10.21105/joss.07753} {\bibfield  {journal} {\bibinfo  {journal} {J. Open Source Softw.}\ }\textbf {\bibinfo {volume} {10}},\ \bibinfo {pages} {7753} (\bibinfo {year} {2025})},\ \Eprint {https://arxiv.org/abs/2409.14143} {arXiv:2409.14143 [astro-ph.IM]} \BibitemShut {NoStop}%
\bibitem [{\citenamefont {Ashton}\ \emph {et~al.}(2019)\citenamefont {Ashton} \emph {et~al.}}]{Ashton:2018jfp}%
  \BibitemOpen
  \bibfield  {author} {\bibinfo {author} {\bibfnamefont {G.}~\bibnamefont {Ashton}} \emph {et~al.},\ }\bibfield  {title} {\bibinfo {title} {{BILBY: A user-friendly Bayesian inference library for gravitational-wave astronomy}},\ }\href {https://doi.org/10.3847/1538-4365/ab06fc} {\bibfield  {journal} {\bibinfo  {journal} {Astrophys. J. Suppl.}\ }\textbf {\bibinfo {volume} {241}},\ \bibinfo {pages} {27} (\bibinfo {year} {2019})},\ \Eprint {https://arxiv.org/abs/1811.02042} {arXiv:1811.02042 [astro-ph.IM]} \BibitemShut {NoStop}%
\bibitem [{\citenamefont {Speagle}(2020)}]{Speagle_2020}%
  \BibitemOpen
  \bibfield  {author} {\bibinfo {author} {\bibfnamefont {J.~S.}\ \bibnamefont {Speagle}},\ }\bibfield  {title} {\bibinfo {title} {{dynesty: a dynamic nested sampling package for estimating Bayesian posteriors and evidences}},\ }\href {https://doi.org/10.1093/mnras/staa278} {\bibfield  {journal} {\bibinfo  {journal} {MNRAS}\ }\textbf {\bibinfo {volume} {493}},\ \bibinfo {pages} {3132} (\bibinfo {year} {2020})},\ \Eprint {https://arxiv.org/abs/https://academic.oup.com/mnras/article-pdf/493/3/3132/32890730/staa278.pdf} {https://academic.oup.com/mnras/article-pdf/493/3/3132/32890730/staa278.pdf} \BibitemShut {NoStop}%
\bibitem [{\citenamefont {Abbott}\ \emph {et~al.}(2020{\natexlab{b}})\citenamefont {Abbott} \emph {et~al.}}]{O4_psds}%
  \BibitemOpen
  \bibfield  {author} {\bibinfo {author} {\bibfnamefont {B.}~\bibnamefont {Abbott}} \emph {et~al.} (\bibinfo {collaboration} {LIGO, Virgo, KAGRA Scientific Collaborations}),\ }\bibfield  {title} {\bibinfo {title} {Noise curves used for simulations in the update of the observing scenarios paper}} (\bibinfo {year} {2020}{\natexlab{b}}),\ \bibinfo {note} {dcc LIGO-T2000012}\BibitemShut {NoStop}%
\bibitem [{\citenamefont {Wysocki}\ \emph {et~al.}(2019)\citenamefont {Wysocki}, \citenamefont {Lange},\ and\ \citenamefont {O'Shaughnessy}}]{Wysocki:2018mpo}%
  \BibitemOpen
  \bibfield  {author} {\bibinfo {author} {\bibfnamefont {D.}~\bibnamefont {Wysocki}}, \bibinfo {author} {\bibfnamefont {J.}~\bibnamefont {Lange}},\ and\ \bibinfo {author} {\bibfnamefont {R.}~\bibnamefont {O'Shaughnessy}},\ }\bibfield  {title} {\bibinfo {title} {Reconstructing phenomenological distributions of compact binaries via gravitational wave observations},\ }\href {https://doi.org/10.1103/PhysRevD.100.043012} {\bibfield  {journal} {\bibinfo  {journal} {Phys. Rev. D}\ }\textbf {\bibinfo {volume} {100}},\ \bibinfo {pages} {043012} (\bibinfo {year} {2019})}\BibitemShut {NoStop}%
\bibitem [{\citenamefont {Edelman}\ \emph {et~al.}(2022)\citenamefont {Edelman}, \citenamefont {Doctor}, \citenamefont {Godfrey},\ and\ \citenamefont {Farr}}]{Edelman:2021zkw}%
  \BibitemOpen
  \bibfield  {author} {\bibinfo {author} {\bibfnamefont {B.}~\bibnamefont {Edelman}}, \bibinfo {author} {\bibfnamefont {Z.}~\bibnamefont {Doctor}}, \bibinfo {author} {\bibfnamefont {J.}~\bibnamefont {Godfrey}},\ and\ \bibinfo {author} {\bibfnamefont {B.}~\bibnamefont {Farr}},\ }\bibfield  {title} {\bibinfo {title} {{Ain\textquoteright{}t No Mountain High Enough: Semiparametric Modeling of LIGO\textendash{}Virgo\textquoteright{}s Binary Black Hole Mass Distribution}},\ }\href {https://doi.org/10.3847/1538-4357/ac3667} {\bibfield  {journal} {\bibinfo  {journal} {Astrophys. J.}\ }\textbf {\bibinfo {volume} {924}},\ \bibinfo {pages} {101} (\bibinfo {year} {2022})},\ \Eprint {https://arxiv.org/abs/2109.06137} {arXiv:2109.06137 [astro-ph.HE]} \BibitemShut {NoStop}%
\bibitem [{\citenamefont {Edelman}\ \emph {et~al.}(2023)\citenamefont {Edelman}, \citenamefont {Farr},\ and\ \citenamefont {Doctor}}]{Edelman:2022ydv}%
  \BibitemOpen
  \bibfield  {author} {\bibinfo {author} {\bibfnamefont {B.}~\bibnamefont {Edelman}}, \bibinfo {author} {\bibfnamefont {B.}~\bibnamefont {Farr}},\ and\ \bibinfo {author} {\bibfnamefont {Z.}~\bibnamefont {Doctor}},\ }\bibfield  {title} {\bibinfo {title} {{Cover Your Basis: Comprehensive Data-driven Characterization of the Binary Black Hole Population}},\ }\href {https://doi.org/10.3847/1538-4357/acb5ed} {\bibfield  {journal} {\bibinfo  {journal} {Astrophys. J.}\ }\textbf {\bibinfo {volume} {946}},\ \bibinfo {pages} {16} (\bibinfo {year} {2023})},\ \Eprint {https://arxiv.org/abs/2210.12834} {arXiv:2210.12834 [astro-ph.HE]} \BibitemShut {NoStop}%
\bibitem [{\citenamefont {Godfrey}\ \emph {et~al.}(2023)\citenamefont {Godfrey}, \citenamefont {Edelman},\ and\ \citenamefont {Farr}}]{Godfrey:2023oxb}%
  \BibitemOpen
  \bibfield  {author} {\bibinfo {author} {\bibfnamefont {J.}~\bibnamefont {Godfrey}}, \bibinfo {author} {\bibfnamefont {B.}~\bibnamefont {Edelman}},\ and\ \bibinfo {author} {\bibfnamefont {B.}~\bibnamefont {Farr}},\ }\bibfield  {title} {\bibinfo {title} {{Cosmic Cousins: Identification of a Subpopulation of Binary Black Holes Consistent with Isolated Binary Evolution}},\ }\href@noop {} {\  (\bibinfo {year} {2023})},\ \Eprint {https://arxiv.org/abs/2304.01288} {arXiv:2304.01288 [astro-ph.HE]} \BibitemShut {NoStop}%
\bibitem [{\citenamefont {Heinzel}(2025)}]{population_error2025github}%
  \BibitemOpen
  \bibfield  {author} {\bibinfo {author} {\bibfnamefont {J.}~\bibnamefont {Heinzel}},\ }\href {https://github.com/jack-heinzel/population-error} {\bibinfo {title} {{population-error}: compute information loss in approximate hierarchical analyses}} (\bibinfo {year} {2025})\BibitemShut {NoStop}%
\bibitem [{\citenamefont {Bradbury}\ \emph {et~al.}(2018)\citenamefont {Bradbury}, \citenamefont {Frostig}, \citenamefont {Hawkins}, \citenamefont {Johnson}, \citenamefont {Leary}, \citenamefont {Maclaurin}, \citenamefont {Necula}, \citenamefont {Paszke}, \citenamefont {Vander{P}las}, \citenamefont {Wanderman-{M}ilne},\ and\ \citenamefont {Zhang}}]{jax2018github}%
  \BibitemOpen
  \bibfield  {author} {\bibinfo {author} {\bibfnamefont {J.}~\bibnamefont {Bradbury}}, \bibinfo {author} {\bibfnamefont {R.}~\bibnamefont {Frostig}}, \bibinfo {author} {\bibfnamefont {P.}~\bibnamefont {Hawkins}}, \bibinfo {author} {\bibfnamefont {M.~J.}\ \bibnamefont {Johnson}}, \bibinfo {author} {\bibfnamefont {C.}~\bibnamefont {Leary}}, \bibinfo {author} {\bibfnamefont {D.}~\bibnamefont {Maclaurin}}, \bibinfo {author} {\bibfnamefont {G.}~\bibnamefont {Necula}}, \bibinfo {author} {\bibfnamefont {A.}~\bibnamefont {Paszke}}, \bibinfo {author} {\bibfnamefont {J.}~\bibnamefont {Vander{P}las}}, \bibinfo {author} {\bibfnamefont {S.}~\bibnamefont {Wanderman-{M}ilne}},\ and\ \bibinfo {author} {\bibfnamefont {Q.}~\bibnamefont {Zhang}},\ }\href {http://github.com/google/jax} {\bibinfo {title} {{JAX}: composable transformations of {P}ython+{N}um{P}y programs}} (\bibinfo {year} {2018})\BibitemShut {NoStop}%
\bibitem [{\citenamefont {Rinaldi}\ and\ \citenamefont {Del~Pozzo}(2021)}]{Rinaldi:2021bhm}%
  \BibitemOpen
  \bibfield  {author} {\bibinfo {author} {\bibfnamefont {S.}~\bibnamefont {Rinaldi}}\ and\ \bibinfo {author} {\bibfnamefont {W.}~\bibnamefont {Del~Pozzo}},\ }\bibfield  {title} {\bibinfo {title} {{(H)DPGMM: a hierarchy of Dirichlet process Gaussian mixture models for the inference of the black hole mass function}},\ }\href {https://doi.org/10.1093/mnras/stab3224} {\bibfield  {journal} {\bibinfo  {journal} {Mon. Not. Roy. Astron. Soc.}\ }\textbf {\bibinfo {volume} {509}},\ \bibinfo {pages} {5454} (\bibinfo {year} {2021})},\ \Eprint {https://arxiv.org/abs/2109.05960} {arXiv:2109.05960 [astro-ph.IM]} \BibitemShut {NoStop}%
\bibitem [{\citenamefont {Callister}\ \emph {et~al.}(2022)\citenamefont {Callister}, \citenamefont {Miller}, \citenamefont {Chatziioannou},\ and\ \citenamefont {Farr}}]{Callister:2022qwb}%
  \BibitemOpen
  \bibfield  {author} {\bibinfo {author} {\bibfnamefont {T.~A.}\ \bibnamefont {Callister}}, \bibinfo {author} {\bibfnamefont {S.~J.}\ \bibnamefont {Miller}}, \bibinfo {author} {\bibfnamefont {K.}~\bibnamefont {Chatziioannou}},\ and\ \bibinfo {author} {\bibfnamefont {W.~M.}\ \bibnamefont {Farr}},\ }\bibfield  {title} {\bibinfo {title} {{No Evidence that the Majority of Black Holes in Binaries Have Zero Spin}},\ }\href {https://doi.org/10.3847/2041-8213/ac847e} {\bibfield  {journal} {\bibinfo  {journal} {Astrophys. J. Lett.}\ }\textbf {\bibinfo {volume} {937}},\ \bibinfo {pages} {L13} (\bibinfo {year} {2022})},\ \Eprint {https://arxiv.org/abs/2205.08574} {arXiv:2205.08574 [astro-ph.HE]} \BibitemShut {NoStop}%
\bibitem [{\citenamefont {Hussain}\ \emph {et~al.}(2024)\citenamefont {Hussain}, \citenamefont {Isi},\ and\ \citenamefont {Zimmerman}}]{Hussain:2024qzl}%
  \BibitemOpen
  \bibfield  {author} {\bibinfo {author} {\bibfnamefont {A.}~\bibnamefont {Hussain}}, \bibinfo {author} {\bibfnamefont {M.}~\bibnamefont {Isi}},\ and\ \bibinfo {author} {\bibfnamefont {A.}~\bibnamefont {Zimmerman}},\ }\bibfield  {title} {\bibinfo {title} {{Hints of spin-magnitude correlations and a rapidly spinning subpopulation of binary black holes}},\ }\href@noop {} {\  (\bibinfo {year} {2024})},\ \Eprint {https://arxiv.org/abs/2411.02252} {arXiv:2411.02252 [astro-ph.HE]} \BibitemShut {NoStop}%
\bibitem [{\citenamefont {Mancarella}\ and\ \citenamefont {Gerosa}(2025)}]{Mancarella:2025uat}%
  \BibitemOpen
  \bibfield  {author} {\bibinfo {author} {\bibfnamefont {M.}~\bibnamefont {Mancarella}}\ and\ \bibinfo {author} {\bibfnamefont {D.}~\bibnamefont {Gerosa}},\ }\bibfield  {title} {\bibinfo {title} {{Sampling the full hierarchical population posterior distribution in gravitational-wave astronomy}},\ }\href {https://doi.org/10.1103/PhysRevD.111.103012} {\bibfield  {journal} {\bibinfo  {journal} {Phys. Rev. D}\ }\textbf {\bibinfo {volume} {111}},\ \bibinfo {pages} {103012} (\bibinfo {year} {2025})},\ \Eprint {https://arxiv.org/abs/2502.12156} {arXiv:2502.12156 [gr-qc]} \BibitemShut {NoStop}%
\bibitem [{\citenamefont {Lorenzo-Medina}\ and\ \citenamefont {Dent}(2025)}]{Lorenzo-Medina:2024opt}%
  \BibitemOpen
  \bibfield  {author} {\bibinfo {author} {\bibfnamefont {A.}~\bibnamefont {Lorenzo-Medina}}\ and\ \bibinfo {author} {\bibfnamefont {T.}~\bibnamefont {Dent}},\ }\bibfield  {title} {\bibinfo {title} {{A physically modelled selection function for compact binary mergers in the LIGO-Virgo O3 run and beyond}},\ }\href {https://doi.org/10.1088/1361-6382/ad9c0e} {\bibfield  {journal} {\bibinfo  {journal} {Class. Quant. Grav.}\ }\textbf {\bibinfo {volume} {42}},\ \bibinfo {pages} {045008} (\bibinfo {year} {2025})},\ \Eprint {https://arxiv.org/abs/2408.13383} {arXiv:2408.13383 [gr-qc]} \BibitemShut {NoStop}%
\bibitem [{\citenamefont {Talbot}\ and\ \citenamefont {Thrane}(2022)}]{Talbot:2020oeu}%
  \BibitemOpen
  \bibfield  {author} {\bibinfo {author} {\bibfnamefont {C.}~\bibnamefont {Talbot}}\ and\ \bibinfo {author} {\bibfnamefont {E.}~\bibnamefont {Thrane}},\ }\bibfield  {title} {\bibinfo {title} {{Flexible and Accurate Evaluation of Gravitational-wave Malmquist Bias with Machine Learning}},\ }\href {https://doi.org/10.3847/1538-4357/ac4bc0} {\bibfield  {journal} {\bibinfo  {journal} {Astrophys. J.}\ }\textbf {\bibinfo {volume} {927}},\ \bibinfo {pages} {76} (\bibinfo {year} {2022})},\ \Eprint {https://arxiv.org/abs/2012.01317} {arXiv:2012.01317 [gr-qc]} \BibitemShut {NoStop}%
\bibitem [{\citenamefont {Gerosa}\ \emph {et~al.}(2020)\citenamefont {Gerosa}, \citenamefont {Pratten},\ and\ \citenamefont {Vecchio}}]{Gerosa:2020pgy}%
  \BibitemOpen
  \bibfield  {author} {\bibinfo {author} {\bibfnamefont {D.}~\bibnamefont {Gerosa}}, \bibinfo {author} {\bibfnamefont {G.}~\bibnamefont {Pratten}},\ and\ \bibinfo {author} {\bibfnamefont {A.}~\bibnamefont {Vecchio}},\ }\bibfield  {title} {\bibinfo {title} {{Gravitational-wave selection effects using neural-network classifiers}},\ }\href {https://doi.org/10.1103/PhysRevD.102.103020} {\bibfield  {journal} {\bibinfo  {journal} {Phys. Rev. D}\ }\textbf {\bibinfo {volume} {102}},\ \bibinfo {pages} {103020} (\bibinfo {year} {2020})},\ \Eprint {https://arxiv.org/abs/2007.06585} {arXiv:2007.06585 [astro-ph.HE]} \BibitemShut {NoStop}%
\bibitem [{\citenamefont {Callister}\ \emph {et~al.}(2024)\citenamefont {Callister}, \citenamefont {Essick},\ and\ \citenamefont {Holz}}]{Callister:2024qyq}%
  \BibitemOpen
  \bibfield  {author} {\bibinfo {author} {\bibfnamefont {T.~A.}\ \bibnamefont {Callister}}, \bibinfo {author} {\bibfnamefont {R.}~\bibnamefont {Essick}},\ and\ \bibinfo {author} {\bibfnamefont {D.~E.}\ \bibnamefont {Holz}},\ }\bibfield  {title} {\bibinfo {title} {{Neural network emulator of the Advanced LIGO and Advanced Virgo selection function}},\ }\href {https://doi.org/10.1103/PhysRevD.110.123041} {\bibfield  {journal} {\bibinfo  {journal} {Phys. Rev. D}\ }\textbf {\bibinfo {volume} {110}},\ \bibinfo {pages} {123041} (\bibinfo {year} {2024})},\ \Eprint {https://arxiv.org/abs/2408.16828} {arXiv:2408.16828 [astro-ph.HE]} \BibitemShut {NoStop}%
\bibitem [{\citenamefont {Leyde}\ \emph {et~al.}(2024)\citenamefont {Leyde}, \citenamefont {Green}, \citenamefont {Toubiana},\ and\ \citenamefont {Gair}}]{Leyde:2023iof}%
  \BibitemOpen
  \bibfield  {author} {\bibinfo {author} {\bibfnamefont {K.}~\bibnamefont {Leyde}}, \bibinfo {author} {\bibfnamefont {S.~R.}\ \bibnamefont {Green}}, \bibinfo {author} {\bibfnamefont {A.}~\bibnamefont {Toubiana}},\ and\ \bibinfo {author} {\bibfnamefont {J.}~\bibnamefont {Gair}},\ }\bibfield  {title} {\bibinfo {title} {{Gravitational wave populations and cosmology with neural posterior estimation}},\ }\href {https://doi.org/10.1103/PhysRevD.109.064056} {\bibfield  {journal} {\bibinfo  {journal} {Phys. Rev. D}\ }\textbf {\bibinfo {volume} {109}},\ \bibinfo {pages} {064056} (\bibinfo {year} {2024})},\ \Eprint {https://arxiv.org/abs/2311.12093} {arXiv:2311.12093 [gr-qc]} \BibitemShut {NoStop}%
\end{thebibliography}%
